\documentclass[prd,aps,nofootinbib,floatfix,11pt]{revtex4-2}
\usepackage{amsmath}
\usepackage{graphicx}
\usepackage{epsfig}
\usepackage{amssymb}
\usepackage{dsfont}
\usepackage{mathtools}
\usepackage[usenames]{color}
\usepackage{ulem}
\usepackage{bigstrut}
\usepackage{slashed}
\usepackage{multirow}
\usepackage{booktabs}
\usepackage{longtable}

\usepackage[dvipdfmx,colorlinks=true,citecolor=blue,linkcolor=blue,urlcolor=blue]{hyperref}

\allowdisplaybreaks
\usepackage{epstopdf}
\begin{document}
\title{B Meson Semi-Invisible Decays via Perturbative QCD}
\author{Han-Bing Liu}
\author{Ye Xing}
\email{Corresponding author. xingye\_guang@cumt.edu.cn}
\author{Bin Luo}
\affiliation{School of Materials Science and Physics, China University of Mining and Technology, Xuzhou 221000, China}
\begin{abstract}
	This paper focuses on the dark sector decay processes of $B$ mesons ($B\to \mathcal{B}_8\ +$ invisible). Using the perturbative QCD (pQCD) approach, we calculate the upper bounds on the branching ratios for decays from $B$ mesons into light baryons and dark baryons within two distinct $B$-Mesogenesis scenarios. A detailed discussion of the form factor $B\to \mathcal{B}_8$ is presented. Based on the derived form factors and effective couplings, we then reach the final numerical analysis. The results show that these upper bounds can reach sizable, especially for $B^0\to n\psi$  in Type-I scenario, with values of orderis $\mathcal{O}(10^{-5})$ . Such processes are expected to facilitate the search for dark matter at hadron colliders and B factories.
	
\end{abstract}
\maketitle
\section{Introduction}
The $B$-mesogenesis scenario has recently attracted growing attention~\cite{Elor:2018twp,Alonso-Alvarez:2021qfd,Elahi:2021jia,Nelson:2019fln,Miro:2024fid,Zheng:2024tkj,Khodjamirian:2023wol,Lenz:2022rbq}, as it offers a promising framework to simultaneously address two fundamental problems in cosmology: the baryon asymmetry and the origin of dark matter. Unlike conventional high-scale baryogenesis models~\cite{Azatov:2021irb,Bodeker:2020ghk,Alonso-Alvarez:2019fym,Bringmann:2018sbs,Gu:2010ft,Ellis:1978xg,Riotto:1999yt}, which typically involve energy scales much higher than the electroweak scale, approaching the GUT scale, and thus remain challenging to probe experimentally. As a low-scale baryogenesis scenario, the $B$-mesogenesis mechanism involves interactions between the dark baryonic sector and Standard Model particles at scales below the electroweak scale, thereby being directly accessible to experimental probes. Consequently, it can be tested at current hadron colliders and $B$ factories. Indeed, ongoing searches at LHCb and Belle are actively looking for signals of long-lived particles or final states with missing energy in $B$-meson decays~\cite{BESIII:2025sfl,Ai:2024nmn,BaBar:2024qqx,BaBar:2023dtq,BaBar:2023rer,Shi:2022rjn,Belle:2021gmc}.

On the theoretical side, processes where a $B$ meson produces ordinary baryons or dark baryons involve new types of interactions that cannot be described by standard factorizable form factors. There are calculations of $B$-meson dark sector decays using the Light cone sum rule (LCSR) method at leading and higher twists of the nucleon~\cite{Khodjamirian:2022vta,Boushmelev:2023huu,Elor:2022jxy}, as well as higher-twist $B$-meson wave functions~\cite{Biswas:2026oxq}, studies of semi-leptonic invisible $B$-meson decays within the HQE framework~\cite{Shi:2023riy}, and analyses of invisible b-baryon decays based on LCSR~\cite{Shi:2024uqs} or $k_T$ factorizaiton~\cite{Li:2024htn,Xing:2025pfw}. These studies provide valuable references for understanding the $B$-mesogenesis dynamics. In particular, if we adopt the possible \(\Psi\)-triquark interaction proposed in Ref.~\cite{Elor:2018twp,Alonso-Alvarez:2021qfd}, in which the operates at very low temperatures $5\sim 30$ MeV~\cite{Alonso-Alvarez:2021qfd}, and the dark baryon $\psi$ has a modest mass of a few GeV, it can be effectively reduced to a form resembling a four-fermion interaction as the mediator mass much larger than the \(W\)-boson mass. This implies that the transition from B meson to light baryon may involve a large momentum transfer, which is typically much larger than the typical QCD scale. Precisely, the perturbative QCD (pQCD) factorization proves to be an efficient and suitable approach to handle such large recoil processes.

The perturbative QCD (pQCD) method, especially within the $k_T$ factorization framework, provides an effective and well-established approach for hard exclusive processes where the momentum transfer between the initial and final hadrons is large~\cite{Xing:2019xti,Han:2022srw,He:2006ud,Li:2008tk,Rui:2022sdc,Ou-Yang:2025ije,Chang:2025kfa,Zhang:2024rmb,Ren:2023ebq,Chai:2025xuz}. In such regimes, the cross sections or decay amplitudes factorize into hard-scattering kernels calculable in perturbation theory and non-perturbative meson light-cone distribution amplitudes. The $k_T$ factorization further incorporates the transverse wave functions and Sudakov resummation~\cite{Li:1992nu,Li:1994cka,Li:2012cfa}, thereby suppressing soft gluon contributions and extending the applicability of pQCD to lower energy scales as low as a few GeV. This makes the pQCD method particularly suitable for studying $B$-meson to light baryon involving large momentum transfer, where conventional factorization become inadequate.

This paper is structured as follows. Section II presents an  analysis of B-meson dark sector decays. Section III calculates these decays within the pQCD framework. In the numerical analysis of Section IV, we compute the form factors for B-meson decays into light baryons under two possible B-Mesogenesis scenarios. The main results are collected in the tables. A summary is concluded In the final section.
\section{DARK BARYON PRODUCTION IN B-MESOGENESIS }
\begin{figure}
	\centering
	\includegraphics[width=0.60\textwidth]{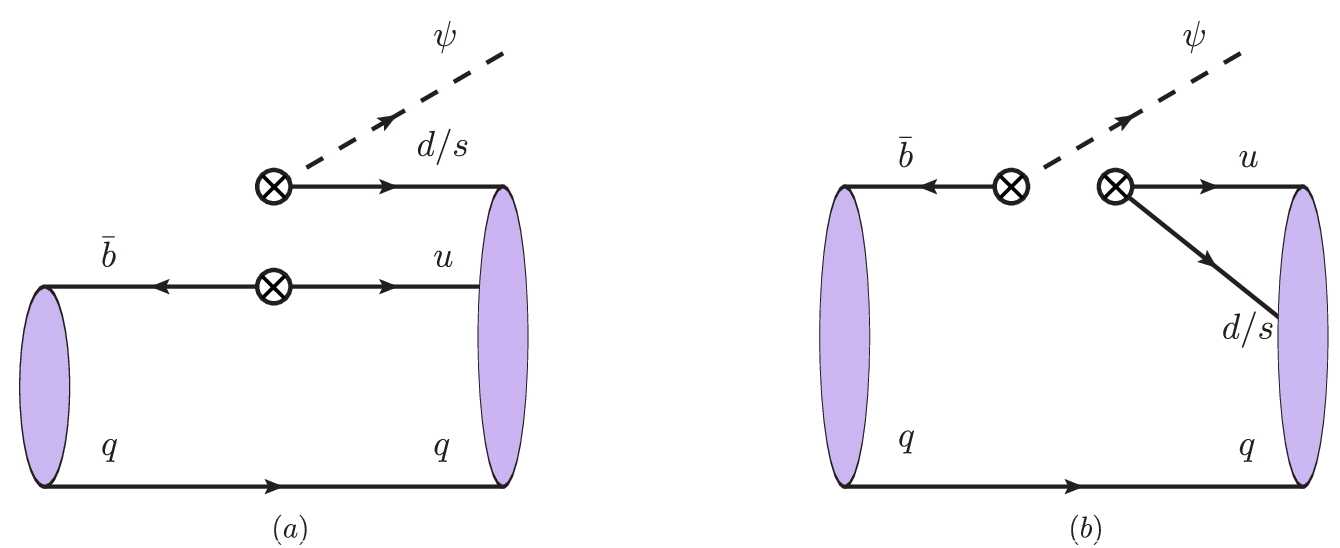}
	\caption{Representative quark-level topologies for
		$B\to\mathcal{B}\psi$ decays in the Type-(a)
		and Type-(b) scenarios, corresponding to
		the Type-I and Type-II models, respectively.
		The crosses denote the effective interaction
		vertices.}
	\label{fig:darkdiag}
\end{figure}

In $B$-Mesogenesis scenarios, a SM quark can couple to a dark baryon carrying baryon number $-1$ via a heavy colored mediator. It can be described by a dimension-6 operator e.g. $\mathcal{O}_{abc}=u_ad_bd_c\psi$. The effective interactions are obtained by integrating out the heavy mediator at the lower energies relevant for hadron decays~\cite{Elor:2018twp}. In the work, we adopt the following two models with scalar mediators (type-I and type-II), in which the dark sector decays of the $b$-quark are described by the factorized dark baryon field $\psi$ in combination with the local three-quark operator,
\begin{align}\label{eq:effhamiltonian}
	&\mathcal{H}_{eff}^{I}=-G_{uq}^{I}\bar{\mathcal{O}}_{uq}^{I}\psi^{c}+h.c.,\quad G_{uq}^{I}=\frac{y_{ub}y_{\phi q}}{M_{Y}^{2}},\quad \bar{\mathcal{O}}_{uq}^{I}=i\epsilon_{ijk}(\bar{b}^j_R C \bar{u}^{iT}_R)\bar{q}^k_R ,\notag\\
	&\mathcal{H}_{eff}^{II}=-G_{uq}^{II}\bar{\mathcal{O}}_{uq}^{II}\psi^{c}+h.c.,\quad G_{uq}^{II}=\frac{y_{\phi b}y_{uq}}{M_{Y}^{2}},\quad \bar{\mathcal{O}}_{uq}^{II}=i\epsilon_{ijk}(\bar{q}^j_R  C \bar{u}^{iT}_R)\bar{b}^k_R,
\end{align}
where the indices $i,j,k$ ensure the operator is antisymmetric in color space. $G^{I,II}_{uq}$$(q=d,s)$  are the effective couplings for the $B$-meson decay into a light baryon and a dark baryon. $q_{R/L}$ denote the left- and right-handed chiral fields, defined as $q_{R/L}=\hat{P}_{R/L} q=(1\pm\gamma^5)/2\ \! q$.
The representative quark-level topologies for the Type-I and Type-II scenarios are shown in Fig.~\ref{fig:darkdiag}, respectively. The possible decay channels are summarized in Table.~\ref{tab:decay_channels}, where (a) and (b) denote the contributions from the Type-I and Type-II scenarios, respectively, while topologies (a)+(b) indicates that the decay channel is allowed in both scenarios. Notably, the deviation between spin-1 $ud$ pair of $\Sigma^0$ and the scalar mediators, process $B_s^0 \to \Sigma^0 \psi$ is forbidden in type-I under angular momentum conservation.
\begin{table}[htbp]
	\centering
	\caption{The allowed $B\to\mathcal B\psi$ decay channels and their corresponding quark-level topologies shown in Fig.~\ref{fig:darkdiag}.}
	\label{tab:decay_channels}
	\begin{tabular}{cccccccc}
		\hline
		Channel & Topology & Channel & Topology
		& Channel & Topology & Channel & Topology \\
		\hline
		$B_s^0 \to \Lambda^0\psi$ & $(a)+(b)$
		& $B_s^0 \to \Xi^0\psi$ & $(a)+(b)$
		& $B^0 \to n\psi$ & $(a)+(b)$
		& $B^+ \to p\psi$ & $(a)+(b)$ \\
		
		$B_s^0 \to \Sigma^0\psi$ & $(b)$
		& $B^0 \to \Lambda^0\psi$ & $(a)+(b)$
		& $B^0 \to \Sigma^0\psi$ & $(a)+(b)$
		& $B^+ \to \Sigma^+\psi$ & $(a)+(b)$ \\
		\hline
	\end{tabular}
\end{table}
\section{perturbative calculation}
\begin{figure}
	\centering
	\includegraphics[width=0.98\textwidth]{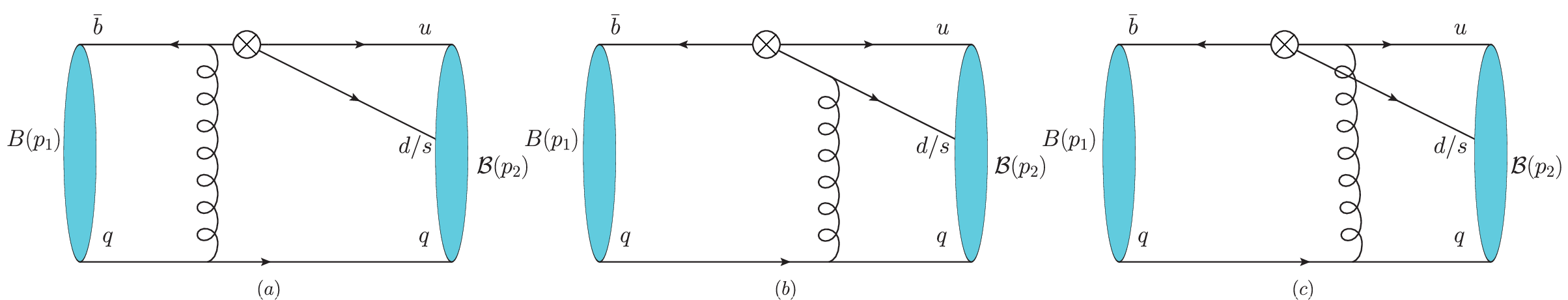}
	\caption{Typical leading-order Feynman diagrams for the transition $B\to \mathcal{B}$. The crosses in the figure represent the effective vertices of the three-quark operator. At large recoil, the leading-order contribution comes from three hard gluons exchange diagrams.}
	\label{fig:feynmandiagram}
\end{figure}

According to the effective Hamiltonian given in Eq.~\ref{eq:effhamiltonian}, the amplitude for the process $B \to \mathcal{B}\psi$ can be written as
\begin{align}
	\mathcal{M}(B\to \mathcal{B}\psi)=G_{uq}^{I/II}\langle \mathcal{B}(p_2,s_{\mathcal{B}})|\bar{\mathcal{O}}_{uq}^{I/II}|B(p_1)\rangle u_{\psi}(q,s_{\psi}),
\end{align}	
where $u_{\psi}$ is the spinor of dark baryon $\psi$, and momentum transfer is defined as $q=p_1-p_2$. The amplitude depends on the form factor of the large recoil transition ($B\to \mathcal{B}$), which can be parameterized as:
\begin{align}
	\langle \mathcal{B}(p_2,s_{\mathcal{B}})|\bar{\mathcal{O}}_{uq}^{I/II}|B(p_1,s_B)\rangle
	= \bar{u}_{\mathcal{B}}(p_2,s_{\mathcal{B}})
	\Bigg[
	\overline{F}_1(q^2)
	+ \overline{F}_2(q^2)\frac{\slashed{p}_1}{m_\mathcal{B}}
	+ \overline{F}_3(q^2)\frac{\slashed{p}_2}{m_\mathcal{B}}
	+ \overline{F}_4(q^2)\frac{\slashed{p}_2 \slashed{p}_1}{m_\mathcal{B}^2}
	\Bigg] \hat{P}_L,
\end{align}
Using the Dirac equation for final-state baryon, only two independent form factors remain, which corresponding with $F_{B\to \mathcal{B}_R}^{(d/b)}(q^2)$ and $\widetilde{F}_{B\to \mathcal{B}_L}^{(d/b)}(q^2)$ in Ref~\cite{Khodjamirian:2022vta},
\begin{align}
\langle \mathcal{B}(p_2,s_{\mathcal{B}})|\bar{\mathcal{O}}_{uq}^{I/II}|B(p_1)\rangle
&=\bar u_R\Big(\overline F_1(q^2)+\overline F_2(q^2)+\overline F_3(q^2)+\overline F_4(q^2)\Big)+\bar u_L\Big(\overline F_2(q^2)+\overline F_4(q^2)\Big)\frac{\slashed{q}}{m_{\mathcal{B}}},\notag\\
&=\bar u_R F_{B\to \mathcal{B}_R}^{(d/b)}(q^2)+\bar u_L \widetilde{F}_{B\to \mathcal{B}_L}^{(d/b)}(q^2)\frac{\slashed{q}}{m_{\mathcal{B}}},
\end{align}
The superscripts (d) and (b) in $F_{B\to \mathcal{B}_{R}}/\widetilde{F}_{B\to \mathcal{B}_{L}}$ correspond to the type-I and type-II models, respectively. To handle the hadronic matrix with large recoil momentum, the perturbative QCD approach with transverse momentum $k_T$ factorization is reliable. Generally, the transition matrix element can be factorized into convolution integrals of the hard scattering kernel  and the non-perturbative wave functions,
\begin{align}
	\mathcal{M}=\int d[x_i] d[\textbf{b}_i] \Phi_B(x_1,\textbf{b}_1) h(x_i,\textbf{b}_i) \Phi_{\mathcal{B}} (x_3,x_4,\textbf{b}_3,\textbf{b}_4) E(t)
\end{align}
Here $h_i(x_i,\textbf{b}_i)$ is the perturbatively calculable hard kernel, which can be computed from relevant Feynman diagrams in Fig.~\ref{fig:feynmandiagram}. $x_i$ and $\textbf{b}_i$ denote the longitudinal momentum fractions and the conjugate variable to the quark's transverse momentum $k_{i\perp}$, respectively. $E(t)$ is correlated with the threshold resummation factor that suppresses the large logarithmic contributions near the endpoint region.

It is convenient to perform the study in light-cone coordinates with two light cone vector $n=(1,0,0_{\perp}),\bar n=(0,1,0_{\perp})$. Specifically, the initial $B$ meson is taken to be at rest, while the final-state baryon with large recoil momentum moves along the light-cone direction n. The momenta of the hadrons and their valence quarks are then taken as follows,
\begin{equation}
\begin{aligned}
	&p_1(B) = \frac{m_{B}}{\sqrt{2}} (1,1,0_\perp),\quad  p_2(\mathcal{B}) = \frac{m_{B}}{\sqrt{2}} (\eta,0,0_\perp), \\
	&k_1 = \frac{m_{B}}{\sqrt{2}} (1,\bar{x}_1,-k_{1\perp}),\quad   k_2 = \frac{m_{B}}{\sqrt{2}} (0,x_1,k_{1\perp}),\\
	& k_3 = \frac{m_{B}}{\sqrt{2}} (x_2\eta,0,k_{2\perp}),\quad
	k_4 = \frac{m_{B}}{\sqrt{2}} (x_3\eta,0,k_{3\perp}),\quad  k_5 = \frac{m_{B}}{\sqrt{2}} (x_4\eta,0,k_{4\perp}) .
\end{aligned}
\end{equation}
where $p_1$ and $p_2$ denote the momentums of $B$ meson and baryon $\mathcal{B}$, $k_1$ is the heavy quark $\bar{b}$ momentum, $k_3$, $k_4$, and $k_5$ are the three light quark momentums in baryon. $x_i$s are their momentum fractions, where $k_{i\perp}$ denote the transverse components in the light-cone decomposition. The parameter $\eta=1-q^2/m_{B}^2$.

To form the energetic final baryon, three possible hard gluon exchange diagrams can be constructed at the leading order, as illustrated in Fig.~\ref{fig:feynmandiagram}. Following the standard pQCD procedure, we can perturbatively calculate the hard kernels corresponding to these three diagrams. For the non-perturbative input wave functions, we adopt the light baryon light cone distribution amplitude (LCDA) up to twist-6 and the leading-order $B$-meson LCDA. After performing the convolution of the hard kernels with the wave functions, we finally obtain the transition element $B\to \mathcal{B}$ expressed in two light-cone vector:
\begin{align}
	\langle \mathcal{B}(p_2,s_{\mathcal{B}})|\bar{\mathcal{O}}_{uq}^{I/II}|B(p_1,s_B)\rangle
	= \bar{u}_{\mathcal{B}}(p_2,s_{\mathcal{B}})
	\Bigg[
	F_1(q^2)
	+ F_2(q^2)\slashed{n}
	+ F_3(q^2)\slashed{\bar{n}}
	+ F_4(q^2)\slashed{\bar{n}}\slashed{n}
	\Bigg] \hat{P}_L.
\end{align}
The explicit forms of $F_i$s can be extracted by matching to the pQCD calculation results in Appendix. In this work, we precisely represent each $F_i$s. The four form factors obtained from the light-cone expansion are not independent, they can also be related to the two independent parameters above,
\begin{align}
	&F_{B\to \mathcal{B}_R}^{(d/b)}(q^2)=F_1(q^2)
	+\frac{\sqrt{2}m_{\mathcal{B}}}{m_B}(1-\frac{1}{\eta}) F_2(q^2)
	+\frac{\sqrt{2}m_{\mathcal{B}}}{m_B \eta} F_3(q^2)
	+\frac{2m_{\mathcal{B}}^2}{m_B^2 \eta}F_4(q^2), \nonumber\\
	&\widetilde{F}_{B\to \mathcal{B}_L}^{(d/b)}(q^2)
	=\frac{\sqrt{2}m_{\mathcal{B}}}{m_B}\, F_2(q^2)
	+\frac{2m_{\mathcal{B}}^2}{m_B^2 \eta}F_4(q^2).
	\label{eq:formfactor_pqcd}
\end{align}
The decay width of $B\to \mathcal{B}\psi$ can then be represented as
\begin{align}
	\Gamma=&\frac{G_{uq}^2\sqrt{m_B^4-2m_B^2m_{\mathcal B}^2+m_{\mathcal B}^4-2m_B^2m_\psi^2-2m_{\mathcal B}^2m_\psi^2+m_\psi^4}}
	{16\pi m_B^3}\notag\\
	&\times\left[4\widetilde{F}^{(d/b)}_LF^{(d/b)}_Rm_\psi^2+(F^{(d/b)}_R)^2(m_B^2-m_{\mathcal B}^2-m_\psi^2)
	-\frac{({\widetilde{F}^{(d/b)}}_L)^2m_\psi^2(-m_B^2+m_{\mathcal B}^2+m_\psi^2)}
	{m_{\mathcal B}^2}\right].
\end{align}

\section{Numerical Analysis}
\subsection{Form factors}
The remarkable hadronic transition $B\to \mathcal{B}$ can be described by form factors $F_i$s in the dark sector decays of $B$ meson. In order to compute the element properly, we will employ the leading-order LCDAs for the $B$ meson, and from twist-3 up to twist-6 for the light baryon. Their explicit expressions are provided in the Appendix.

\begin{figure}[htbp]
	\centering
	\includegraphics[width=0.99\textwidth]{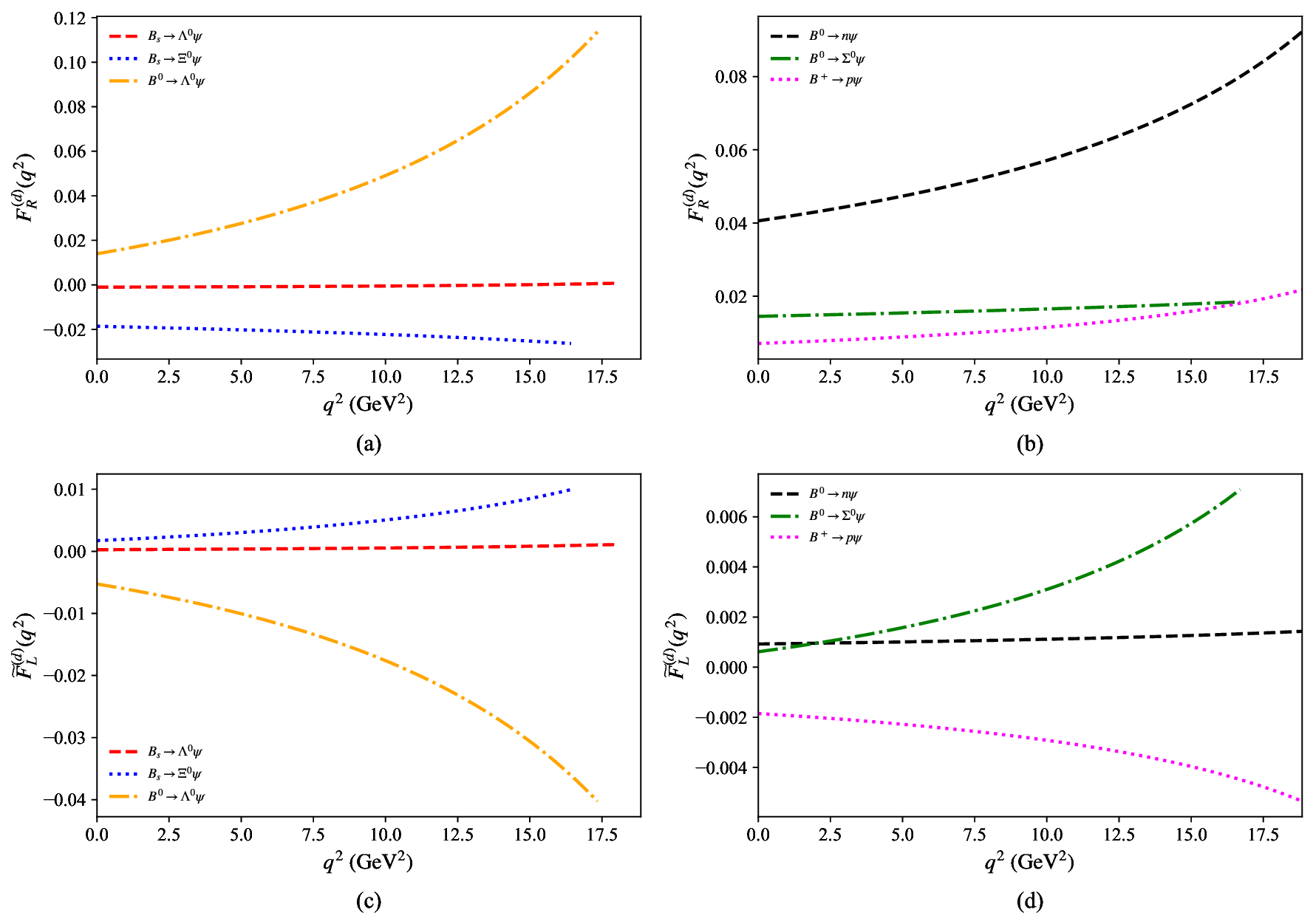}
	\caption{
		The $q^2$ dependence of the form factors $F_{R}^{(d)}(q^2)$ and $\widetilde{F}_{L}^{(d)}(q^2)$  for different processes, with contributions from light-baryon LCDAs of twist-3 to twist-6.}
\label{fig:formfactorq2d}
\end{figure}

\begin{figure}[htbp]
	\centering
	\includegraphics[width=0.99\textwidth]{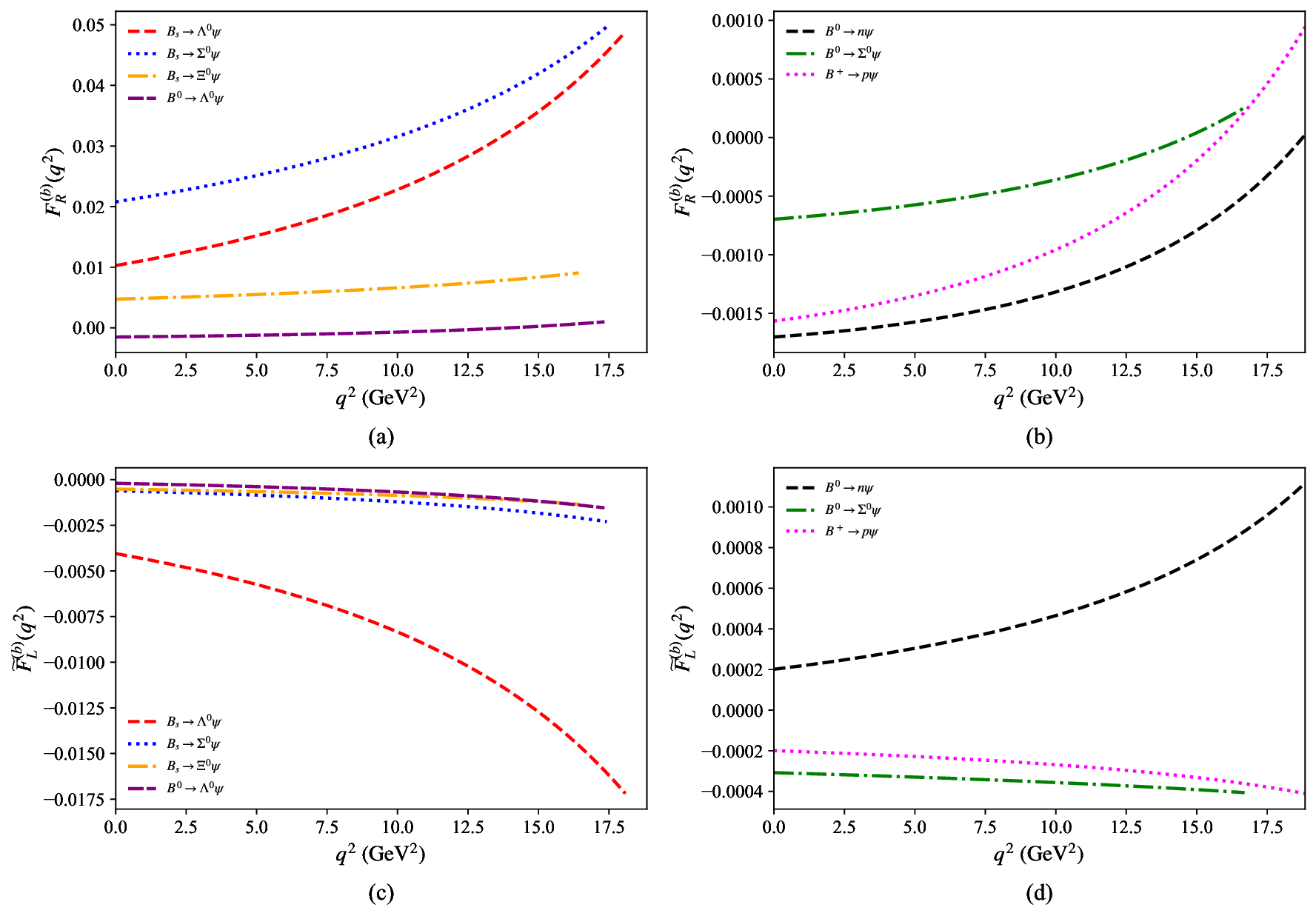}
	\caption{The $q^2$ dependence of the form factors $F_{R}^{(b)}(q^2)$ and $\widetilde{F}_{L}^{(b)}(q^2)$ for different processes, with contributions from light-baryon LCDAs of twist-3 to twist-6.}
\label{fig:formfactorq2b}
\end{figure}

The pQCD approach can reliably handle large-recoil processes (small $q^2$). In our numerical calculation, ten equally spaced values of $\eta$ are taken in the range $0.6\leq\eta\leq1.0$, corresponding to
	$0\leq q^2\leq0.4m_B^2$ through $q^2=m_B^2(1-\eta)$.
	To obtain the form factors in the full kinematic region
	$0\leq q^2\leq q_{\max}^2=(m_B-m_{\mathcal B})^2$, an extrapolation is necessary. We therefore adopt the following BCL $z$-series parametrization~\cite{Boushmelev:2023huu}.

\begin{equation}
		F(q^2)=
		\frac{F(0)}{1-q^2/m_{\Lambda_b}^2}\left\{1+a_1\left[z(q^2)-z(0)+\frac{1}{2}\left(z(q^2)^2-z(0)^2\right)\right]
		\right\}.
\end{equation}
where $F_i$s correspond to the form factors $F_{B\to \mathcal{B}_R}^{(d/b)}, \widetilde{F}_{B\to \mathcal{B}_L}^{(d/b)}$.
The variable $z(q^2)$ is defined as
\begin{equation}
	z(q^2) = \frac{\sqrt{t_+ - q^2} - \sqrt{t_+ - t_0}}
	{\sqrt{t_+ - q^2} + \sqrt{t_+ - t_0}},
\end{equation}
with $t_0 = (m_B + m_{\mathcal{B}})(\sqrt{m_{B}}-\sqrt{m_{\mathcal{B}}})^2$ and
$t_+ = (m_B + m_{\mathcal{B}})^2$. However, with this choice of $t_+$, the isolated $\Lambda_b$ pole at
	$q^2=m_{\Lambda_b}^2$ lies inside the unit disk of the $z$ plane. The form factor at $q^2=0$ can be determined directly from the pQCD
	expressions given above, while the slope parameter $a_1$ is obtained
	from the fitting procedure. For simplicity, we restrict the $z$-series
	expansion to $n=1$, and the fit results of the two scenarios are
	listed in Tab.~\ref{tab:formfactors}. The $q^2$ behavior of form factors
	is also provided, as illustrated in Fig.~\ref{fig:formfactorq2d} and
	Fig.~\ref{fig:formfactorq2b} for the Type-I and Type-II scenarios.
	It can be seen that sizable differences exist between the present pQCD
	results and the available LCSR results for some form factors. Although
	the comparison is performed at the same value of $q^2$, the two approaches
	are based on different theoretical frameworks, approximations, and regions
	of applicability. The pQCD calculation is based on $k_T$ factorization
	and is most reliable in the large-recoil region, whereas the LCSR approach
	treats the nonperturbative dynamics through a light-cone expansion of
	correlation functions. In addition, the two approaches adopt different
	baryon LCDAs and treatments of higher-twist effects, which can further
	contribute to the sizable differences in the form factors.
	
We compare our form factors at $q^2=0$ with the predictions from previous light-cone sum rule (LCSR) studies, as summarized in Table II. For the $B^+\to p\psi$ decay, the pQCD results obtained by including proton LCDAs from twist-3 to twist-6 are compared with three sets of LCSR predictions below. The numerical values of the form factors at $q^2=0$ (in units of $10^{-2}\,\mathrm{GeV}^2$) are given by
\begin{eqnarray}
&\text{LCSR}\cite{Khodjamirian:2022vta,Boushmelev:2023huu}
: &F_{{B\to p\psi}_R}^{(d)}=2.2,\ F_{{B\to p\psi}_R}^{(b)}=-4.05,\ \widetilde{F}_{{B\to p\psi}_L}^{(d)}=0.55,\ \widetilde{F}_{{B\to p\psi}_L}^{(b)}=-0.65,\notag\\
&\text{LCSR}\cite{Biswas:2026oxq}
: &F_{{B\to p\psi}_R}^{(d)}=1.3,\ F_{{B\to p\psi}_R}^{(b)}=2.6,\ \widetilde{F}_{{B\to p\psi}_L}^{(d)}=0.245,\ \widetilde{F}_{{B\to p\psi}_L}^{(b)}=0.49,\\
&\text{LCSR}\cite{Elor:2022jxy}
: &F_{{B\to p\psi}_R}^{(d)}=1.94,\ F_{{B\to p\psi}_R}^{(b)}=-0.92,\ \widetilde{F}_{{B\to p\psi}_L}^{(d)}=0.059,\ \widetilde{F}_{{B\to p\psi}_L}^{(b)}=0.059,\notag\\
&\text{pQCD}: &F_{{B\to p\psi}_R}^{(d)}=0.708,\ F_{{B\to p\psi}_R}^{(b)}=-0.157,\ \widetilde{F}_{{B\to p\psi}_L}^{(d)}=-0.185,\ \widetilde{F}_{{B\to p\psi}_L}^{(b)}=-0.020.\notag
\end{eqnarray}

\begin{figure}[h]
	\centering
	\includegraphics[width=0.99\textwidth]{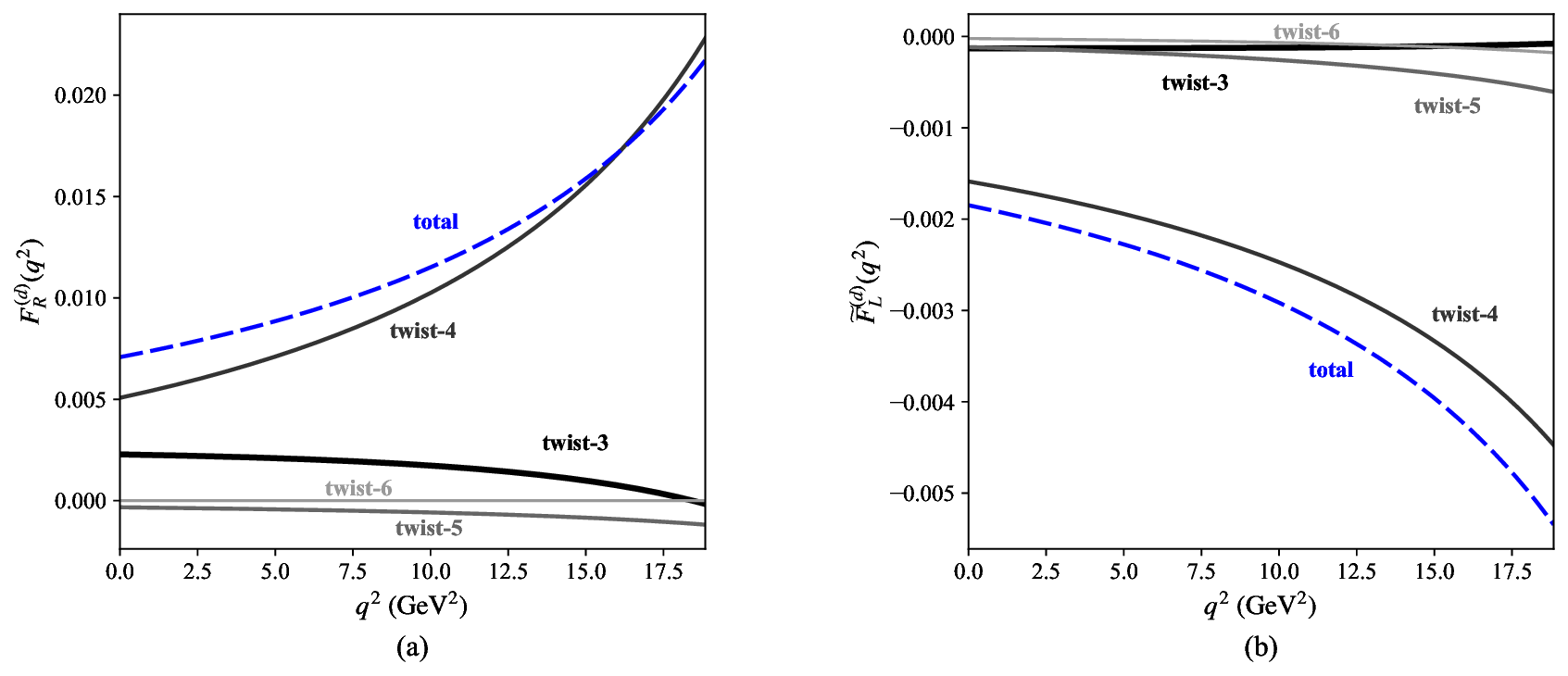}
\caption{The $q^2$ dependence of the form factors $F_{R}^{(d)}(q^2)$ and $\widetilde{F}_{L}^{(d)}(q^2)$ for $B^+\to p\psi$. The solid curves represent individual twist contributions, and the blue dashed curves represent the total results.}
\label{fig:Btopff}
\end{figure}

To further identify the contributions from different twist, we present the $q^2$ dependence of the $B^+\to p\psi$ form factors in Fig.~\ref{fig:Btopff} and their individual contributions at $q^2=0$ in Tab.~\ref{tab:Btopff}. The results show that twist-3 and twist-4 LCDAs provide the dominant contributions, while the endpoint behavior of the convolution integrals is suggested to play an important role~\cite{Han:2022srw}.
For instance, defining $x_2=1-x_3-x_4$, the twist-3 and twist-4 contributions of form factor can be deduced as
$F^{\rm tw3}\sim x_2x_3x_4 P_3$,
$F^{\rm tw4}\sim x_2x_3 P_4,$
where $P_3$ and $P_4$ denote the corresponding combinations of proton LCDAs and hard-scattering coefficients after factoring out the explicit endpoint dependence. Consequently,
$\frac{F^{\rm tw4}}{F^{\rm tw3}}\sim
\frac{1}{x_4}\frac{P_4}{P_3}$,
up to mass-dependent power factors. The relevant twist-4 structures are therefore less suppressed than the twist-3 contribution in the small-\(x_4\) region. This motivates a closer examination of the momentum-fraction dependence of the hard-scattering kernels.
\begin{table}[t]
	\centering
	\caption{Form factors at $q^2=0$ (in units of $10^{-2}\,\mathrm{GeV}^2$) and fitted
			slope parameters $a_1$ for the decay channels in the Type-I and Type-II scenarios.
			The uncertainties arise from the variation of $\Lambda_{\rm QCD}$ and
			the wave function parameters.}
	\label{tab:formfactors}
	\begin{tabular}{cc|ccc|c|c}
		\hline\hline
		\multicolumn{2}{c|}{ }
		& \multicolumn{3}{|c|}{this work }
		& LCSR\cite{Elor:2022jxy}
		& LCSR\cite{Khodjamirian:2022vta,Boushmelev:2023huu,Abri:2026rqj} \\
		\cline{3-5}\cline{6-7}
		\multicolumn{2}{c|}{ }
		& twist(3) & twist(3-6) & $a_1$
		& twist(3) & twist(3-6) \\
		\hline
		
		\multirow{4}{*}{$B^+\to p$}
		& $F_{B\to p_R}^{(d)}(0)$
		& $0.228_{-0.097}^{+0.095}$
		& $0.708_{-0.145}^{+0.141}$
		& $-1.419_{-0.868}^{+0.735}$
		& 1.94 & 2.2 \\
		
		& $\widetilde{F}_{B\to p_L}^{(d)}(0)$
		& $-0.013_{-0.006}^{+0.005}$
		& $-0.185_{-0.020}^{+0.027}$
		& $-0.998_{-0.617}^{+0.519}$
		& 0.059 & 0.55 \\
		
		& $F_{B\to p_R}^{(b)}(0)$
		& $0.052_{-0.007}^{+0.006}$
		& $-0.157_{-0.096}^{+0.090}$
		& $7.456_{-0.193}^{+0.840}$
		& $-0.92$ & $-4.05$ \\
		
		& $\widetilde{F}_{B\to p_L}^{(b)}(0)$
		& $-0.004_{-0.001}^{+0.001}$
		& $-0.020_{-0.001}^{+0.001}$
		& $1.018_{-0.857}^{+1.005}$
		& 0.059 & $-0.65$ \\
		\hline
		
		\multirow{2}{*}{$B_s^0 \to \Sigma^0$}
		& $F_{B\to \Sigma^0_R}^{(b)}(0)$
		& $1.139_{-0.031}^{+0.023}$
		& $2.078_{-0.142}^{+0.104}$
		& $-0.551_{-0.041}^{+0.018}$
		& $-$ & $-$ \\
		
		& $\widetilde{F}_{B\to \Sigma^0_L}^{(b)}(0)$
		& $-0.045_{-0.001}^{+0.001}$
		& $-0.061_{-0.001}^{+0.002}$
		& $-5.408_{-0.642}^{+0.627}$
		& $-$ & $-$ \\
		\hline
		
		\multirow{4}{*}{$B_s^0 \to \Xi^0$}
		& $F_{B\to \Xi^0_R}^{(d)}(0)$
		& $-1.697_{-0.143}^{+0.165}$
		& $-1.858_{-0.142}^{+0.168}$
		& $2.797_{-0.180}^{+0.390}$
		& 4.69 & $-$ \\
		
		& $\widetilde{F}_{B\to \Xi^0_L}^{(d)}(0)$
		& $0.159_{-0.009}^{+0.007}$
		& $0.171_{-0.023}^{+0.031}$
		& $-15.537_{-1.752}^{+1.854}$
		& 0.095 & $-$ \\
		
		& $F_{B\to \Xi^0_R}^{(b)}(0)$
		& $0.210_{-0.014}^{+0.013}$
		& $0.473_{-0.012}^{+0.007}$
		& $0.707_{-0.253}^{+0.222}$
		& $-1.83$ & $-$ \\
		
		& $\widetilde{F}_{B\to \Xi^0_L}^{(b)}(0)$
		& $-0.028_{-0.002}^{+0.002}$
		& $-0.051_{-0.003}^{+0.004}$
		& $-2.208_{-0.800}^{+0.541}$
		& 0.095 & $-$ \\
		\hline
		
		\multirow{4}{*}{$B^0 \to \Lambda$}
		& $F_{B\to \Lambda_R}^{(d)}(0)$
		& $-1.285_{-0.009}^{+0.012}$
		& $1.397_{-0.280}^{+0.286}$
		& $-19.352_{-2.389}^{+1.141}$
		& 4.09 & $-4.51$ \\
		
		& $\widetilde{F}_{B\to \Lambda_L}^{(d)}(0)$
		& $0.097_{-0.005}^{+0.003}$
		& $-0.526_{-0.072}^{+0.073}$
		& $-17.795_{-1.189}^{+0.030}$
		& $-$ & $0.115$ \\
		
		& $F_{B\to \Lambda_R}^{(b)}(0)$
		& $0.055_{-0.007}^{+0.007}$
		& $-0.152_{-0.003}^{+0.004}$
		& $9.384_{-0.001}^{+0.198}$
		& 4.09 & $-0.462$ \\
		
		& $\widetilde{F}_{B\to \Lambda_L}^{(b)}(0)$
		& $-0.002_{-0.001}^{+0.001}$
		& $-0.020_{-0.003}^{+0.002}$
		& $-18.027_{-1.224}^{+0.904}$
		& $-$ & 0.110 \\
		\hline
		
		\multirow{4}{*}{$B^0 \to n$}
		& $F_{B\to n_R}^{(d)}(0)$
		& $1.070_{-0.149}^{+0.135}$
		& $4.059_{-0.911}^{+0.840}$
		& $0.501_{-0.044}^{+0.556}$
		& 1.94 & $-$ \\
		
		& $\widetilde{F}_{B\to n_L}^{(d)}(0)$
		& $-0.054_{-0.001}^{+0.001}$
		& $0.092_{-0.045}^{+0.034}$
		& $2.256_{-1.689}^{+9.556}$
		& $-$ & $-$ \\
		
		& $F_{B\to n_R}^{(b)}(0)$
		& $0.078_{-0.018}^{+0.019}$
		& $-0.170_{-0.115}^{+0.108}$
		& $6.035_{-1.578}^{+8.143}$
		& 1.94 & $-$ \\
		
		& $\widetilde{F}_{B\to n_L}^{(b)}(0)$
		& $-0.006_{-0.001}^{+0.001}$
		& $0.020_{-0.008}^{+0.007}$
		& $-7.447_{-0.676}^{+2.028}$
		& $-$ & $-$ \\
		\hline
		
		\multirow{4}{*}{$B^0 \to \Sigma^0$}
		& $F_{B\to \Sigma^0_R}^{(d)}(0)$
		& $0.043_{-0.002}^{+0.002}$
		& $1.449_{-0.082}^{+0.079}$
		& $3.157_{-0.121}^{+0.748}$
		& $-$ & $-$ \\
		
		& $\widetilde{F}_{B\to \Sigma^0_L}^{(d)}(0)$
		& $-0.003_{-0.001}^{+0.001}$
		& $0.061_{-0.047}^{+0.062}$
		& $-34.902_{-95.263}^{+14.792}$
		& $-$ & $-$ \\
		
		& $F_{B\to \Sigma^0_R}^{(b)}(0)$
		& $0.031_{-0.016}^{+0.017}$
		& $-0.070_{-0.012}^{+0.008}$
		& $9.235_{-3.135}^{+1.704}$
		& $-$ & $-$ \\
		
		& $\widetilde{F}_{B\to \Sigma^0_L}^{(b)}(0)$
		& $-0.007_{-0.002}^{+0.002}$
		& $-0.031_{-0.002}^{+0.004}$
		& $2.979_{-0.903}^{+1.469}$
		& $-$ & $-$ \\
		\hline
		
		\multirow{4}{*}{$B_s^0 \to \Lambda$}
		& $F_{B\to \Lambda_R}^{(d)}(0)$
		& $0.002_{-0.001}^{+0.001}$
		& $-0.103_{-0.039}^{+0.047}$
		& $9.344_{-2.592}^{+11.313}$
		& $-$ & $-$ \\
		
		& $\widetilde{F}_{B\to \Lambda_L}^{(d)}(0)$
		& $-0.001_{-0.001}^{+0.001}$
		& $0.025_{-0.002}^{+0.001}$
		& $-6.067_{-0.321}^{+0.657}$
		& $-$ & $-$ \\
		
		& $F_{B\to \Lambda_R}^{(b)}(0)$
		& $0.061_{-0.003}^{+0.004}$
		& $1.028_{-0.057}^{+0.089}$
		& $-7.344_{-0.094}^{+0.001}$
		& $-$ & $-$ \\
		
		& $\widetilde{F}_{B\to \Lambda_L}^{(b)}(0)$
		& $-0.002_{-0.000}^{+0.000}$
		& $-0.405_{-0.018}^{+0.017}$
		& $-5.801_{-1.624}^{+0.982}$
		& $-$ & $-$ \\
		\hline
		
		\multirow{4}{*}{$B^+ \to \Sigma^+$}
		& $F_{B\to \Sigma^+_R}^{(d)}(0)$
		& $0.844_{-0.056}^{+0.035}$
		& $1.444_{-0.076}^{+0.086}$
		& $3.135_{-0.068}^{+0.774}$
		& 3.12 & $-$ \\
		
		& $\widetilde{F}_{B\to \Sigma^+_L}^{(d)}(0)$
		& $-0.061_{-0.005}^{+0.008}$
		& $0.061_{-0.047}^{+0.064}$
		& $-34.669_{-102.387}^{+15.212}$
		& 0.093 & $-$ \\
		
		& $F_{B\to \Sigma^+_R}^{(b)}(0)$
		& $0.033_{-0.014}^{+0.013}$
		& $-0.068_{-0.014}^{+0.009}$
		& $9.904_{-4.218}^{+1.601}$
		& $-1.73$ & $-$ \\
		
		& $\widetilde{F}_{B\to \Sigma^+_L}^{(b)}(0)$
		& $-0.007_{-0.002}^{+0.002}$
		& $-0.031_{-0.002}^{+0.004}$
		& $2.926_{-0.934}^{+1.595}$
		& 0.093 & $-$ \\
		\hline\hline
	\end{tabular}
\end{table}
\begin{table}[htbp]
	\centering
\caption{Contributions from proton light-cone distribution
	amplitudes (LCDAs) of twists 3--6 to the form factors
	$F_{B\to p_R}^{(d/b)}(0)$ and
	$\widetilde{F}_{B\to p_L}^{(d/b)}(0)$
	(in units of $10^{-2}\,\mathrm{GeV}^2$)
	for the $B^+\to p\psi$ decay in the
	Type-I and Type-II scenarios.}
	\label{tab:Btopff}
	\begin{tabular}{c c c c c c}
		\hline\hline
		& Twist-3 & Twist-4 & Twist-5 & Twist-6 & Total \\
		\hline
		$F^{(d)}_{B \to p_R}(0)$
		& 0.2282 & 0.5073 & -0.0326 & 0.0 & 0.7080 \\
		$\widetilde{F}^{(d)}_{B \to p_L}(0)$
		& -0.0130 & -0.1587 & -0.0116 & -0.0022 & -0.1848 \\
		$F^{(b)}_{B \to p_R}(0)$
		& 0.0524 & -0.3037 & 0.0966 & 0.0 & -0.1566 \\
		$\widetilde{F}^{(b)}_{B \to p_L}(0)$
		& -0.0039 & -0.0097 & -0.0066 & 0.0002 & -0.0199 \\
		\hline\hline
	\end{tabular}
\end{table}
\subsection{Branching ratios}
The effective coupling $G_{uq}^{I/II}$s are defined by short distance Wilson coefficients for the effective Hamiltonian in Eq.~\ref{eq:effhamiltonian}. Usually, they can  be constrained by LHC experiments, which provide upper limits on them~\cite{Alonso-Alvarez:2021qfd}.
\begin{equation}	
\begin{aligned}
	\text{Type-I:}\quad
	G_{ud}^2 &< (1.024 \pm 0.10)\times 10^{-13}\ \mathrm{GeV}^{-4}, \\
	G_{us}^2 &< (1.024 \pm 0.10)\times 10^{-13}\ \mathrm{GeV}^{-4}, \\
	\text{Type-II:}\quad
    G_{ud}^2 &< (4.624 \pm 0.45)\times 10^{-15}\ \mathrm{GeV}^{-4}, \\
    G_{us}^2 &< (3.610 \pm 0.36)\times 10^{-14}\ \mathrm{GeV}^{-4}.
\end{aligned}	\label{eq:G}
\end{equation}

\begin{figure}[htbp]
	\centering
	\includegraphics[width=0.9\textwidth]{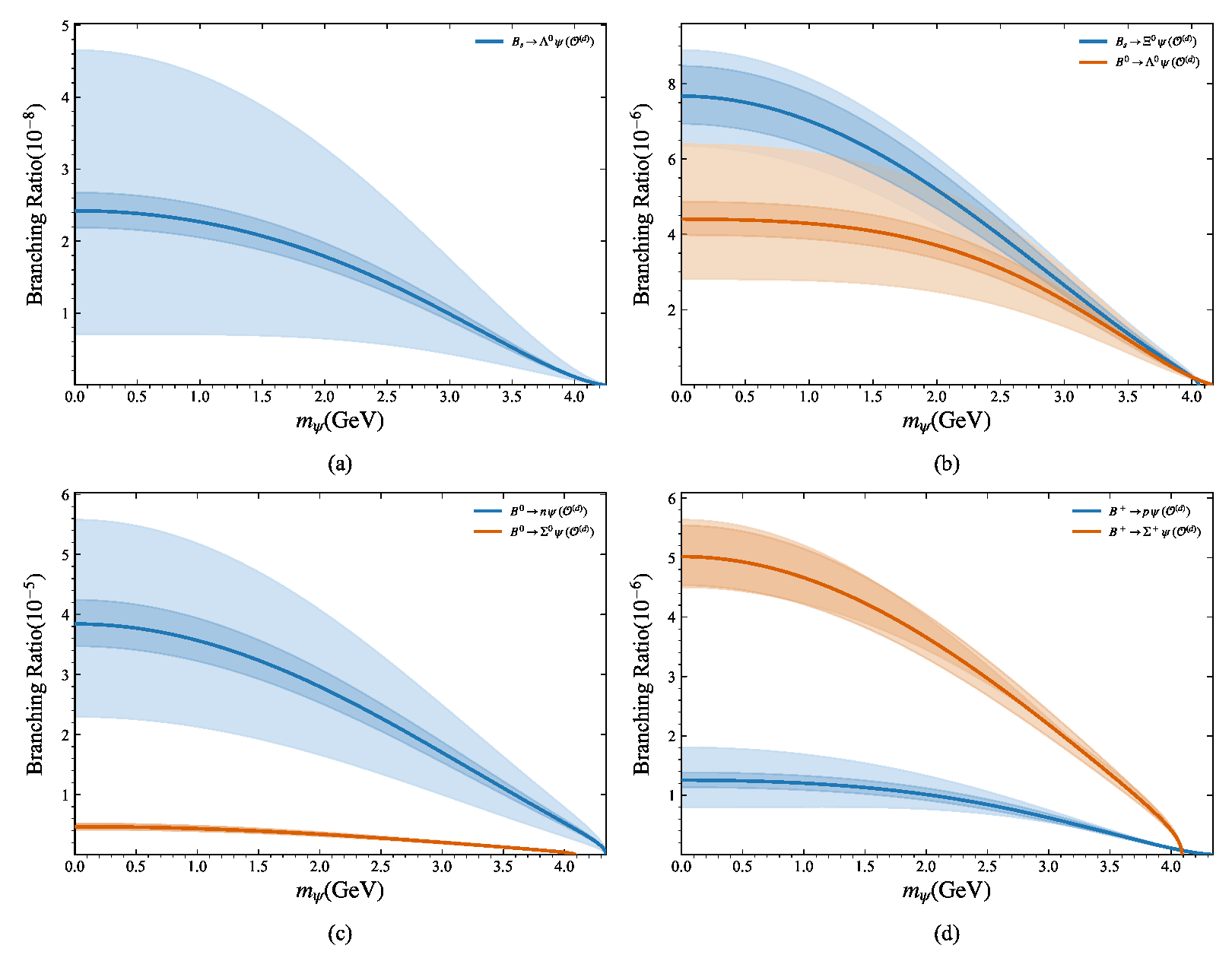}
	\caption{The dark baryon mass $m_{\psi}$-dependence of the upper bounds on the branching ratios for different processes in type-I scenario. The dark- and light-shaded regions correspond to the uncertainty from effective couplings $G_{uq}^{I}$ and form factors $F_{L/R}^{(d)}$, respectively.}
\label{fig:branchingratios1}
\end{figure}

\begin{figure}[htbp]
	\centering
	\includegraphics[width=0.9\textwidth]{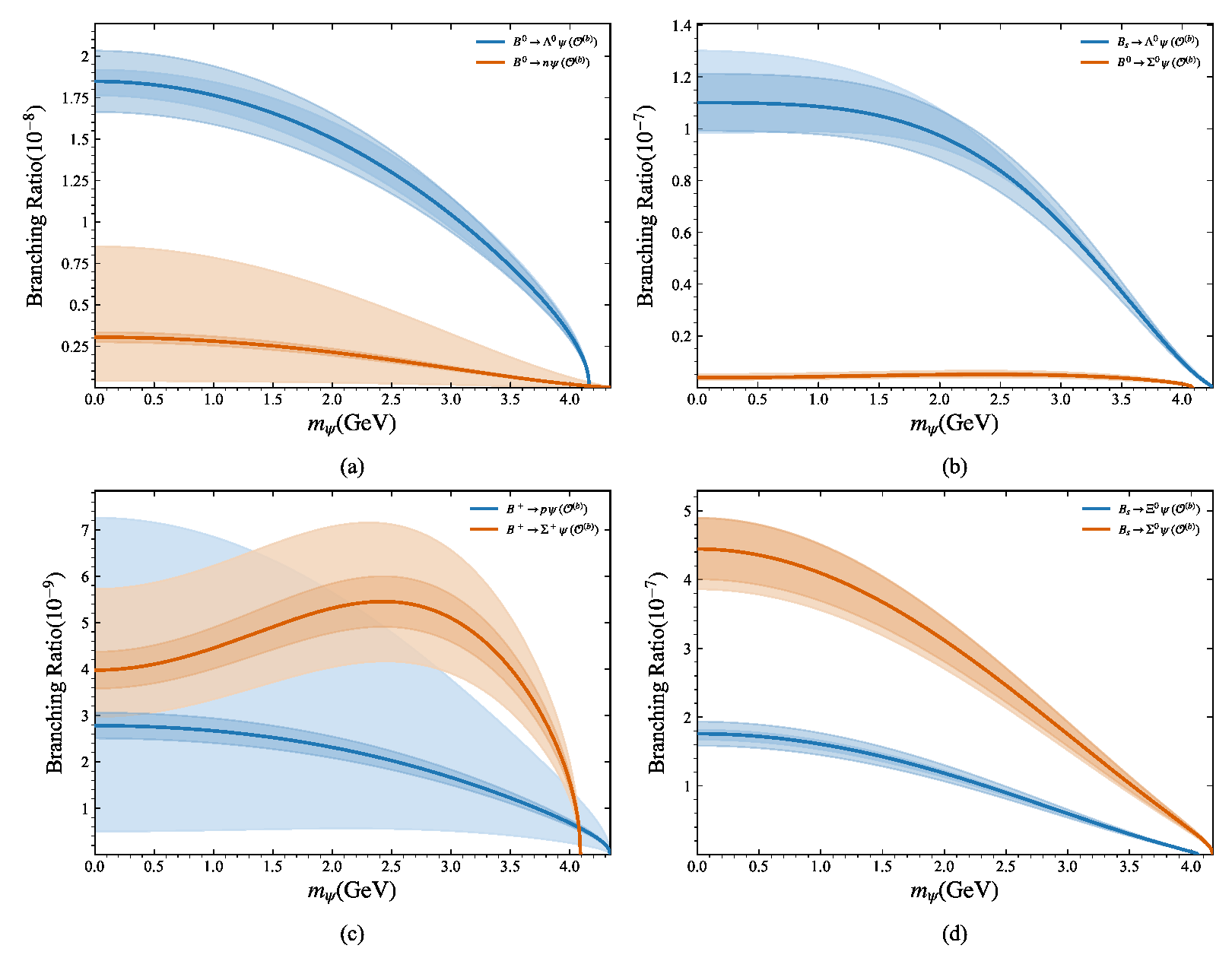}
	\caption{The dark baryon mass $m_{\psi}$-dependence of the upper bounds on the branching ratios for different processes in type-II scenario. The dark- and light-shaded regions correspond to the uncertainty from effective couplings $G_{uq}^{II}$ and form factors $F_{L/R}^{(b)}$, respectively.}
\label{fig:branchingratios2}
\end{figure}

Using the upper limits on the effective couplings in Eq.~\ref{eq:G} and the calculated form factors, we obtain upper bounds on the branching ratios of the corresponding decay modes. In the numerical analysis, we take \(\Lambda_{\rm QCD}=0.25\pm0.025~\mathrm{GeV}\) and consider the dark-baryon mass range $(0\leq m_\psi\leq m_B-m_{\mathcal B})$. The results for the two scenarios are shown in Figs.~\ref{fig:branchingratios1} and \ref{fig:branchingratios2}, respectively.

These predictions can be compared with existing experimental searches for baryonic B-meson decays with invisible final states. For example, the BABAR search for $B^+\to p\psi$ constrains the branching ratio at the $10^{-7}-10^{-5}$ level over the mass range $1.0<m_{\psi}<4.3~\mathrm{GeV}/c^2$~\cite{BaBar:2023dtq}. For the $B^0\to\Lambda\psi$ channel, BABAR obtains upper limits of $(0.13-5.2)\times10^{-5}$ for $1.0<m_{\psi}<4.2~\mathrm{GeV}/c^2$~\cite{BaBar:2023rer}. More recent Belle/Belle II searches for B decays with an invisible particle have reached sensitivities down to the $10^{-6}$ level~\cite{Belle:2026kum}. Therefore, the upper limits obtained in the type-I scenario, which can reach the $10^{-5}$ level, are within the range currently accessible to experiment. By contrast, the type-II results, typically of order $10^{-9}-10^{-6}$, lie mostly below the present experimental sensitivity.

\clearpage
\section{Conclusions}
We study the possible dark sector decays of $B$ meson within the perturbative QCD framework. Under the effective Hamiltonian at two types Mesogenesis scenario, we determine the decay branching ratios for $B$ meson decays into the light baryon octet $\mathcal{B}$ and dark baryon $\psi$. Including the light baryon LCDAs contributions from twist-3 to twist-6, we systematically compute the form factors for $B\to \mathcal{B}$ light baryon transitions. The dominant contribution comes from twist-3 and twist-4 LCDAs of light baryons. Adopting the LHC experimental upper bound on the effective couplings, we obtain the branching ratios for B decays into dark baryons can reach to $10^{-5}$.
\clearpage
\appendix
\section{light-cone wave functions}
The light-cone wave functions of mesons are introduced to describe the non-perturbative hadronic effects. For a heavy pseudoscalar meson such as the $B$ meson, the light-cone matrix element can be written as
\begin{align}
	& \int \frac{d^{4}z}{(2\pi)^4}e^{ik_{1}\cdot z}\langle0|q_{\beta}(z)\bar{b}_{\alpha}(0)|\bar B(P_{B})\rangle=\frac{i}{\sqrt{2N_c}}\Big\{({P\!\!\!\!/}_{B}+m_{B})\gamma_{5}\phi_{B}(k_{1})\Big\}_{\beta\alpha}.
\end{align}
The distribution amplitude $\phi_{B_{(s)}}$ satisfies
\begin{equation}
	\int \frac{d^4k_1}{(2\pi)^4} \phi_{B}(k_1) = \frac{f_{B}}{2\sqrt{6}}.
\end{equation}
It is usually expressed in the momentum fraction $x$ and the conjugate transverse coordinate $\textbf{b}$ as
\begin{equation}
	\phi_{B}(x, \textbf{b}) = N_{B} x^2 (1 - x)^2 \exp\left[-\frac{M_{B}^2 x^2}{2\omega_{b}^2} - \frac{1}{2}(\omega_{b} \textbf{b})^2\right],
\end{equation}
where $N_B$ is the normalization constant determined by the normalization
	condition of the distribution amplitude, and $\omega_B$ is the shape
	parameter characterizing the momentum distribution of the light spectator
	quark inside the $B$ meson. In the numerical analysis, we take
	$\omega_B=0.40~\mathrm{GeV}$ and $\omega_{B_s}=0.50~\mathrm{GeV}$,
	with the corresponding normalization constants $N_B=91.7835$ and
	$N_{B_s}=63.5761$, while the decay constants are chosen as
	$f_B=0.190~\mathrm{GeV}$ and $f_{B_s}=0.230~\mathrm{GeV}$.

For the light baryon moving along the light cone direction n, its light-cone wave function can be defined as~\cite{King:1986wi,Braun:2000kw,RQCD:2019hps,Liu:2013bxa},
\begin{align}
	(\bar{Y}_{\mathcal{B}})_{\alpha\beta\gamma}(x_i,\mu)
	&= -\frac{1}{8\sqrt{2}N_c} \Big\{
	S_1 m_{\mathcal{B}} C_{\beta\alpha} (\bar{\mathcal{B}}^{+}\gamma_5)_\gamma
	+ S_2 m_{\mathcal{B}} C_{\beta\alpha} (\bar{\mathcal{B}}^{-}\gamma_5)_\gamma  \notag\\
	&\quad + P_1 m_{\mathcal{B}} (C\gamma_5)_{\beta\alpha}\,\bar{\mathcal{B}}^{+}_\gamma
	+ P_2 m_{\mathcal{B}} (C\gamma_5)_{\beta\alpha}\,\bar{\mathcal{B}}^{-}_\gamma  \notag\\
	&\quad + V_1 (C\slashed{P}^{\prime})_{\beta\alpha} (\bar{\mathcal{B}}^{+}\gamma_5)_\gamma
	+ V_2 (C\slashed{P}^{\prime})_{\beta\alpha} (\bar{\mathcal{B}}^{-}\gamma_5)_\gamma  \notag\\
	&\quad + V_3 \frac{m_{\mathcal{B}}}{2} (C\gamma_\perp)_{\beta\alpha}
	(\bar{\mathcal{B}}^{+}\gamma_5\gamma^\perp)_\gamma
	+ V_4 \frac{m_{\mathcal{B}}}{2} (C\gamma_\perp)_{\beta\alpha}
	(\bar{\mathcal{B}}^{-}\gamma_5\gamma^\perp)_\gamma  \notag\\
	&\quad + V_5 \frac{m_{\mathcal{B}}^2}{2P^{\prime}z} (C\slashed{z})_{\beta\alpha}
	(\bar{\mathcal{B}}^{+}\gamma_5)_\gamma
	+ V_6 \frac{m_{\mathcal{B}}^2}{2P^{\prime}z} (C\slashed{z})_{\beta\alpha}
	(\bar{\mathcal{B}}^{-}\gamma_5)_\gamma  \notag\\
	&\quad + A_1 (C\gamma_5\slashed{P}^{\prime})_{\beta\alpha}
	(\bar{\mathcal{B}}^{+})_\gamma
	+ A_2 (C\gamma_5\slashed{P}^{\prime})_{\beta\alpha}
	(\bar{\mathcal{B}}^{-})_\gamma  \notag\\
	&\quad + A_3 \frac{m_{\mathcal{B}}}{2} (C\gamma_5\gamma_\perp)_{\beta\alpha}
	(\bar{\mathcal{B}}^{+}\gamma^\perp)_\gamma
	+ A_4 \frac{m_{\mathcal{B}}}{2} (C\gamma_5\gamma_\perp)_{\beta\alpha}
	(\bar{\mathcal{B}}^{-}\gamma^\perp)_\gamma  \notag\\
	&\quad + A_5 \frac{m_{\mathcal{B}}^2}{2P^{\prime}z} (C\gamma_5\slashed{z})_{\beta\alpha}
	(\bar{\mathcal{B}}^{+})_\gamma
	+ A_6 \frac{m_{\mathcal{B}}^2}{2P^{\prime}z} (C\gamma_5\slashed{z})_{\beta\alpha}
	(\bar{\mathcal{B}}^{-})_\gamma  \notag\\
	&\quad - T_1 (iC\sigma_{\perp P^{\prime}})_{\beta\alpha}
	(\bar{\mathcal{B}}^{+}\gamma_5\gamma^\perp)_\gamma
	- T_2 (iC\sigma_{\perp P^{\prime}})_{\beta\alpha}
	(\bar{\mathcal{B}}^{-}\gamma_5\gamma^\perp)_\gamma  \notag\\
	&\quad - T_3 \frac{m_{\mathcal{B}}}{P^{\prime}z} (iC\sigma_{P^{\prime}z})_{\beta\alpha}
	(\bar{\mathcal{B}}^{+}\gamma_5)_\gamma
	- T_4 \frac{m_{\mathcal{B}}}{P^{\prime}z} (iC\sigma_{P^{\prime}z})_{\beta\alpha}
	(\bar{\mathcal{B}}^{-}\gamma_5)_\gamma  \notag\\
	&\quad - T_5 \frac{m_{\mathcal{B}}^2}{2P^{\prime}z} (iC\sigma_{\perp z})_{\beta\alpha}
	(\bar{\mathcal{B}}^{+}\gamma_5\gamma^\perp)_\gamma
	- T_6 \frac{m_{\mathcal{B}}^2}{2P^{\prime}z} (iC\sigma_{\perp z})_{\beta\alpha}
	(\bar{\mathcal{B}}^{-}\gamma_5\gamma^\perp)_\gamma  \notag\\
	&\quad + T_7 \frac{m_{\mathcal{B}}}{2} (C\sigma_{\perp\perp'})_{\beta\alpha}
	(\bar{\mathcal{B}}^{+}\gamma_5\sigma^{\perp\perp'})_\gamma + T_8 \frac{m_{\mathcal{B}}}{2} (C\sigma_{\perp\perp'})_{\beta\alpha}
	(\bar{\mathcal{B}}^{-}\gamma_5\sigma^{\perp\perp'})_\gamma
	\Big\}.
\end{align}
The ``large'' and ``small'' light-cone components are defined as
\begin{align}
	\mathcal{B}^{+} = \frac{{P\!\!\!/}^{\prime} z\!\!\!/}{2P^{\prime}\cdot z}\mathcal{B}, \qquad
	\mathcal{B}^{-} = \frac{z\!\!\!/ {P\!\!\!/}^{\prime}}{2P^{\prime}\cdot z}\mathcal{B},
\end{align}
where the two light-cone vectors $P^{\prime}$ and $z$ is defined by the baryon momentum $p_2$,
\begin{align}
P^{\prime}_{\mu}=p_{2\mu}-\frac{1}{2}z_{\mu}\frac{m_{\mathcal{B}}^2}{P^{\prime}\cdot z},
\end{align}
The shorthand $\sigma_{P^{\prime}z}=\sigma^{\mu\nu}P^{\prime}_{\mu}z_{\nu}$, and $\perp$ stands for the projection transverse to $z$, $P^{\prime}$, e.g. $\gamma_{\perp}\gamma^{\perp}=\gamma_{\mu} g^{\mu\nu}_{\perp} \gamma_{\nu}$ with
$g^{\mu\nu}_{\perp} = g^{\mu\nu} - (P^{'\mu} z^\nu + z^\mu P^{'\nu})/(P^{\prime}\cdot z)$. $V_i$, $A_i$, $T_i$, $S_i$, $P_i$ are the light-cone distribution amplitudes with different Lorentz structure. Below we present their explicit forms.

For proton or neutron,
\begin{itemize}
	\item Twist-3 LCDAs\\
	{\footnotesize
	\begin{equation}
		\begin{aligned}
			V_1(x_i)=&120x_1x_2x_3[\phi_3^0+\phi_3^+(1-3x_3)],\\
			A_1(x_i)=&120x_1x_2x_3(x_2-x_1)\phi_3^-,\\
			T_1(x_i)=&120x_1x_2x_3[\phi_3^0+\frac{1}{2}(\phi_3^--\phi_3^+)(1-3x_3)].
	\end{aligned}
\end{equation}}
	\item Twist-4 LCDAs\\
	{\footnotesize
			\begin{equation}
		\begin{aligned}
			V_2(x_i)=&24x_1x_2[\phi_4^0+\phi_4^+(1-5x_3)],\\
			V_3(x_i)=&12x_3[\psi_4^0(1-x_3)+\psi_4^-(x_1^2+x_2^2-x_3(1-x_3))+\psi_4^+(1-x_3-10x_1x_2)],\\
			A_2(x_i)=&24x_1x_2(x_2-x_1)\phi_4^-,\\
			A_3(x_i)=&12x_3(x_2-x_1)[(\psi_4^0+\psi_4^+)+\psi_4^-(1-2x_3)],\\
			T_2(x_i)=&24x_1x_2[\xi_4^0+\xi_4^+(1-5x_3)],\\
			T_3(x_i)=&6x_3[(\xi_4^0+\phi_4^0+\psi_4^0)(1-x_3)+(\xi_4^-+\phi_4^--\psi_4^-)(x_1^2+x_2^2-x_3(1-x_3))\\
			&+(\xi_4^++\phi_4^++\psi_4^+)(1-x_3-10x_1x_2)],\\
			T_7(x_i)=&6x_3[(-\xi_4^0+\phi_4^0+\psi_4^0)(1-x_3)+(-\xi_4^-+\phi_4^--\psi_4^-)(x_1^2+x_2^2-x_3(1-x_3))\\
			&+(-\xi_4^++\phi_4^++\psi_4^+)(1-x_3-10x_1x_2)],\\
			S_1(x_i)=&6x_3(x_2-x_1)[(\xi_4^0+\phi_4^0+\psi_4^0+\xi_4^++\phi_4^++\psi_4^+)+(\xi_4^-+\phi_4^--\psi_4^-)(1-2x_3)],\\
			P_1(x_i)=&6x_3(x_2-x_1)[(\xi_4^0-\phi_4^0-\psi_4^0+\xi_4^+-\phi_4^+-\psi_4^+)+(\xi_4^--\phi_4^-+\psi_4^-)(1-2x_3)].
	\end{aligned}
\end{equation}}
	\item Twist-5 LCDAs\\
	{\footnotesize
	\begin{equation}
		\begin{aligned}
			V_4(x_i)=&3[\psi_5^0(1-x_3)+\psi_5^-(2x_1x_2-x_3(1-x_3))+\psi_5^+(1-x_3-2(x_1^2+x_2^2))],\\
			V_5(x_i)=&6x_3[\phi_5^0+\phi_5^+(1-2x_3)],\\
			A_4(x_i)=&3(x_2-x_1)[-\psi_5^0+\psi_5^-x_3+\psi_5^+(1-2x_3)],\\
			A_5(x_i)=&6x_3(x_2-x_1)\phi_5^-,\\
			T_4(x_i)=&\frac{3}{2}[(\xi_5^0+\psi_5^0+\phi_5^0)(1-x_3)+(\xi_5^-+\phi_5^--\psi_5^-)(2x_1x_2-x_3(1-x_3))\\
			&+(\xi_5^++\phi_5^++\psi_5^+)(1-x_3-2(x_1^2+x_2^2))],\\
			T_5(x_i)=&6x_3[\xi_5^0+\xi_5^+(1-2x_3)],\\
			T_8(x_i)=&\frac{3}{2}[(\psi_5^0+\phi_5^0-\xi_5^0)(1-x_3)+(\phi_5^--\psi_5^--\xi_5^-)(2x_1x_2-x_3(1-x_3))\\
			&+(\phi_5^++\psi_5^+-\xi_5^+)(1-x_3-2(x_1^2+x_2^2))],\\
			S_2(x_i)=&\frac{3}{2}(x_2-x_1)[-(\psi_5^0+\phi_5^0+\xi_5^0)+(\xi_5^-+\phi_5^--\psi_5^0)x_3+(\xi_5^++\phi_5^++\psi_5^0)(1-2x_3)],\\
			P_2(x_i)=&\frac{3}{2}(x_2-x_1)[(\psi_5^0+\phi_5^0-\xi_5^0)+(\xi_5^--\phi_5^-+\psi_5^0)x_3+(\xi_5^+-\phi_5^+-\psi_5^0)(1-2x_3)].
	\end{aligned}
\end{equation}}
	\item Twist-6 LCDAs\\
	{\footnotesize
		\begin{equation}
		\begin{aligned}
			V_6(x_i)=&2[\phi_6^0+\phi_6^+(1-3x_3)],\\
			A_6(x_i)=&2(x_2-x_1)\phi_6^-,\\
			T_6(x_i)=&2[\phi_6^0+\frac{1}{2}(\phi_6^--\phi_6^+)(1-3x_3)],
	\end{aligned}
\end{equation}}
\end{itemize}
For $\Lambda$ baryon,
\begin{itemize}
	\item Twist-3 LCDAs\\
	{\footnotesize
		\begin{equation}
		\begin{aligned}
			V_1(x_i)=&120x_1x_2x_3(x_1-x_2)\left(\frac{21\sqrt{6}}{4}\phi_{11}-\frac{7\sqrt{6}}{4}\phi_{10}\right),\\
			A_1(x_i)=&120x_1x_2x_3\left[-f_\Lambda+\frac{7\sqrt{6}}{4}(\phi_{11}+\phi_{10})(x_1+x_2)-\frac{14\sqrt{6}}{4}x_3(\phi_{11}+\phi_{10})\right],\\
			T_1(x_i)=&120x_1x_2x_3\left[\frac{7\sqrt{6}}{2}\pi_{10}(x_1-x_2)\right].
	\end{aligned}	\end{equation}}
	\item Twist-4 LCDAs\\
	{\footnotesize
			\begin{equation}
		\begin{aligned}
			S_1(x_i)=&6x_3(1-x_3)(\xi_4^0+\xi_4'^{0}),\\
			P_1(x_i)=&6x_3(1-x_3)(\xi_4^0-\xi_4'^{0}),\\
			V_2(x_i)=&0,\\
			A_2(x_i)=&-24x_1x_2\phi_4^0,\\
			V_3(x_i)=&12(x_1-x_2)x_3\psi_4^0,\\
			A_3(x_i)=&-12x_3(1-x_3)\psi_4^0,\\
			T_2(x_i)=&0,\\
			T_3(x_i)=&6(x_2-x_1)x_3(-\xi_4^0+\xi_4'^{0}),\\
			T_7(x_i)=&-6(x_1-x_2)x_3(\xi_4^0+\xi_4'^{0}).
	\end{aligned}
	\end{equation}}
	\item Twist-5 LCDAs\\
	{\footnotesize
			\begin{equation}
		\begin{aligned}
			S_2(x_i)=&\frac{3}{2}(x_1+x_2)(\xi_5^0+\xi_5'^{0}),\\
			P_2(x_i)=&\frac{3}{2}(x_1+x_2)(\xi_5^0-\xi_5'^{0}),\\
			V_4(x_i)=&3(x_2-x_1)\psi_5^0,\\
			A_4(x_i)=&-3(1-x_3)\psi_5^0,\\
			V_5(x_i)=&0,\\
			A_5(x_i)=&-6x_3\phi_5^0,\\
			T_4(x_i)=&-\frac{3}{2}(x_1-x_2)(\xi_5^0+\xi_5'^{0}),\\
			T_5(x_i)=&0,\\
			T_8(x_i)=&-\frac{3}{2}(x_1-x_2)(\xi_5^0-\xi_5'^{0}).
	\end{aligned}
\end{equation}}
	\item Twist-6 LCDAs\\
	{\footnotesize
		\begin{equation}
		\begin{aligned}
			V_6(x_i)=&0,\\
			A_6(x_i)=&-2\phi_6^0,\\
			T_6(x_i)=&0.
	\end{aligned}
\end{equation}}
\end{itemize}
For $\Sigma$ or $\Xi$ baryon,
\begin{itemize}
	\item Twist-3 LCDAs\\
	{\footnotesize
	\begin{equation}
		\begin{aligned}
			V_1(x_i)=&120x_1x_2x_3\,\phi_3^0,\\
			A_1(x_i)=&0,\\
			T_1(x_i)=&120x_1x_2x_3\,\phi_3^{\prime 0}.
	\end{aligned}
\end{equation}}
	\item Twist-4 LCDAs\\
	{\footnotesize
		\begin{equation}
		\begin{aligned}
			S_1(x_i)=&6(x_2-x_1)x_3(\xi_4^0+\xi_4^{\prime 0}),\\
			P_1(x_i)=&6(x_2-x_1)x_3(\xi_4^0-\xi_4^{\prime 0}),\\
			V_2(x_i)=&24x_1x_2\phi_4^0,\\
			A_2(x_i)=&0,\\
			V_3(x_i)=&12x_3(1-x_3)\psi_4^0,\\
			A_3(x_i)=&-12x_3(x_1-x_2)\psi_4^0,\\
			T_2(x_i)=&24x_1x_2\phi_4^{\prime 0},\\
			T_3(x_i)=&6x_3(1-x_3)(\xi_4^0+\xi_4^{\prime 0}),\\
			T_7(x_i)=&6x_3(1-x_3)(\xi_4^{\prime 0}-\xi_4^0).
	\end{aligned}
\end{equation}}
	\item Twist-5 LCDAs\\
	{\footnotesize
			\begin{equation}
		\begin{aligned}
			S_2(x_i)=&\frac{3}{2}(x_1-x_2)(\xi_5^0+\xi_5^{\prime 0}),\\
			P_2(x_i)=&\frac{3}{2}(x_1-x_2)(\xi_5^0-\xi_5^{\prime 0}),\\
			V_4(x_i)=&3(1-x_3)\psi_5^0,\\
			A_4(x_i)=&3(x_1-x_2)\psi_5^0,\\
			V_5(x_i)=&6x_3\phi_5^0,\\
			A_5(x_i)=&0,\\
			T_4(x_i)=&-\frac{3}{2}(x_1+x_2)(\xi_5^{\prime 0}+\xi_5^0),\\
			T_5(x_i)=&6x_3\phi_5^{\prime 0},\\
			T_8(x_i)=&\frac{3}{2}(x_1+x_2)(\xi_5^{\prime 0}-\xi_5^0).
	\end{aligned}
\end{equation}}
	\item Twist-6 LCDAs\\
	{\footnotesize
	\begin{equation}
		\begin{aligned}
			V_6(x_i)=&2\phi_6^0,\\
			A_6(x_i)=&0,\\
			T_6(x_i)=&2\phi_6^{\prime 0}.
	\end{aligned}
	\end{equation}}
\end{itemize}
By an explicit calculation one obtains the LCDAs depend on the scale $\mu$,
\begin{eqnarray}
\phi_i(x_i,\mu)=\Big(\frac{\alpha_s(\mu)}{\alpha_s(\mu_0)}\Big)^{2/(3\beta_0)}\phi_i(x_i,\mu_0),
\end{eqnarray}
where $\beta_0=11-2/3n_f$ and $n_f=4$ for $\mu=2$ GeV, the scale dependence is calculated using one-loop anomalous dimensions.
\begin{table}[htbp]
	\centering
	\caption{Parameters of the light-cone distribution amplitudes (LCDAs) for the baryons $n/p$, $\Lambda$, $\Sigma$, and $\Xi$ from twist-3 ($i=3$) to twist-6 ($i=6$), in units of $10^{-3}\,\mathrm{GeV}^2$.}
	\setlength{\tabcolsep}{2pt}
	\renewcommand{\arraystretch}{1.15}
		\begin{tabular}{cccccccccc}
			\hline\hline
			$n/p$ & $\phi_i^0$ & $\phi_i^-$ & $\phi_i^+$& $\psi_i^0$ & $\psi_i^-$ & $\psi_i^+$ & $\xi_i^0$ & $\xi_i^-$ & $\xi_i^+$ \\
			\hline
			twist-3  & $5.3^{+0.5}_{-0.5}$ & 21.1 & 5.7 & & & & & & \\
			twist-4  & $-10.8^{+4.7}_{-4.7}$ & 32.2 & 21.2 & $16.1^{+4.7}_{-4.7}$ & $-61.3$ & 9.9 & $8.5^{+3.1}_{-3.1}$ & 27.9 & 5.6 \\
			twist-5  & $-10.8^{+4.7}_{-4.7}$ & $-20.1$ & 14.2 & $16.1^{+4.7}_{-4.7}$ & $-9.8$ & $-9.9$ & $8.5^{+3.1}_{-3.1}$ & $-9.5$ & 4.6 \\
			twist-6  & $5.3^{+0.5}_{-0.5}$ & 30.9 & $-2.5$ & & & & & & \\\hline
	
		$\Lambda$ & $f_\Lambda$ & $\phi_{11}$ & $\phi_{10}$ & $\pi_{10}$ & $\phi_i^0$ & $\psi_i^0$ & $\xi_i^0$ & $\xi_i^{\prime 0}$&  \\
			\hline
			twist-3
			& $6.20^{+0.11}_{-0.09}$
			& $0.25^{+0.06}_{-0.02}$
			& $0.72^{+0.04}_{-0.04}$
			& $0.25^{+0.04}_{-0.03}$
			& & & & & \\
		twist-4
			& & & &
			& $-10.78^{+0.66}_{-0.67}$
			& $-4.57^{+0.77}_{-0.76}$
			& $-18.02^{+0.94}_{-0.94}$
			& $19.07^{+1.84}_{-1.84}$ &\\
			twist-5
			& & & &
			& $-10.78^{+0.66}_{-0.67}$
			& $-4.57^{+0.77}_{-0.76}$
			& $-18.02^{+0.94}_{-0.94}$
			& $19.07^{+1.84}_{-1.84}$ &\\
		twist-6
			& & & &
			& $6.20^{+0.11}_{-0.09}$ & & & &\\
			\hline\hline
		$\Sigma$ & $\phi_i^0$ & $\phi_i^{\prime 0}$ & $\psi_i^0$ & $\xi_i^0$ & $\xi_i^{\prime 0}$ & & & & \\
		\hline
		twist-3  & $5.32^{+0.12}_{-0.12}$ & $6.00^{+0.50}_{-0.50}$ & & & & & & &\\
		twist-4  & $-9.84^{+0.44}_{-0.44}$ & $4.67^{+0.83}_{-0.83}$ & $15.16^{+0.56}_{-0.56}$ & $4.67^{+0.83}_{-0.83}$ & $3.33^{+1.17}_{-1.17}$ & & & &\\
		twist-5  & $-9.84^{+0.44}_{-0.44}$ & $7.33^{+0.17}_{-0.17}$ & $15.16^{+0.56}_{-0.56}$ & $-6.00^{+0.50}_{-0.50}$ & $-7.33^{+1.17}_{-1.17}$ & & & &\\
		twist-6  & $5.32^{+0.12}_{-0.12}$ & $6.00^{+0.50}_{-0.50}$ & & & & & & &\\
		\hline\hline
		$\Xi$& $\phi_i^0$ & $\phi_i^{\prime 0}$ & $\psi_i^0$ & $\xi_i^0$ & $\xi_i^{\prime 0}$ & & & &\\
		\hline
		twist-3  & $9.90^{+0.40}_{-0.40}$ & $2.67^{+0.33}_{-0.33}$ & & & & & & &\\
		twist-4  & $-9.05^{+0.30}_{-0.30}$ & $-3.33^{+0.33}_{-0.33}$ & $18.95^{+0.70}_{-0.70}$ & $-3.33^{+0.33}_{-0.33}$ & $-9.33^{+0.33}_{-0.33}$ & & & &\\
		twist-5  & $-9.05^{+0.30}_{-0.30}$ & $8.67^{+0.33}_{-0.33}$ & $18.95^{+0.70}_{-0.70}$ & $-2.67^{+0.33}_{-0.33}$ & $-8.67^{+0.33}_{-0.33}$ & & & &\\
		twist-6  & $9.90^{+0.40}_{-0.40}$ & $2.67^{+0.33}_{-0.33}$ & & & & & & &\\
		\hline\hline
	\end{tabular}%
\end{table}
\section{pQCD calculation}
For the $B$ meson semi-invisible decays, such as the process $B^+ \to p \psi$, the decay amplitudes are governed by the two possible effective operators describing the dark sector interaction. In the type-I scenario, the amplitude of Fig.~\ref{fig:feynmandiagram}(a,b,c) respectively given as
\begin{align}
	\mathcal M_a={}& \frac{C_F\Lambda_{\mathrm{QCD}}}{72\sqrt{2}\,\pi}\frac{\phi_{B^+}}{\eta^{2}}\int_0^1dx_1\,dx_3\,dx_4\int_0^{1/\Lambda_{\mathrm{QCD}}}db_1\,db_3\,db_4\notag \\
	& b_1b_4\,E(t_a)\,h_1(x_1,x_4,b_1,b_4)\notag \\
	&\times\Biggl\{
	80 x_2 x_3 x_4 \bigl(\sqrt{6} \eta m_{B^+} - 2 \sqrt{3} m_p\bigr)\,\bigl(-m_p^{2} + \eta^{2} m_{B^+}^{2}\notag \\
	&+ \eta m_p^{2} x_4\bigr)\,\bigl(\phi_{3}^{0} + \phi_{3}^{-} \bigl(x_3 + 2 x_4 - 1\bigr) + \phi_{3}^{+} \bigl(1 - 3 x_3\bigr)\bigr)\notag \\
	& +\;\bigl[-\;8\sqrt{3} m_p\,\bigl(4 m_p^{2} x_2 x_4 + \eta m_{B^+} m_p x_2^{2} + \eta m_{B^+} m_p x_4^{2} - \eta m_{B^+} m_p x_2\notag \\
	& - \eta m_{B^+} m_p x_4 - 4 \eta m_p^{2} x_2 x_4^{2} - 4 \eta^{2} m_{B^+}^{2} x_2 x_4 + 2 \eta m_{B^+} m_p x_2 x_4\bigr)\notag \\
	& +\;4\sqrt{6} \eta^{2} m_{B^+}^{2} m_p x_3 \bigl(- x_2 - x_4\bigr)\bigr]\,\phi_{4}^{0} +\;\bigl[\;8\sqrt{3} m_p\,\bigl(-20 m_p^{2} x_2 x_4^{2} - 20 m_p^{2} x_2^{2} x_4\notag \\
	& + 16 m_p^{2} x_2 x_4 + \eta m_{B^+} m_p x_2 + \eta m_{B^+} m_p x_4 - \eta m_{B^+} m_p x_2^{2}\notag \\
	& - \eta m_{B^+} m_p x_4^{2} - 16 \eta m_p^{2} x_2 x_4^{2} - 16 \eta^{2} m_{B^+}^{2} x_2 x_4 + 20 \eta m_p^{2} x_2 x_4^{3}\notag \\
	& + 20 \eta m_p^{2} x_2^{2} x_4^{2} + 20 \eta^{2} m_{B^+}^{2} x_2 x_4^{2}\notag \\
	& + 20 \eta^{2} m_{B^+}^{2} x_2^{2} x_4 - 12 \eta m_{B^+} m_p x_2 x_4 + 10 \eta m_{B^+} m_p x_2 x_4^{2}\notag \\
	& + 10 \eta m_{B^+} m_p x_2^{2} x_4\bigr) -\;4\sqrt{6} \eta^{2} m_{B^+}^{2} m_p \bigl(x_2 + x_4 - 1\bigr) \bigl(10 x_2 x_4 - x_2 - x_4\bigr)\bigr]\,\phi_{4}^{+}\notag \\
	& +\;\bigl[-\;8\sqrt{3} m_p\,\bigl(-4 m_p^{2} x_2^{2} x_4 + 4 m_p^{2} x_2 x_4^{2} + \eta m_{B^+} m_p x_2 + \eta m_{B^+} m_p x_4\notag \\
	& - 4 \eta m_p^{2} x_2 x_4^{3} - 4 \eta^{2} m_{B^+}^{2} x_2 x_4^{2} - 3 \eta m_{B^+} m_p x_2^{2} - 3 \eta m_{B^+} m_p x_4^{2}\notag \\
	& + 2 \eta m_{B^+} m_p x_2^{3} + 2 \eta m_{B^+} m_p x_4^{3} + 4 \eta m_p^{2} x_2^{2} x_4^{2}\notag \\
	& + 4 \eta^{2} m_{B^+}^{2} x_2^{2} x_4 - 4 \eta m_{B^+} m_p x_2 x_4 + 4 \eta m_{B^+} m_p x_2 x_4^{2} + 4 \eta m_{B^+} m_p x_2^{2} x_4\bigr)\notag \\
	& +\;4\sqrt{6} \eta^{2} m_{B^+}^{2} m_p \bigl(x_2 + x_4 - 1\bigr)\,\bigl(-x_2 - x_4 + 2 x_2^{2} + 2 x_4^{2} + 2 x_2 x_4\bigr)\bigr]\,\phi_{4}^{-}\notag \\
	& -\;\bigl[-\;2\sqrt{3} m_p^{2}\,\bigl(\eta m_{B^+} - 4 m_p x_3 - \eta m_{B^+} x_3 - 4 \eta m_{B^+} x_4 + 2 \eta^{2} m_{B^+} x_4^{2} + 4 \eta m_p x_3 x_4\bigr)\notag \\
	&+\;4\sqrt{6} \eta m_{B^+} m_p^{2} x_3 \bigl(\eta x_4 - 1\bigr)\bigr]\,\phi_{5}^{0}\notag \\
	&-\;\bigl[\;2\sqrt{3} m_p^{2} \bigl(\eta x_4 - 1\bigr)\,\bigl(\eta m_{B^+} - 4 m_p x_3 + 8 m_p x_3^{2} - 4 \eta m_{B^+} x_4\notag \\
	& - 3 \eta m_{B^+} x_3 + 2 \eta m_{B^+} x_3^{2} + 4 \eta m_{B^+} x_4^{2} + 4 \eta m_{B^+} x_3 x_4\bigr)\notag \\
	&-\;4\sqrt{6} \eta m_{B^+} m_p^{2} x_3 \bigl(2 x_3 - 1\bigr) \bigl(\eta x_4 - 1\bigr)\bigr]\,\phi_{5}^{+}\notag \\
	&+\;\bigl[\;2\sqrt{3} m_p^{2} \bigl(\eta x_4 - 1\bigr)\,\bigl(-4 m_p x_3 + 4 m_p x_3^{2} + \eta m_{B^+} x_3^{2}\notag \\
	& - \eta m_{B^+} x_3 - 2 \eta m_{B^+} x_4^{2} + 2 \eta m_{B^+} x_4 + 8 m_p x_3 x_4 - 2 \eta m_{B^+} x_3 x_4\bigr)\notag \\
	& -\;4\sqrt{6} \eta m_{B^+} m_p^{2} x_3 \bigl(- x_2 + x_4\bigr) \bigl(\eta x_4 - 1\bigr)\bigr]\,\phi_{5}^{-}\notag \\
	& -\;\frac{8 \sqrt{3} m_p^{3} \bigl(\eta x_4 - 1\bigr)}{3}\,\bigl(\phi_{6}^{0} + \phi_{6}^{-} \bigl(x_3 + 2 x_4 - 1\bigr) + \phi_{6}^{+} \bigl(1 - 3 x_3\bigr)\bigr)\notag \\
	& -\;4 \eta m_{B^+} m_p x_3 \bigl(x_2 + x_4\bigr) \bigl(\sqrt{6} \eta m_{B^+} - 2 \sqrt{3} m_p\bigr)\,\psi_{4}^{0}\notag \\
	& -\;4 \eta m_{B^+} m_p x_3 \bigl(\sqrt{6} \eta m_{B^+} - 2 \sqrt{3} m_p\bigr)\,\bigl(1 - x_3 - 10 x_4 + 10 x_4^{2} + 10 x_3 x_4\bigr)\,\psi_{4}^{+}\notag \\
	& +\;4 \eta m_{B^+} m_p x_3 \bigl(\sqrt{6} \eta m_{B^+} - 2 \sqrt{3} m_p\bigr)\,\bigl(1 - 3 x_3 - 2 x_4 + 2 x_3^{2} + 2 x_4^{2} + 2 x_3 x_4\bigr)\,\psi_{4}^{-}\notag \\
	&+\;2 \sqrt{3} \eta m_{B^+} m_p^{2} \bigl(x_2 + x_4\bigr) \bigl(\eta x_4 - 1\bigr)\,\psi_{5}^{0}\notag \\
	&-\;2 \sqrt{3} \eta m_{B^+} m_p^{2} \bigl(\eta x_4 - 1\bigr)\,\bigl(-x_2 - x_4 + 2 x_2^{2} + 2 x_4^{2}\bigr)\,\psi_{5}^{+}\notag \\
	& -\;2 \sqrt{3} \eta m_{B^+} m_p^{2} \bigl(\eta x_4 - 1\bigr)\,\bigl(x_2^{2} + x_4^{2} - x_2 - x_4 + 4 x_2 x_4\bigr)\,\psi_{5}^{-}\notag \\
	& +\;4 \eta m_{B^+} m_p \bigl(x_2 + x_4 - 1\bigr) \bigl(2 \eta x_2 x_4 - 3 x_2 + x_4\bigr)\,\bigl(-2 \sqrt{3} m_p + \sqrt{6} \eta m_{B^+}\bigr)\,\xi_{4}^{0}\notag \\
	&-\;4 \eta m_{B^+} m_p \bigl(\sqrt{6} \eta m_{B^+} - 2 \sqrt{3} m_p\bigr) \bigl(x_2 + x_4\notag \\
	&- 1\bigr)\,\bigl(-x_4 + 3 x_2 - 10 x_2 x_4 - 2 \eta x_2 x_4 + 10 \eta x_2 x_4^{2}\bigr)\,\xi_{4}^{+}\notag \\
	&+\;4 \eta m_{B^+} m_p \bigl(\sqrt{6} \eta m_{B^+} - 2 \sqrt{3} m_p\bigr) \bigl(x_2\notag \\
	&+ x_4 - 1\bigr)\,\bigl(-x_4 - 6 x_2^{2} + 2 x_4^{2} + 3 x_2 - 2 x_2 x_4 - 2 \eta x_2 x_4\notag \\
	& + 2 \eta x_2 x_4^{2} + 4 \eta x_2^{2} x_4\bigr)\,\xi_{4}^{-} +\;2 \sqrt{3} \eta m_{B^+} m_p^{2} \bigl(x_2 + x_4\bigr)\,\xi_{5}^{0}\notag \\
	& -\;2 \sqrt{3} \eta m_{B^+} m_p^{2}\,\bigl(x_2 - 3 x_4 - 2 x_2^{2} + 6 x_4^{2} + \eta x_4^{2} - 2 \eta x_4^{3} - \eta x_2 x_4 + 2 \eta x_2^{2} x_4\bigr)\,\xi_{5}^{+}\notag \\
	& +\;2 \sqrt{3} \eta m_{B^+} m_p^{2}\,\bigl(x_3 - x_3^{2} - 2 x_4^{2} + 2 x_4 - 6 x_3 x_4 + \eta x_3^{2} x_4\notag \\
	& - \eta x_3 x_4 + 2 \eta x_3 x_4^{2}\bigr)\,\xi_{5}^{-}\Biggr\}.
\end{align}

\begin{align}
	\mathcal M_b={}& -\frac{\pi C_F\Lambda_{\mathrm{QCD}}}{18\sqrt{2}}\frac{\phi_{B^+}}{\eta}\int_0^1dx_1\,dx_3\,dx_4\int_0^{1/\Lambda_{\mathrm{QCD}}}db_1\,db_3\,db_4\notag \\
	& b_3b_4\,E(t_b)\,h_2(x_1,x_3,x_4,b_1,b_3,b_4)\notag \\
	&\times\Biggl\{
	\bigl[-\;80\sqrt{3} \eta m_{B^+}^{2} x_2 x_3 x_4 \bigl(\eta m_{B^+} + m_p\bigr) +\;20\sqrt{6} m_{B^+} x_2 x_3 x_4\,\bigl(-m_p^{2} + \eta^{2} m_{B^+}^{2} + \eta m_p^{2} x_4\notag \\
	& + 4 \eta m_{B^+} m_p\bigr)\bigr]\,\phi_{3}^{0} -\;20 m_{B^+} x_2 x_3 x_4 \bigl(3 x_3 - 1\bigr)\,\bigl(-\sqrt{6} m_p^{2} + \sqrt{6} \eta^{2} m_{B^+}^{2}\notag \\
	& + 2 \sqrt{3} \eta^{2} m_{B^+}^{2} + \sqrt{6} \eta m_p^{2} x_4 - 4 \sqrt{3} \eta m_{B^+} m_p - 2 \sqrt{6} \eta m_{B^+} m_p\bigr)\,\phi_{3}^{+}\notag \\
	&+\;\bigl[\;40\sqrt{3} \eta m_{B^+}^{2} x_2 x_4 \bigl(x_2 + x_4 - 1\bigr)\,\bigl(-2 \eta m_{B^+}\notag \\
	&- 2 m_p x_2 + 2 m_p x_4 + 3 \eta m_{B^+} x_2 + 3 \eta m_{B^+} x_4\bigr)\notag \\
	& +\;20\sqrt{6} m_{B^+} x_2 x_4 \bigl(x_2 + x_4 - 1\bigr)\,\bigl(m_p^{2} x_4 - m_p^{2} x_2 + \eta^{2} m_{B^+}^{2} x_2 - \eta m_p^{2} x_4^{2} - \eta^{2} m_{B^+}^{2} x_4\notag \\
	& + 4 \eta m_{B^+} m_p + \eta m_p^{2} x_2 x_4 - 6 \eta m_{B^+} m_p x_2 - 6 \eta m_{B^+} m_p x_4\bigr)\bigr]\,\phi_{3}^{-}\notag \\
	& +\;\bigl[\;16\sqrt{3} \eta m_{B^+}^{2} m_p x_2 x_4 +\;2\sqrt{6} \eta m_{B^+}^{2} m_p x_3 \bigl(- x_2 - x_4\bigr) \bigl(\eta x_4 - 1\bigr)\bigr]\,\phi_{4}^{0}\notag \\
	&-\;\bigl[\;16\sqrt{3} \eta m_{B^+}^{2} m_p x_2 x_4 \bigl(5 x_3 - 1\bigr) +\;2\sqrt{6} \eta m_{B^+}^{2} m_p x_3 \bigl(\eta x_4\notag \\
	&- 1\bigr)\,\bigl(1 - x_3 - 10 x_4 + 10 x_4^{2} + 10 x_3 x_4\bigr)\bigr]\,\phi_{4}^{+}\notag \\
	& -\;\bigl[\;16\sqrt{3} \eta m_{B^+}^{2} m_p x_2 x_4 \bigl(x_2 - x_4\bigr)\notag \\
	& +\;2\sqrt{6} \eta m_{B^+}^{2} m_p x_3 \bigl(\eta x_4 - 1\bigr)\,\bigl(1 - 3 x_3 - 2 x_4 + 2 x_3^{2} + 2 x_4^{2} + 2 x_3 x_4\bigr)\bigr]\,\phi_{4}^{-}\notag \\
	& -\;m_{B^+} m_p^{2} \bigl(\sqrt{6} \eta x_3 x_4 - \sqrt{6} x_3 - 4 \sqrt{3} x_4\bigr)\,\phi_{5}^{0} +\;\bigl[-\;2\sqrt{3} m_{B^+} m_p^{2} \bigl(2 x_2^{2} - x_2 + 2 x_4^{2} - x_4\bigr)\notag \\
	& +\;\sqrt{6}\,m_{B^+} m_p^{2} x_3 \bigl(2 x_3 - 1\bigr) \bigl(\eta x_4 - 1\bigr)\bigr]\,\phi_{5}^{+} -\;\bigl[-\;2\sqrt{3} m_{B^+} m_p^{2} \bigl(x_2^{2} + 4 x_2 x_4 - x_2 + x_4^{2} - x_4\bigr)\notag \\
	& +\;\sqrt{6}\,m_{B^+} m_p^{2} x_3 \bigl(- x_2 + x_4\bigr) \bigl(\eta x_4 - 1\bigr)\bigr]\,\phi_{5}^{-} -\;\bigl[\;8\sqrt{3} \eta m_{B^+}^{2} m_p x_2 \bigl(\eta x_4 - 2\bigr) \bigl(x_2 + x_4 - 1\bigr)\notag \\
	& -\;2\sqrt{6} m_{B^+} m_p \bigl(\eta x_4 - 1\bigr) \bigl(x_2 + x_4 - 1\bigr)\,\bigl(4 m_p x_2 + \eta m_{B^+} x_2 + \eta m_{B^+} x_4\bigr)\bigr]\,\psi_{4}^{0}\notag \\
	& +\;\bigl[\;8\sqrt{3} \eta m_{B^+}^{2} m_p x_2 \bigl(5 x_4 - 1\bigr) \bigl(\eta x_4 - 2\bigr)\,\bigl(-1 + x_2 + x_4\bigr)\notag \\
	& -\;2\sqrt{6} m_{B^+} m_p \bigl(\eta x_4 - 1\bigr) \bigl(x_2 + x_4 - 1\bigr)\,\bigl(-4 m_p x_2 - \eta m_{B^+} x_2 - \eta m_{B^+} x_4 + 20 m_p x_2 x_4\notag \\
	& + 10 \eta m_{B^+} x_2 x_4\bigr)\bigr]\,\psi_{4}^{+} -\;\bigl[\;8\sqrt{3} \eta m_{B^+}^{2} m_p x_2 \bigl(\eta x_4 - 2\bigr) \bigl(x_2 + x_4 - 1\bigr)\,\bigl(-1 + x_4 + 2 x_2\bigr)\notag \\
	& +\;2\sqrt{6} m_{B^+} m_p \bigl(\eta x_4 - 1\bigr) \bigl(x_2 + x_4 - 1\bigr)\,\bigl(-8 m_p x_2^{2} + 4 m_p x_2 - \eta m_{B^+} x_2 - \eta m_{B^+} x_4\notag \\
	& - 4 m_p x_2 x_4 + 2 \eta m_{B^+} x_2^{2} + 2 \eta m_{B^+} x_4^{2} + 2 \eta m_{B^+} x_2 x_4\bigr)\bigr]\,\psi_{4}^{-}\notag \\
	& -\;2 m_{B^+} m_p^{2} \bigl(- \sqrt{6} \eta x_4^{2} + \sqrt{3} x_3 + \sqrt{6} x_4 - \sqrt{3}\bigr)\,\psi_{5}^{0} -\;\bigl[\;2\sqrt{3} m_{B^+} m_p^{2} \bigl(2 x_2^{2} - x_2 + 2 x_4^{2} - x_4\bigr)\notag \\
	& +\;2\sqrt{6} m_{B^+} m_p^{2} x_4 \bigl(2 x_4 - 1\bigr) \bigl(\eta x_4 - 1\bigr)\bigr]\,\psi_{5}^{+} -\;\bigl[\;2\sqrt{3} m_{B^+} m_p^{2} \bigl(x_2^{2} + 4 x_2 x_4 - x_2 + x_4^{2} - x_4\bigr)\notag \\
	& -\;2\sqrt{6} m_{B^+} m_p^{2} x_4 \bigl(\eta x_4 - 1\bigr) \bigl(2 x_2 + x_4 - 1\bigr)\bigr]\,\psi_{5}^{-} -\;\bigl[-\;16\sqrt{3} m_{B^+} m_p^{2} x_2 x_3\notag \\
	& +\;2\sqrt{6} \eta m_{B^+}^{2} m_p\,\bigl(-8 x_4^{2} - 3 x_3^{2} + 3 x_3 + 8 x_4 - 12 x_3 x_4 + \eta x_3^{2} x_4 - \eta x_3 x_4 + 2 \eta x_3 x_4^{2}\bigr)\bigr]\,\xi_{4}^{0}\notag \\
	& -\;\bigl[\;16\sqrt{3} m_{B^+} m_p^{2} x_2 x_3 \bigl(5 x_4 - 1\bigr) +\;2\sqrt{6} \eta m_{B^+}^{2} m_p\,\bigl(-8 x_4^{2} - 3 x_3^{2} + 3 x_3 + 8 x_4 - 62 x_3 x_4\notag \\
	& + 50 x_3 x_4^{2} + 50 x_3^{2} x_4 + \eta x_3^{2} x_4 - \eta x_3 x_4 + 2 \eta x_3 x_4^{2}\bigr)\bigr]\,\xi_{4}^{+}\notag \\
	& -\;\bigl[\;16\sqrt{3} m_{B^+} m_p^{2} x_2 \bigl(x_2 + x_4 - 1\bigr) \bigl(2 x_2 + x_4 - 1\bigr)\notag \\
	&+\;2\sqrt{6} \eta m_{B^+}^{2} m_p \bigl(x_2 + x_4 - 1\bigr)\,\bigl(-x_4 - 6 x_2^{2} + 2 x_4^{2} + 3 x_2\notag \\
	&+ \eta x_4^{2} - 2 \eta x_4^{3} - 2 x_2 x_4 - \eta x_2 x_4 + 2 \eta x_2^{2} x_4\bigr)\bigr]\,\xi_{4}^{-}\notag \\
	& +\;4 \sqrt{3} m_{B^+} m_p^{2} x_3 \bigl(\eta x_4 - 1\bigr)\,\xi_{5}^{0} -\;2 \sqrt{3} m_{B^+} m_p^{2} \bigl(2 x_3 - 1\bigr) \bigl(2 \eta x_3 x_4 - x_3 + 2 x_4 - 1\bigr)\,\xi_{5}^{+}\notag \\
	& -\;2 \sqrt{3} m_{B^+} m_p^{2} x_3 \bigl(x_2 - x_4\bigr)\,\xi_{5}^{-}\Biggr\}.
\end{align}

\begin{align}
	\mathcal M_c={}& -\frac{\pi C_F\Lambda_{\mathrm{QCD}}}{18\sqrt{2}}\frac{\phi_{B^+}}{\eta^{2}}\int_0^1dx_1\,dx_3\,dx_4\int_0^{1/\Lambda_{\mathrm{QCD}}}db_1\,db_3\,db_4\notag \\
	& b_3b_4\,E(t_c)\,h_3(x_1,x_3,x_4,b_1,b_3,b_4)\notag \\
	&\times\Biggl\{-\;40 x_2 x_3 x_4 \bigl(\sqrt{6} \eta m_{B^+} - 2 \sqrt{3} m_p\bigr)\,\bigl(-m_p^{2} + \eta^{2} m_{B^+}^{2}\notag \\
	&+ \eta m_p^{2} x_4\bigr)\,\bigl(\phi_{3}^{0} + \phi_{3}^{-} \bigl(x_3 + 2 x_4 - 1\bigr) + \phi_{3}^{+} \bigl(1 - 3 x_3\bigr)\bigr)\notag \\
	& -\;16 \sqrt{3} m_p x_2 x_4 \bigl(\eta^{2} m_{B^+}^{2} + \eta m_p^{2} x_4 - m_p^{2}\bigr)\,\bigl(\phi_{4}^{0} + \phi_{4}^{-} \bigl(x_3 + 2 x_4 - 1\bigr) + \phi_{4}^{+} \bigl(1 - 5 x_3\bigr)\bigr)\notag \\
	& -\;\bigl[\;4\sqrt{3} m_p^{2}\,\bigl(-m_p x_3 + 2 \eta m_{B^+} - \eta^{2} m_{B^+} x_4 - 4 \eta m_{B^+} x_4 - 2 \eta m_{B^+} x_3\notag \\
	& + 2 \eta^{2} m_{B^+} x_4^{2} + \eta m_p x_3 x_4 + \eta^{2} m_{B^+} x_3 x_4\bigr)\notag \\
	& -\;2\sqrt{6} \eta m_{B^+} m_p^{2} x_3 \bigl(\eta x_4 - 1\bigr)\bigr]\,\phi_{5}^{0} -\;2 m_p^{2} x_3 \bigl(2 x_3 - 1\bigr) \bigl(\eta x_4 - 1\bigr)\,\bigl(-2 \sqrt{3} m_p + \sqrt{6} \eta m_{B^+}\bigr)\,\phi_{5}^{+}\notag \\
	& -\;2 m_p^{2} x_3 \bigl(x_2 - x_4\bigr) \bigl(\eta x_4 - 1\bigr)\,\bigl(-2 \sqrt{3} m_p + \sqrt{6} \eta m_{B^+}\bigr)\,\phi_{5}^{-}\notag \\
	& +\;\frac{4 \sqrt{3} m_p^{3} \bigl(\eta x_4 - 1\bigr)}{3}\,\bigl(\phi_{6}^{0} + \phi_{6}^{-} \bigl(x_3 + 2 x_4 - 1\bigr) + \phi_{6}^{+} \bigl(1 - 3 x_3\bigr)\bigr)\notag \\
	& +\;8 \eta m_{B^+} m_p x_3 \bigl(x_2 - x_4\bigr) \bigl(\eta x_4 - 2\bigr)\,\bigl(-2 \sqrt{3} m_p + \sqrt{6} \eta m_{B^+}\bigr)\,\bigl(\xi_{4}^{0} + \xi_{4}^{-} \bigl(1 - 2 x_3\bigr) + \xi_{4}^{+}\bigr)\notag \\
	& -\;4 \sqrt{3} \eta m_{B^+} m_p^{2} \bigl(x_2 - x_4\bigr) \bigl(2 x_3 - 1\bigr)\,\bigl(-2 + \eta x_4\bigr)\,\xi_{5}^{+}\notag \\
	& +\;4 \sqrt{3} \eta m_{B^+} m_p^{2} x_3 \bigl(x_2 - x_4\bigr) \bigl(\eta x_4 - 2\bigr)\,\xi_{5}^{-}\Biggr\}.
\end{align}
where $C_F$ denotes the color factor, and $h_i(x_i,\textbf{b}_i)$ are the
perturbatively calculable hard-scattering kernels generated by hard-gluon
exchange. The factors $E(t_i)$ include the running coupling and the Sudakov
evolution effects associated with the participating hadron wave functions. The decay amplitudes in the type-II scenario are given as,
\begin{equation}
	\begin{aligned}
	\mathcal M_a' ={}&\frac{\Lambda_{\mathrm{QCD}} C_F}{72\sqrt{2}\pi}\int_0^1 dx_1\,dx_3\,dx_4\int_0^{1/\Lambda_{\mathrm{QCD}}} db_1\,db_3\,db_4 b_1b_4\,E(t_a)\phi_{B^+}\, h_1(x_1,x_4,b_1,b_4)\,\\
	&\quad\times \Bigl\{\eta^{2}\,m_{B^+}^{3}\bigl[\,40 \sqrt{3} x_{2} x_{3} x_{4} \bigl(x_{2} - 1\bigr)\,\bigl(2 \phi_{3}^{0} + \bigl(1 - 3 x_{3}\bigr) \bigl(\phi_{3}^{-} - \phi_{3}^{+}\bigr)\bigr)\,\bigr]\\
	&\quad {}+\eta\,m_{B^+}^{3}\,\bigl[\,40 \sqrt{6} x_{1} x_{2} x_{3} x_{4}\,\bigl(\phi_{3}^{0} + \phi_{3}^{-} \bigl(- x_{2} + x_{4}\bigr) + \phi_{3}^{+} \bigl(1 - 3 x_{3}\bigr)\bigr)\,\bigr]\\
	&\quad {}+\eta\,m_{B^+}^{2}\,m_p\,\bigl[-\,40 \sqrt{6} x_{2} x_{3} x_{4} \bigl(x_{2} - 1\bigr)\,\bigl(2 \phi_{3}^{0} - \bigl(\phi_{3}^{-} - \phi_{3}^{+}\bigr) \bigl(3 x_{3} - 1\bigr)\bigr)\\
	&\quad {}+16 \sqrt{3} x_{2} x_{3} \bigl(1 - x_{2}\bigr)\,\psi_{4}^{0} -\,16 \sqrt{3} x_{2} x_{3} \bigl(x_{2} - 1\bigr) \bigl(x_{2} - x_{3}\bigr)\,\psi_{4}^{-}\\
	&\quad {}+16 \sqrt{3} x_{2} x_{3} \bigl(x_{2} - 1\bigr) \bigl(5 x_{4} - 1\bigr)\,\psi_{4}^{+} +\,4 \sqrt{6} x_{2} \bigl(x_{2} - 1\bigr) \bigl(x_{3} + 4 x_{4}\bigr)\,\xi_{4}^{0}\\
	&\quad {}+4 \sqrt{6} x_{2} x_{3} \bigl(x_{2} - 1\bigr) \bigl(x_{2} - x_{3}\bigr)\,\xi_{4}^{-} +\,4 \sqrt{6} x_{2} \bigl(1 - x_{2}\bigr) \bigl(25 x_{3} x_{4} - x_{3} - 4 x_{4}\bigr)\,\xi_{4}^{+}\,\bigr]\\
	&\quad {}+m_{B^+}^{2}\,m_p\,\bigl[\,160 \sqrt{3} x_{1} x_{2} x_{3} x_{4}\,\bigl(- \phi_{3}^{0} + \phi_{3}^{-} \bigl(x_{2} - x_{4}\bigr) + \phi_{3}^{+} \bigl(3 x_{3} - 1\bigr)\bigr)\\
	&\quad {}+2 x_{1} \bigl(16 \sqrt{3} x_{2} x_{4} + \sqrt{6} x_{3}^{2} - \sqrt{6} x_{3}\bigr)\,\phi_{4}^{0} -\,32 \sqrt{3} x_{1} x_{2} x_{4} \bigl(x_{2} - x_{4}\bigr)\,\phi_{4}^{-}\\
	&\quad {}-2 \sqrt{6} x_{1} x_{3} \bigl(x_{2}^{2} + x_{3}^{2} - x_{3} + x_{4}^{2}\bigr)\,\phi_{4}^{-} -\,32 \sqrt{3} x_{1} x_{2} x_{4} \bigl(5 x_{3} - 1\bigr)\,\phi_{4}^{+}\\
	&\quad {}+2 \sqrt{6} x_{1} x_{3} \bigl(10 x_{2} x_{4} + x_{3} - 1\bigr)\,\phi_{4}^{+} +\,2 x_{1} x_{3} \bigl(\sqrt{6} x_{3} + 8 \sqrt{3} x_{3} + 8 \sqrt{3} x_{4} - 8 \sqrt{3} - \sqrt{6}\bigr)\,\psi_{4}^{0}\\
	&\quad {}+16 \sqrt{3} x_{1} x_{2} x_{3} \bigl(- x_{2} + x_{3}\bigr)\,\psi_{4}^{-} +\,2 \sqrt{6} x_{1} x_{3} \bigl(2 x_{3}^{2} + 2 x_{3} x_{4} - 3 x_{3} + 2 x_{4}^{2} - 2 x_{4} + 1\bigr)\,\psi_{4}^{-}\\
	&\quad {}+16 \sqrt{3} x_{1} x_{2} x_{3} \bigl(5 x_{4} - 1\bigr)\,\psi_{4}^{+} +\,2 \sqrt{6} x_{1} x_{3} \bigl(10 x_{2} x_{4} + x_{3} - 1\bigr)\,\psi_{4}^{+}\\
	&\quad {}+2 \sqrt{6} x_{1} x_{3} \bigl(x_{2} - x_{4}\bigr)\,\bigl(\xi_{4}^{0} + \xi_{4}^{+}\bigr) -\,2 \sqrt{6} x_{1} x_{3} \bigl(x_{2} - x_{4}\bigr) \bigl(2 x_{3} - 1\bigr)\,\xi_{4}^{-}\,\bigr]\\
	&\quad {}+m_{B^+}\,m_p^{2}\,\bigl[\,40 \sqrt{6} x_{2} x_{3} x_{4} \bigl(x_{2} - 1\bigr)\,\bigl(- \phi_{3}^{0} + \phi_{3}^{-} \bigl(x_{2} - x_{4}\bigr) + \phi_{3}^{+} \bigl(3 x_{3} - 1\bigr)\bigr)\\
	&\quad {}+16 \sqrt{6} x_{2} x_{3} \bigl(x_{2} - 1\bigr)\,\psi_{4}^{0} +\,16 \sqrt{6} x_{2} x_{3} \bigl(x_{2} - 1\bigr) \bigl(x_{2} - x_{3}\bigr)\,\psi_{4}^{-}\\
	&\quad {}+16 \sqrt{6} x_{2} x_{3} \bigl(1 - x_{2}\bigr) \bigl(5 x_{4} - 1\bigr)\,\psi_{4}^{+} +\,2 \sqrt{6} x_{3} \bigl(x_{2} - 1\bigr)\,\phi_{5}^{0}\\
	&\quad {}+2 \sqrt{6} x_{3} \bigl(x_{2} - x_{4}\bigr) \bigl(x_{3} + x_{4}\bigr)\,\phi_{5}^{-} +\,2 \sqrt{6} x_{3} \bigl(1 - x_{2}\bigr) \bigl(2 x_{3} - 1\bigr)\,\phi_{5}^{+}\\
	&\quad {}+4 \sqrt{6} x_{4} \bigl(1 - x_{2}\bigr)\,\psi_{5}^{0} -\,4 \sqrt{6} x_{4} \bigl(x_{2} - 1\bigr) \bigl(x_{2} - x_{3}\bigr)\,\psi_{5}^{-}\\
	&\quad {}+4 \sqrt{6} x_{4} \bigl(x_{2} - 1\bigr) \bigl(2 x_{4} - 1\bigr)\,\psi_{5}^{+}\,\bigr]\\
	&\quad {}+\eta^{-1}\,m_{B^+}\,m_p^{2}\,\bigl[-\,8 \sqrt{3} x_{1} x_{3} \bigl(x_{3} - 1\bigr)\,\bigl(\phi_{4}^{0} + \psi_{4}^{0}\bigr)\\
	&\quad {}+8 \sqrt{3} x_{1} x_{3} \bigl(x_{2}^{2} + x_{3}^{2} - x_{3} + x_{4}^{2}\bigr)\,\bigl(\phi_{4}^{-} - \psi_{4}^{-}\bigr)\\
	&\quad {}-8 \sqrt{3} x_{1} x_{3} \bigl(10 x_{2} x_{4} + x_{3} - 1\bigr)\,\bigl(\phi_{4}^{+} + \psi_{4}^{+}\bigr) -\,8 \sqrt{3} x_{1} x_{3} \bigl(x_{2} - x_{4}\bigr)\,\bigl(\xi_{4}^{0} + \xi_{4}^{+}\bigr)\\
	&\quad {}+8 \sqrt{3} x_{1} x_{3} \bigl(x_{2} - x_{4}\bigr) \bigl(2 x_{3} - 1\bigr)\,\xi_{4}^{-} +\,2 \sqrt{3} x_{1} \bigl(x_{2} - x_{4}\bigr)\,\phi_{5}^{0} +\,2 \sqrt{3} x_{1} \bigl(x_{3} + 1\bigr)\,\xi_{5}^{0}\\
	&\quad {}+4 \sqrt{3} x_{1} x_{4} \bigl(x_{2} - x_{3}\bigr)\,\xi_{5}^{-} -\,4 \sqrt{3} x_{1} \bigl(x_{2} + 2 x_{3}^{2} + 2 x_{4}^{2} - 1\bigr)\,\xi_{5}^{+}\,\bigr]\,\Bigr\}.
\end{aligned}
\end{equation}

\begin{equation}
	\begin{aligned}
	\mathcal M_b' ={}&-\frac{\pi\Lambda_{\mathrm{QCD}} C_F}{18\sqrt{2}}\int_0^1 dx_1\,dx_3\,dx_4\int_0^{1/\Lambda_{\mathrm{QCD}}} db_1\,db_3\,db_4 b_3b_4\,E(t_b)\,\phi_{B^+}\, h_2(x_1,x_3,x_4,b_1,b_3,b_4)\,\\
	&\quad\times \Bigl\{\eta\,m_{B^+}^{3}\,\bigl[\,40 x_{1} x_{2} x_{3} x_{4}\,\bigl(\phi_{3}^{0} \bigl(- 2 \sqrt{3} + \sqrt{6}\bigr)\\
	&\quad {}+\phi_{3}^{-} \bigl(\sqrt{6} x_{3} + 3 \sqrt{3} x_{3} + 2 \sqrt{6} x_{4} - \sqrt{6} - \sqrt{3}\bigr) + \phi_{3}^{+} \bigl(- 3 \sqrt{6} x_{3} - 3 \sqrt{3} x_{3} + \sqrt{3} + \sqrt{6}\bigr)\bigr)\,\bigr]\\
	&\quad {}+\eta\,m_{B^+}^{2}\,m_p\,\bigl[\,2 \sqrt{6} x_{3} \bigl(1 - x_{2}\bigr) \bigl(x_{3} - 1\bigr)\,\bigl(\phi_{4}^{0} + \psi_{4}^{0}\bigr)\\
	&\quad {}-2 \sqrt{6} x_{3} \bigl(x_{3} + x_{4}\bigr) \bigl(x_{2}^{2} + x_{3}^{2} - x_{3} + x_{4}^{2}\bigr)\,\bigl(\phi_{4}^{-} - \psi_{4}^{-}\bigr)\\
	&\quad {}+2 \sqrt{6} x_{3} \bigl(x_{3} + x_{4}\bigr) \bigl(10 x_{2} x_{4} + x_{3} - 1\bigr)\,\bigl(\phi_{4}^{+} + \psi_{4}^{+}\bigr)\\
	&\quad {}+2 \sqrt{6} x_{3} \bigl(x_{2} - 1\bigr) \bigl(x_{2} - x_{4}\bigr)\,\bigl(\xi_{4}^{0} + \xi_{4}^{+}\bigr) +\,2 \sqrt{6} x_{3} \bigl(x_{2} - x_{4}\bigr) \bigl(x_{3} + x_{4}\bigr) \bigl(2 x_{3} - 1\bigr)\,\xi_{4}^{-}\,\bigr]\\
	&\quad {}+m_{B^+}^{2}\,m_p\,\bigl[\,40 x_{1} x_{2} x_{3} x_{4}\,\bigl(\phi_{3}^{0} \bigl(- 2 \sqrt{3} + 2 \sqrt{6}\bigr)\\
	&\quad {}+\phi_{3}^{-} \bigl(2 \sqrt{3} x_{2} - 3 \sqrt{6} x_{3} - 2 \sqrt{3} x_{4} + \sqrt{6}\bigr) + \phi_{3}^{+} \bigl(3 \sqrt{6} x_{3} + 6 \sqrt{3} x_{3} - 2 \sqrt{3} - \sqrt{6}\bigr)\bigr)\\
	&\quad {}+16 \sqrt{3} x_{1} x_{2} x_{4}\,\phi_{4}^{0} -\,16 \sqrt{3} x_{1} x_{2} x_{4} \bigl(x_{2} - x_{4}\bigr)\,\phi_{4}^{-} -\,16 \sqrt{3} x_{1} x_{2} x_{4} \bigl(5 x_{3} - 1\bigr)\,\phi_{4}^{+}\\
	&\quad {}+4 \sqrt{6} x_{1} x_{4} \bigl(3 x_{3} + 4 x_{4} - 4\bigr)\,\xi_{4}^{0} +\,4 \sqrt{6} x_{1} x_{3} x_{4} \bigl(x_{3} - x_{4}\bigr)\,\xi_{4}^{-}\\
	&\quad {}-4 \sqrt{6} x_{1} x_{4} \bigl(25 x_{3}^{2} + 25 x_{3} x_{4} - 28 x_{3} - 4 x_{4} + 4\bigr)\,\xi_{4}^{+}\,\bigr]\\
	&\quad {}+m_{B^+}\,m_p^{2}\,\bigl[\,40 \sqrt{6} x_{2} x_{3} x_{4} \bigl(x_{2} - 1\bigr)\,\bigl(- \phi_{3}^{0} + \phi_{3}^{-} \bigl(x_{2} - x_{4}\bigr) + \phi_{3}^{+} \bigl(3 x_{3} - 1\bigr)\bigr)\\
	&\quad {}+8 \sqrt{3} x_{3} \bigl(x_{2} - 1\bigr) \bigl(x_{3} - 1\bigr)\,\bigl(\phi_{4}^{0} + \psi_{4}^{0}\bigr)\\
	&\quad {}+8 \sqrt{3} x_{3} \bigl(x_{3} + x_{4}\bigr) \bigl(x_{2}^{2} + x_{3}^{2} - x_{3} + x_{4}^{2}\bigr)\,\bigl(\phi_{4}^{-} - \psi_{4}^{-}\bigr)\\
	&\quad {}-8 \sqrt{3} x_{3} \bigl(x_{3} + x_{4}\bigr) \bigl(10 x_{2} x_{4} + x_{3} - 1\bigr)\,\bigl(\phi_{4}^{+} + \psi_{4}^{+}\bigr)\\
	&\quad {}+8 \sqrt{3} x_{3} \bigl(x_{2} - x_{4}\bigr) \bigl(x_{3} + x_{4}\bigr)\,\bigl(\xi_{4}^{0} + \xi_{4}^{+}\bigr) +\,8 \sqrt{3} x_{3} \bigl(x_{2} - 1\bigr) \bigl(x_{2} - x_{4}\bigr) \bigl(2 x_{3} - 1\bigr)\,\xi_{4}^{-}\\
	&\quad {}+2 \bigl(1 - x_{2}\bigr) \bigl(- \sqrt{6} x_{3} + \sqrt{3} x_{3} + 2 \sqrt{3} x_{4} - \sqrt{3}\bigr)\,\phi_{5}^{0}\\
	&\quad {}+2 \sqrt{6} x_{3} \bigl(x_{2} - x_{4}\bigr) \bigl(x_{3} + x_{4}\bigr)\,\phi_{5}^{-}\\
	&\quad {}+2 \sqrt{6} x_{3} \bigl(1 - x_{2}\bigr) \bigl(2 x_{3} - 1\bigr)\,\phi_{5}^{+} +\,2 \sqrt{3} \bigl(1 - x_{2}\bigr) \bigl(x_{3} + 1\bigr)\,\xi_{5}^{0}\\
	&\quad {}+4 \sqrt{3} x_{2} \bigl(x_{2} - 1\bigr) \bigl(x_{3} - x_{4}\bigr)\,\xi_{5}^{-} +\,4 \sqrt{3} \bigl(x_{2} - 1\bigr) \bigl(4 x_{3}^{2} + 4 x_{3} x_{4} - 4 x_{3} + 2 x_{4}^{2} - 3 x_{4} + 1\bigr)\,\xi_{5}^{+}\,\bigr]\\
	&\quad {}+\eta^{-1}\,m_p^{3}\,\bigl[\,80 \sqrt{3} x_{2} x_{3} x_{4}\,\bigl(\phi_{3}^{0} \bigl(x_{2} - 1\bigr) + \phi_{3}^{-} \bigl(x_{2} x_{3} + x_{2} x_{4} - x_{3} x_{4} - x_{4}^{2}\bigr)\\
	&\quad {}+\phi_{3}^{+} \bigl(- 3 x_{2} x_{3} + x_{2} + 3 x_{3} - 1\bigr)\bigr)\\
	&\quad {}+16 \sqrt{3} x_{2} x_{4} \bigl(1 - x_{2}\bigr)\,\phi_{4}^{0} +\,16 \sqrt{3} x_{2} x_{4} \bigl(x_{2} - 1\bigr) \bigl(x_{2} - x_{4}\bigr)\,\phi_{4}^{-}\\
	&\quad {}+16 \sqrt{3} x_{2} x_{4} \bigl(x_{2} - 1\bigr) \bigl(5 x_{3} - 1\bigr)\,\phi_{4}^{+} +\,16 \sqrt{3} x_{2} x_{3} \bigl(1 - x_{2}\bigr)\,\psi_{4}^{0}\\
	&\quad {}-16 \sqrt{3} x_{2} x_{3} \bigl(x_{2} - 1\bigr) \bigl(x_{2} - x_{3}\bigr)\,\psi_{4}^{-} +\,16 \sqrt{3} x_{2} x_{3} \bigl(x_{2} - 1\bigr) \bigl(5 x_{4} - 1\bigr)\,\psi_{4}^{+}\\
	&\quad {}+4 \sqrt{3} x_{3} \bigl(1 - x_{2}\bigr)\,\phi_{5}^{0} +\,4 \sqrt{3} x_{3} \bigl(x_{2} - 1\bigr) \bigl(x_{2} - x_{4}\bigr)\,\phi_{5}^{-}\\
	&\quad {}+4 \sqrt{3} x_{3} \bigl(x_{2} - 1\bigr) \bigl(2 x_{3} - 1\bigr)\,\phi_{5}^{+} +\,4 \sqrt{3} \bigl(x_{2} - 1\bigr) / 3\,\phi_{6}^{0}\\
	&\quad {}+4 \sqrt{3} \bigl(x_{2} - x_{4}\bigr) \bigl(x_{3} + x_{4}\bigr) / 3\,\phi_{6}^{-} +\,4 \sqrt{3} \bigl(1 - x_{2}\bigr) \bigl(3 x_{3} - 1\bigr) / 3\,\phi_{6}^{+}\,\bigr]\,\Bigr\}.
\end{aligned}
\end{equation}
\begin{equation}
\begin{aligned}
	\mathcal M_c' ={}&\frac{\pi\Lambda_{\mathrm{QCD}} C_F}{18\sqrt{2}}\int_0^1 dx_1\,dx_3\,dx_4\int_0^{1/\Lambda_{\mathrm{QCD}}} db_1\,db_3\,db_4 b_3b_4\,E(t_c)\,\phi_{B^+}\, h_3\,\\
	&\quad \times\Bigl\{\eta\,m_{B^+}^{3}\,\bigl[\,20 \sqrt{6} x_{1} x_{2} x_{3} x_{4}\,\bigl(- \phi_{3}^{0} + \phi_{3}^{-} \bigl(x_{2} - x_{4}\bigr) + \phi_{3}^{+} \bigl(3 x_{3} - 1\bigr)\bigr)\,\bigr]\\
	&\quad {}+\eta\,m_{B^+}^{2}\,m_p\,\bigl[\,8 \sqrt{3} x_{2} x_{3} \bigl(1 - x_{2}\bigr)\,\psi_{4}^{0} -\,8 \sqrt{3} x_{2} x_{3} \bigl(x_{2} - 1\bigr) \bigl(x_{2} - x_{3}\bigr)\,\psi_{4}^{-}\\
	&\quad {}+8 \sqrt{3} x_{2} x_{3} \bigl(x_{2} - 1\bigr) \bigl(5 x_{4} - 1\bigr)\,\psi_{4}^{+} -\,4 \sqrt{6} x_{3} \bigl(x_{2} - 1\bigr) \bigl(x_{2} - x_{4}\bigr)\,\bigl(\xi_{4}^{0} + \xi_{4}^{+}\bigr)\\
	&\quad {}+4 \sqrt{6} x_{3} \bigl(x_{2} - 1\bigr) \bigl(x_{2} - x_{4}\bigr) \bigl(2 x_{3} - 1\bigr)\,\xi_{4}^{-}\\
	&\quad {}-\sqrt{6} \bigl(x_{2} - 1\bigr) \bigl(x_{2} - x_{4}\bigr) / 2\,\bigl(\phi_{5}^{0} + \psi_{5}^{0} - \xi_{5}^{0}\bigr)\\
	&\quad {}-\sqrt{6} x_{3} \bigl(x_{2} - 1\bigr) \bigl(x_{2} - x_{4}\bigr) / 2\,\bigl(- \phi_{5}^{-} + \psi_{5}^{-} + \xi_{5}^{-}\bigr)\\
	&\quad {}-\sqrt{6} \bigl(x_{2} - 1\bigr) \bigl(x_{2} - x_{4}\bigr) \bigl(2 x_{3} - 1\bigr) / 2\,\bigl(\phi_{5}^{+} + \psi_{5}^{+} - \xi_{5}^{+}\bigr)\,\bigr]\\
	&\quad {}+m_{B^+}^{2}\,m_p\,\bigl[-\,8 \sqrt{3} x_{1} x_{2} x_{3}\,\psi_{4}^{0} -\,8 \sqrt{3} x_{1} x_{2} x_{3} \bigl(x_{2} - x_{3}\bigr)\,\psi_{4}^{-} +\,8 \sqrt{3} x_{1} x_{2} x_{3} \bigl(5 x_{4} - 1\bigr)\,\psi_{4}^{+}\\
	&\quad {}-4 \sqrt{6} x_{1} x_{3} \bigl(x_{2} - x_{4}\bigr)\,\bigl(\xi_{4}^{0} + \xi_{4}^{+}\bigr) +\,4 \sqrt{6} x_{1} x_{3} \bigl(x_{2} - x_{4}\bigr) \bigl(2 x_{3} - 1\bigr)\,\xi_{4}^{-}\\
	&\quad {}-\sqrt{6} x_{1} \bigl(x_{2} - x_{4}\bigr) / 2\,\bigl(\phi_{5}^{0} + \psi_{5}^{0} - \xi_{5}^{0}\bigr) -\,\sqrt{6} x_{1} x_{3} \bigl(x_{2} - x_{4}\bigr) / 2\,\bigl(- \phi_{5}^{-} + \psi_{5}^{-} + \xi_{5}^{-}\bigr)\\
	&\quad {}-\sqrt{6} x_{1} \bigl(x_{2} - x_{4}\bigr) \bigl(2 x_{3} - 1\bigr) / 2\,\bigl(\phi_{5}^{+} + \psi_{5}^{+} - \xi_{5}^{+}\bigr)\,\bigr]\\
	&\quad {}+m_{B^+}\,m_p^{2}\,\bigl[\,20 \sqrt{6} x_{2} x_{3} x_{4}\,\bigl(\phi_{3}^{0} \bigl(x_{2} - 1\bigr) + \phi_{3}^{-} \bigl(x_{2} x_{3} + x_{2} x_{4} - x_{3} x_{4} - x_{4}^{2}\bigr)\\
	&\quad {}+\phi_{3}^{+} \bigl(- 3 x_{2} x_{3} + x_{2} + 3 x_{3} - 1\bigr)\bigr)\\
	&\quad {}+8 \sqrt{6} x_{2} x_{3} \bigl(x_{2} - 1\bigr)\,\psi_{4}^{0} +\,8 \sqrt{6} x_{2} x_{3} \bigl(x_{2} - 1\bigr) \bigl(x_{2} - x_{3}\bigr)\,\psi_{4}^{-}\\
	&\quad {}+8 \sqrt{6} x_{2} x_{3} \bigl(1 - x_{2}\bigr) \bigl(5 x_{4} - 1\bigr)\,\psi_{4}^{+} +\,16 \sqrt{3} x_{3} \bigl(x_{2} - 1\bigr) \bigl(x_{2} - x_{4}\bigr)\,\bigl(\xi_{4}^{0} + \xi_{4}^{+}\bigr)\\
	&\quad {}-16 \sqrt{3} x_{3} \bigl(x_{2} - 1\bigr) \bigl(x_{2} - x_{4}\bigr) \bigl(2 x_{3} - 1\bigr)\,\xi_{4}^{-} +\,\sqrt{6} x_{3} \bigl(1 - x_{2}\bigr)\,\phi_{5}^{0}\\
	&\quad {}-x_{3} \bigl(- \sqrt{6} + 4 \sqrt{3}\bigr) \bigl(x_{2} - 1\bigr) \bigl(x_{2} - x_{4}\bigr)\,\phi_{5}^{-} +\,4 \sqrt{3} \bigl(1 - x_{2}\bigr) \bigl(- x_{2} + x_{4}\bigr) \bigl(2 x_{3} - 1\bigr)\,\phi_{5}^{+}\\
	&\quad {}+\sqrt{6} x_{3} \bigl(x_{2} - 1\bigr) \bigl(2 x_{3} - 1\bigr)\,\phi_{5}^{+} +\,2 \bigl(1 - x_{2}\bigr) \bigl(2 \sqrt{3} x_{3} + \sqrt{6} x_{4} + 4 \sqrt{3} x_{4} - 2 \sqrt{3}\bigr)\,\psi_{5}^{0}\\
	&\quad {}+4 \sqrt{3} x_{3} \bigl(1 - x_{2}\bigr) \bigl(- x_{2} + x_{4}\bigr)\,\psi_{5}^{-} -\,2 \sqrt{6} x_{4} \bigl(1 - x_{2}\bigr) \bigl(- x_{2} + x_{3}\bigr)\,\psi_{5}^{-}\\
	&\quad {}+4 \sqrt{3} \bigl(1 - x_{2}\bigr) \bigl(- x_{2} + x_{4}\bigr) \bigl(2 x_{3} - 1\bigr)\,\psi_{5}^{+} +\,2 \sqrt{6} x_{4} \bigl(x_{2} - 1\bigr) \bigl(2 x_{4} - 1\bigr)\,\psi_{5}^{+}\\
	&\quad {}-4 \sqrt{3} \bigl(x_{2} - 1\bigr) \bigl(x_{2} - x_{4}\bigr)\,\xi_{5}^{0}\,\bigr]\\
	&\quad {}+\eta^{-1}\,m_p^{3}\,\bigl[\,80 \sqrt{3} x_{2} x_{3} x_{4} \bigl(x_{2} - 1\bigr)\,\bigl(- \phi_{3}^{0} + \phi_{3}^{-} \bigl(x_{2} - x_{4}\bigr) + \phi_{3}^{+} \bigl(3 x_{3} - 1\bigr)\bigr)\\
	&\quad {}+16 \sqrt{3} x_{2} x_{4} \bigl(x_{2} - 1\bigr)\,\phi_{4}^{0} +\,16 \sqrt{3} x_{2} x_{4} \bigl(x_{2} - x_{4}\bigr) \bigl(x_{3} + x_{4}\bigr)\,\phi_{4}^{-}\\
	&\quad {}+16 \sqrt{3} x_{2} x_{4} \bigl(1 - x_{2}\bigr) \bigl(5 x_{3} - 1\bigr)\,\phi_{4}^{+} +\,16 \sqrt{3} x_{2} x_{3} \bigl(x_{2} - 1\bigr)\,\psi_{4}^{0}\\
	&\quad {}+16 \sqrt{3} x_{2} x_{3} \bigl(x_{2} - 1\bigr) \bigl(x_{2} - x_{3}\bigr)\,\psi_{4}^{-} +\,16 \sqrt{3} x_{2} x_{3} \bigl(1 - x_{2}\bigr) \bigl(5 x_{4} - 1\bigr)\,\psi_{4}^{+}\\
	&\quad {}+4 \sqrt{3} x_{3} \bigl(x_{2} - 1\bigr)\,\phi_{5}^{0} +\,4 \sqrt{3} x_{3} \bigl(x_{2} - x_{4}\bigr) \bigl(x_{3} + x_{4}\bigr)\,\phi_{5}^{-}\\
	&\quad {}+4 \sqrt{3} x_{3} \bigl(1 - x_{2}\bigr) \bigl(2 x_{3} - 1\bigr)\,\phi_{5}^{+}\\
	&\quad {}+\sqrt{3} \bigl(x_{2} - 1\bigr)\,\bigl(- 4 \phi_{6}^{0} / 3 + \phi_{6}^{-} \bigl(4 x_{2} / 3 - 4 x_{4} / 3\bigr)\\
	&\quad {}+\phi_{6}^{+} \bigl(4 x_{3} - 4 / 3\bigr)\bigr)\,\bigr]\,\Bigr\}.
\end{aligned}
\end{equation}
The hard functions $h_i(x_i,\mathbf{b}_i)$s are obtained by performing the Fourier transformation from momentum space to impact-parameter
space, which given as follows:
\begin{equation}
	\begin{aligned}
		&h_1(x_1,x_4,\textbf{b}_1,\textbf{b}_4)=(2\pi)^2 K_0(\beta_1|\textbf{b}_1|)K_0(\alpha_1 |\textbf{b}_4-\textbf{b}_1|)S_t(x_4), \\
		&	h_2(x_1,x_3,x_4,\textbf{b}_1,\textbf{b}_3,\textbf{b}_4)= K_0(\alpha_2 \textbf{b}_4)K_0(\beta_2(\textbf{b}_1+\textbf{b}_3))|_{\textbf{b}_1=\textbf{b}_4} S_t(x_4),\\
		&h_3(x_1,x_3,x_4,\textbf{b}_1,\textbf{b}_3,\textbf{b}_4)= K_0(\alpha_3 \textbf{b}_3)K_0(\beta_3 \textbf{b}_4)|_{\textbf{b}_1=\textbf{b}_3+\textbf{b}_4} S_t(x_1 )S_t(x_4),
	\end{aligned}
\end{equation}
where $K_0$ and $H_0^{(1)} = J_0 + i Y_0$ are Bessel functions, and $\alpha_i^2$ and $\beta_i^2$ are the virtualities of the internal gluon and quark, with the explicit forms giving:
\begin{align}
	&\alpha_1^2=m_B^2(1-\eta x_4),\,\ \alpha_2^2=m_B^2(\eta x_3+\eta x_4)x_1,\,\ \alpha^2_3=m_B^2(\eta x_2+\eta x_4)x_1,\\\nonumber
	&\beta_1^2=m_B^2x_1 x_4\eta, \,\ \beta_2^2=m_B^2\eta x_4x_1,\,\beta_3^2=m_B^2x_1 x_4\eta.
\end{align}
The threshold resummation factor $S_t$ suppressing large endpoint logarithms to ensure reliable perturbative calculations, and evolution factors $E(t)$ resuming the logarithms in non-perturbative wave function evolution,
\begin{align}
	&S_t(x) = \frac{2^{1+2c}\Gamma(3/2+c)}{\sqrt{\pi}\Gamma(1+c)}[x(1-x)]^c,\\
    &E(t)=\alpha_s(t) exp\left[-S_B(t)-S_{\mathcal{B}}(t)\right],
\end{align}
here $c=0.4$, $\gamma_q=-\alpha_s/\pi$ is the quark anomalous dimension. $\alpha_s(t)$ is the QCD running coupling at the hard scale $t$, and $S_B(t)$ and $S_{\mathcal{B}}(t)$ denote the Sudakov exponents for the B meson and the light baryon, respectively.
\begin{equation}
\begin{aligned}
	&S_{B}(t) = s \left(\frac{m_{B}}{\sqrt{2}}x_1, b_1 \right) + \frac{5}{3} \int_{1/b_1}^{t} \frac{d\bar{\mu}}{\bar{\mu}} \gamma_q(\alpha_s(\bar{\mu})),\\
	&S_{\mathcal{B}}(t) = s \left(\frac{m_{B}}{\sqrt{2}} x_3, b_2 \right) + s \left(\frac{m_{B}}{\sqrt{2}} x_4, b_2 \right) + 3\int_{1/b_2}^t \frac{d\bar{\mu}}{\bar{\mu}} \gamma_q(\alpha_s(\bar{\mu})).
\end{aligned}
\end{equation}
The explicit form for the function $s(Q,b)$ can be found in Appendix A of ref~\cite{Ali:2007ff}.

The form factor $F_{B^+\to p_R}^{(d)}(q^2)$ and 	$\widetilde{F}_{B^+\to p_L}^{(d)}(q^2)$ can be extracted by matching to the Lorentz structures of Eq.~\ref{eq:formfactor_pqcd}, as follows:
	\begin{flalign}
	F_{B^+\to p_R}^{(d)}(q^2)={}&\frac{\pi\sqrt{3}\Lambda_{\mathrm{QCD}}C_F}{18}\int_0^1 dx_1\,dx_3\,dx_4\int_0^{1/\Lambda_{\mathrm{QCD}}}db_1\,db_3\,db_4\nonumber\\
	& \times\phi_{B^+}\,\biggl(\frac{b_1b_4}{4\pi^2}\,E(t_a)h_1(x_1,x_4,b_1,b_4)\bigl[\nonumber\\
	&+\,\phi_3^{0}\,\frac{80\,x_2\,x_3\,x_4\,\left(\eta^{2}\,m_{B^+}^{2} - 2\,\eta\,m_p^{2} + 2\,m_p^{2}\right)\,\left(\eta^{2}\,m_{B^+}^{2} + \eta\,m_p^{2}\,x_4 - m_p^{2}\right)}{\eta^{3}\,m_{B^+}}\nonumber\\[1pt]
	&+\,\phi_3^{-}\,\frac{80\,x_2\,x_3\,x_4\,\left(- x_2 + x_4\right)\,\left(\eta^{2}\,m_{B^+}^{2} - 2\,\eta\,m_p^{2} + 2\,m_p^{2}\right)\,\left(\eta^{2}\,m_{B^+}^{2} + \eta\,m_p^{2}\,x_4 - m_p^{2}\right)}{\eta^{3}\,m_{B^+}}\nonumber\\[1pt]
	&-\,\phi_3^{+}\,\frac{80\,x_2\,x_3\,x_4\,\left(3\,x_3 - 1\right)\,\left(\eta^{2}\,m_{B^+}^{2} - 2\,\eta\,m_p^{2} + 2\,m_p^{2}\right)\,\left(\eta^{2}\,m_{B^+}^{2} + \eta\,m_p^{2}\,x_4 - m_p^{2}\right)}{\eta^{3}\,m_{B^+}}\nonumber\\[1pt]
	&+\,\phi_4^{0}\,\frac{4\,m_p}{\eta^{3}\,m_{B^+}}\,\left(\begin{aligned}
		&{}+\eta^{3}\,m_{B^+}^{3}\,x_3^{2} - \eta^{3}\,m_{B^+}^{3}\,x_3 + 8\,m_p^{3}\,x_2\,x_4 - 8\,\eta\,m_p^{3}\,x_2\,x_4\\
		&{}-8\,\eta\,m_p^{3}\,x_2\,x_4^{2} - 2\,\eta\,m_{B^+}\,m_p^{2}\,x_3 - 2\,\eta^{2}\,m_{B^+}\,m_p^{2}\,x_3^{2}\\
		&{}+2\,\eta\,m_{B^+}\,m_p^{2}\,x_3^{2} + 2\,\eta^{2}\,m_{B^+}\,m_p^{2}\,x_3 + 8\,\eta^{2}\,m_p^{3}\,x_2\,x_4^{2}\\
		&{}-8\,\eta^{2}\,m_{B^+}^{2}\,m_p\,x_2\,x_4 + 8\,\eta^{3}\,m_{B^+}^{2}\,m_p\,x_2\,x_4
	\end{aligned}\right)\nonumber\\[1pt]
	&-\,\phi_4^{-}\,\frac{4\,m_p}{\eta^{3}\,m_{B^+}}\,\left(\begin{aligned}
		&{}+\eta^{3}\,m_{B^+}^{3}\,x_3^{3} - \eta^{3}\,m_{B^+}^{3}\,x_3^{2} - 8\,m_p^{3}\,x_2\,x_4^{2}\\
		&{}+8\,m_p^{3}\,x_2^{2}\,x_4 + \eta^{3}\,m_{B^+}^{3}\,x_2^{2}\,x_3 + \eta^{3}\,m_{B^+}^{3}\,x_3\,x_4^{2}\\
		&{}-8\,\eta\,m_p^{3}\,x_2^{2}\,x_4 - 8\,\eta\,m_p^{3}\,x_2^{2}\,x_4^{2} - 8\,\eta^{2}\,m_p^{3}\,x_2\,x_4^{3}\\
		&{}-2\,\eta\,m_{B^+}\,m_p^{2}\,x_3^{2} - 2\,\eta^{2}\,m_{B^+}\,m_p^{2}\,x_3^{3} + 2\,\eta\,m_{B^+}\,m_p^{2}\,x_3^{3}\\
		&{}+2\,\eta^{2}\,m_{B^+}\,m_p^{2}\,x_3^{2} + 8\,\eta\,m_p^{3}\,x_2\,x_4^{2} + 8\,\eta\,m_p^{3}\,x_2\,x_4^{3}\\
		&{}+8\,\eta^{2}\,m_p^{3}\,x_2^{2}\,x_4^{2} - 8\,\eta^{3}\,m_{B^+}^{2}\,m_p\,x_2\,x_4^{2}\\
		&{}-8\,\eta^{2}\,m_{B^+}^{2}\,m_p\,x_2^{2}\,x_4 - 2\,\eta^{2}\,m_{B^+}\,m_p^{2}\,x_2^{2}\,x_3\\
		&{}-2\,\eta^{2}\,m_{B^+}\,m_p^{2}\,x_3\,x_4^{2} + 2\,\eta\,m_{B^+}\,m_p^{2}\,x_2^{2}\,x_3\\
		&{}+2\,\eta\,m_{B^+}\,m_p^{2}\,x_3\,x_4^{2} + 8\,\eta^{2}\,m_{B^+}^{2}\,m_p\,x_2\,x_4^{2}\\
		&{}+8\,\eta^{3}\,m_{B^+}^{2}\,m_p\,x_2^{2}\,x_4
	\end{aligned}\right)\nonumber\\[1pt]
	&+\,\phi_4^{+}\,\frac{4\,m_p}{\eta^{3}\,m_{B^+}}\,\left(\begin{aligned}
		&{}+\eta^{3}\,m_{B^+}^{3}\,x_3^{2} - \eta^{3}\,m_{B^+}^{3}\,x_3 + 8\,m_p^{3}\,x_2\,x_4\\
		&{}-40\,m_p^{3}\,x_2\,x_3\,x_4 - 8\,\eta\,m_p^{3}\,x_2\,x_4 - 8\,\eta\,m_p^{3}\,x_2\,x_4^{2}\\
		&{}-2\,\eta\,m_{B^+}\,m_p^{2}\,x_3 - 2\,\eta^{2}\,m_{B^+}\,m_p^{2}\,x_3^{2} + 2\,\eta\,m_{B^+}\,m_p^{2}\,x_3^{2}\\
		&{}+2\,\eta^{2}\,m_{B^+}\,m_p^{2}\,x_3 + 8\,\eta^{2}\,m_p^{3}\,x_2\,x_4^{2} - 40\,\eta^{2}\,m_p^{3}\,x_2\,x_3\,x_4^{2}\\
		&{}-8\,\eta^{2}\,m_{B^+}^{2}\,m_p\,x_2\,x_4 + 8\,\eta^{3}\,m_{B^+}^{2}\,m_p\,x_2\,x_4\\
		&{}+10\,\eta^{3}\,m_{B^+}^{3}\,x_2\,x_3\,x_4 + 40\,\eta\,m_p^{3}\,x_2\,x_3\,x_4 + 40\,\eta\,m_p^{3}\,x_2\,x_3\,x_4^{2}\\
		&{}-40\,\eta^{3}\,m_{B^+}^{2}\,m_p\,x_2\,x_3\,x_4 - 20\,\eta^{2}\,m_{B^+}\,m_p^{2}\,x_2\,x_3\,x_4\\
		&{}+20\,\eta\,m_{B^+}\,m_p^{2}\,x_2\,x_3\,x_4 + 40\,\eta^{2}\,m_{B^+}^{2}\,m_p\,x_2\,x_3\,x_4
	\end{aligned}\right)\nonumber\\[1pt]
	&+\,\psi_4^{0}\,\frac{4\,m_p\,x_3\,\left(x_3 - 1\right)\,\left(\eta^{2}\,m_{B^+}^{2} - 2\,\eta\,m_p^{2} + 2\,m_p^{2}\right)}{\eta^{2}}\nonumber\\[1pt]
	&+\,\psi_4^{-}\,\frac{4\,m_p\,x_3\,\left(\eta^{2}\,m_{B^+}^{2} - 2\,\eta\,m_p^{2} + 2\,m_p^{2}\right)\,\left(x_2^{2} + x_3^{2} - x_3 + x_4^{2}\right)}{\eta^{2}}\nonumber\\[1pt]
	&+\,\psi_4^{+}\,\frac{4\,m_p\,x_3\,\left(\eta^{2}\,m_{B^+}^{2} - 2\,\eta\,m_p^{2} + 2\,m_p^{2}\right)\,\left(10\,x_2\,x_4 + x_3 - 1\right)}{\eta^{2}}\nonumber\\[1pt]
	&+\,\xi_4^{0}\,\frac{4\,m_p\,x_3\,\left(\eta^{2}\,m_{B^+}^{2} - 2\,\eta\,m_p^{2} + 2\,m_p^{2}\right)\,\left(3\,x_2-x_4-2\,\eta\,x_2\,x_4\right)}{\eta^{2}}\nonumber\\[1pt]
	&-\,\xi_4^{-}\,\frac{4\,m_p\,x_3\,\left(\eta^{2}\,m_{B^+}^{2} - 2\,\eta\,m_p^{2} + 2\,m_p^{2}\right)}{\eta^{2}}\,\left(\begin{aligned}
		&{}-3 - 6\,x_3^{2} - 2\,x_4^{2} + 6\,x_4 + 9\,x_3\\
		&{}-10\,x_3\,x_4 - 4\,\eta\,x_4^{2} + 2\,\eta\,x_4 + 2\,\eta\,x_4^{3}\\
		&{}-6\,\eta\,x_3\,x_4 + 4\,\eta\,x_3^{2}\,x_4 + 6\,\eta\,x_3\,x_4^{2}
	\end{aligned}\right)\nonumber\\[1pt]
	&+\,\xi_4^{+}\,\frac{4\,m_p\,x_3\,\left(\eta^{2}\,m_{B^+}^{2} - 2\,\eta\,m_p^{2} + 2\,m_p^{2}\right)}{\eta^{2}}\,\left(\begin{aligned}
		&{}+1 - x_3 - 2\,x_4 + 2\,x_2 + \eta\,x_4^{2} - \eta\,x_4\\
		&{}-10\,x_2\,x_4 + \eta\,x_3\,x_4 - \eta\,x_2\,x_4 + 10\,\eta\,x_2\,x_4^{2}
	\end{aligned}\right)\nonumber\\[1pt]
	&-\,\phi_5^{0}\,\frac{2\,m_p^{2}}{\eta^{3}\,m_{B^+}}\,\left(\begin{aligned}
		&{}-4\,m_p^{2}\,x_3 + \eta\,m_{B^+}\,m_p - \eta^{2}\,m_{B^+}\,m_p - 2\,\eta^{2}\,m_{B^+}^{2}\,x_3\\
		&{}+4\,\eta\,m_p^{2}\,x_3 + \eta^{2}\,m_{B^+}\,m_p\,x_3 - \eta\,m_{B^+}\,m_p\,x_3 - 4\,\eta\,m_{B^+}\,m_p\,x_4\\
		&{}-4\,\eta^{2}\,m_p^{2}\,x_3\,x_4 - 2\,\eta^{3}\,m_{B^+}\,m_p\,x_4^{2} + 2\,\eta^{2}\,m_{B^+}\,m_p\,x_4^{2}\\
		&{}+2\,\eta^{3}\,m_{B^+}^{2}\,x_3\,x_4 + 4\,\eta\,m_p^{2}\,x_3\,x_4 + 4\,\eta^{2}\,m_{B^+}\,m_p\,x_4
	\end{aligned}\right)\nonumber\\[1pt]
	&+\,\phi_5^{-}\,\frac{2\,m_p^{2}\,\left(\eta\,x_4 - 1\right)}{\eta^{3}\,m_{B^+}}\,\left(\begin{aligned}
		&{}-4\,m_p^{2}\,x_3\,x_4 + 4\,m_p^{2}\,x_2\,x_3 + \eta\,m_{B^+}\,m_p\,x_3 + \eta^{2}\,m_{B^+}\,m_p\,x_3^{2}\\
		&{}-\eta\,m_{B^+}\,m_p\,x_3^{2} - \eta^{2}\,m_{B^+}\,m_p\,x_3 - 4\,\eta\,m_p^{2}\,x_2\,x_3\\
		&{}-2\,\eta^{2}\,m_{B^+}^{2}\,x_3\,x_4 + 2\,\eta^{2}\,m_{B^+}^{2}\,x_2\,x_3 + 4\,\eta\,m_p^{2}\,x_3\,x_4\\
		&{}-2\,\eta\,m_{B^+}\,m_p\,x_2\,x_4 + 2\,\eta^{2}\,m_{B^+}\,m_p\,x_2\,x_4
	\end{aligned}\right)\nonumber\\[1pt]
	&+\,\phi_5^{+}\,\frac{2\,m_p^{2}\,\left(\eta\,x_4 - 1\right)}{\eta^{3}\,m_{B^+}}\,\left(\begin{aligned}
		&{}-4\,m_p^{2}\,x_3 + 8\,m_p^{2}\,x_3^{2} + \eta^{2}\,m_{B^+}\,m_p - \eta\,m_{B^+}\,m_p\\
		&{}-8\,\eta\,m_p^{2}\,x_3^{2} - 2\,\eta^{2}\,m_{B^+}^{2}\,x_3 + 4\,\eta\,m_p^{2}\,x_3 + 4\,\eta^{2}\,m_{B^+}^{2}\,x_3^{2}\\
		&{}+\eta\,m_{B^+}\,m_p\,x_3 - \eta^{2}\,m_{B^+}\,m_p\,x_3 - 2\,\eta^{2}\,m_{B^+}\,m_p\,x_2^{2}\\
		&{}-2\,\eta^{2}\,m_{B^+}\,m_p\,x_4^{2} + 2\,\eta\,m_{B^+}\,m_p\,x_2^{2} + 2\,\eta\,m_{B^+}\,m_p\,x_4^{2}
	\end{aligned}\right)\nonumber\\[1pt]
	&-\,\psi_5^{0}\,\frac{2\,m_p^{3}\,\left(\eta - 1\right)\,\left(x_3 - 1\right)\,\left(\eta\,x_4 - 1\right)}{\eta^{2}}\nonumber\\[1pt]
	&-\,\psi_5^{-}\,\frac{2\,m_p^{3}\,\left(\eta - 1\right)\,\left(\eta\,x_4 - 1\right)\,\left(2\,x_2\,x_4 + x_3^{2} - x_3\right)}{\eta^{2}}\nonumber\\[1pt]
	&-\,\psi_5^{+}\,\frac{2\,m_p^{3}\,\left(\eta - 1\right)\,\left(\eta\,x_4 - 1\right)\,\left(2\,x_2^{2} + x_3 + 2\,x_4^{2} - 1\right)}{\eta^{2}}\nonumber\\[1pt]
	&-\,\xi_5^{0}\,\frac{2\,m_p^{3}\,\left(\eta - 1\right)\,\left(x_3 - 1\right)}{\eta^{2}}\nonumber\\[1pt]
	&+\,\xi_5^{-}\,\frac{2\,m_p^{3}\,\left(\eta - 1\right)\,\left(- \eta\,x_2\,x_3\,x_4 + \eta\,x_3\,x_4^{2} + 2\,x_2\,x_3 + 2\,x_2\,x_4 + x_3^{2} - 2\,x_3\,x_4 - x_3\right)}{\eta^{2}}\nonumber\\[1pt]
	&-\,\xi_5^{+}\,\frac{2\,m_p^{3}\,\left(\eta - 1\right)}{\eta^{2}}\,\left(\begin{aligned}
		&{}-1 + x_3 - 2\,x_2 + 2\,x_4 + 2\,x_2^{2} + 2\,x_4^{2} - \eta\,x_4^{2} - 4\,x_3\,x_4\\
		&{}+4\,x_2\,x_3 + \eta\,x_2\,x_4 + 2\,\eta\,x_3\,x_4^{2} - 2\,\eta\,x_2\,x_3\,x_4
	\end{aligned}\right)\nonumber\\[1pt]
	&-\,\phi_6^{0}\,\frac{8\,m_p^{4}\,\left(\eta - 1\right)\,\left(\eta\,x_4 - 1\right)}{3\,\eta^{3}\,m_{B^+}}\nonumber\\[1pt]
	&-\,\phi_6^{-}\,\frac{8\,m_p^{4}\,\left(\eta - 1\right)\,\left(- x_2 + x_4\right)\,\left(\eta\,x_4 - 1\right)}{3\,\eta^{3}\,m_{B^+}}\nonumber\\[1pt]
	&+\,\phi_6^{+}\,\frac{8\,m_p^{4}\,\left(\eta - 1\right)\,\left(3\,x_3 - 1\right)\,\left(\eta\,x_4 - 1\right)}{3\,\eta^{3}\,m_{B^+}}\bigr]\nonumber\\[5pt]
	&-2\,m_p\,b_3b_4\,E(t_b)h_2(x_1,x_3,x_4,b_1,b_3,b_4)\bigl[\nonumber\\
	&+\,\phi_3^{0}\,\frac{20\,x_2\,x_3\,x_4}{\eta^{2}\,m_{B^+}}\,\left(\begin{aligned}
		&{}+m_p^{3} + 2\,\eta^{2}\,m_{B^+}^{3} + \eta\,m_{B^+}^{2}\,m_p - \eta\,m_p^{3}\,x_4 - \eta^{2}\,m_{B^+}^{2}\,m_p\\
		&{}-4\,\eta\,m_{B^+}\,m_p^{2} + \eta^{2}\,m_{B^+}^{2}\,m_p\,x_4
	\end{aligned}\right)\nonumber\\[1pt]
	&-\,\phi_3^{-}\,\frac{20\,x_2\,x_3\,x_4}{\eta^{2}\,m_{B^+}}\,\left(\begin{aligned}
		&{}+m_p^{3}\,x_2 - m_p^{3}\,x_4 - \eta^{2}\,m_{B^+}^{3} + \eta\,m_p^{3}\,x_4^{2} + \eta^{2}\,m_{B^+}^{2}\,m_p\\
		&{}+2\,\eta\,m_{B^+}\,m_p^{2} + 3\,\eta^{2}\,m_{B^+}^{3}\,x_3 + \eta\,m_{B^+}^{2}\,m_p\,x_2 - \eta\,m_{B^+}^{2}\,m_p\,x_4\\
		&{}-\eta\,m_p^{3}\,x_2\,x_4 - \eta^{2}\,m_{B^+}^{2}\,m_p\,x_3 - \eta^{2}\,m_{B^+}^{2}\,m_p\,x_4^{2}\\
		&{}-6\,\eta\,m_{B^+}\,m_p^{2}\,x_3 - 2\,\eta^{2}\,m_{B^+}^{2}\,m_p\,x_2 + \eta^{2}\,m_{B^+}^{2}\,m_p\,x_2\,x_4
	\end{aligned}\right)\nonumber\\[1pt]
	&-\,\phi_3^{+}\,\frac{20\,x_2\,x_3\,x_4\,\left(3\,x_3 - 1\right)}{\eta^{2}\,m_{B^+}}\,\left(\begin{aligned}
		&{}+m_p^{3} - \eta^{2}\,m_{B^+}^{3} + \eta\,m_{B^+}^{2}\,m_p - \eta\,m_p^{3}\,x_4 - \eta^{2}\,m_{B^+}^{2}\,m_p\\
		&{}+2\,\eta\,m_{B^+}\,m_p^{2} + \eta^{2}\,m_{B^+}^{2}\,m_p\,x_4
	\end{aligned}\right)\nonumber\\[1pt]
	&+\,\phi_4^{0}\,1 / \eta\,\left(\begin{aligned}
		&{}-2\,m_p^{2}\,x_3 + 2\,m_p^{2}\,x_3^{2} - 2\,\eta\,m_{B^+}^{2}\,x_3^{2} + 2\,\eta\,m_{B^+}^{2}\,x_3\\
		&{}-8\,m_{B^+}\,m_p\,x_2\,x_4 - 2\,\eta\,m_p^{2}\,x_3^{2}\,x_4 - 2\,\eta^{2}\,m_{B^+}^{2}\,x_3\,x_4\\
		&{}+2\,\eta\,m_p^{2}\,x_3\,x_4 + 2\,\eta^{2}\,m_{B^+}^{2}\,x_3^{2}\,x_4 + 8\,\eta\,m_{B^+}\,m_p\,x_2\,x_4
	\end{aligned}\right)\nonumber\\[1pt]
	&-\,\phi_4^{-}\,1 / \eta\,\left(\begin{aligned}
		&{}-2\,m_p^{2}\,x_3^{2} + 2\,m_p^{2}\,x_3^{3} - 2\,\eta\,m_{B^+}^{2}\,x_3^{3} + 2\,\eta\,m_{B^+}^{2}\,x_3^{2}\\
		&{}+2\,m_p^{2}\,x_2^{2}\,x_3 + 2\,m_p^{2}\,x_3\,x_4^{2} - 8\,m_{B^+}\,m_p\,x_2^{2}\,x_4\\
		&{}-2\,\eta\,m_{B^+}^{2}\,x_2^{2}\,x_3 - 2\,\eta\,m_{B^+}^{2}\,x_3\,x_4^{2} - 2\,\eta\,m_p^{2}\,x_3\,x_4^{3}\\
		&{}-2\,\eta\,m_p^{2}\,x_3^{3}\,x_4 - 2\,\eta^{2}\,m_{B^+}^{2}\,x_3^{2}\,x_4 + 2\,\eta\,m_p^{2}\,x_3^{2}\,x_4\\
		&{}+2\,\eta^{2}\,m_{B^+}^{2}\,x_3\,x_4^{3} + 2\,\eta^{2}\,m_{B^+}^{2}\,x_3^{3}\,x_4 + 8\,m_{B^+}\,m_p\,x_2\,x_4^{2}\\
		&{}-8\,\eta\,m_{B^+}\,m_p\,x_2\,x_4^{2} - 2\,\eta\,m_p^{2}\,x_2^{2}\,x_3\,x_4\\
		&{}+2\,\eta^{2}\,m_{B^+}^{2}\,x_2^{2}\,x_3\,x_4 + 8\,\eta\,m_{B^+}\,m_p\,x_2^{2}\,x_4
	\end{aligned}\right)\nonumber\\[1pt]
	&+\,\phi_4^{+}\,1 / \eta\,\left(\begin{aligned}
		&{}-2\,m_p^{2}\,x_3 + 2\,m_p^{2}\,x_3^{2} - 2\,\eta\,m_{B^+}^{2}\,x_3^{2} + 2\,\eta\,m_{B^+}^{2}\,x_3\\
		&{}-8\,m_{B^+}\,m_p\,x_2\,x_4 - 2\,\eta\,m_p^{2}\,x_3^{2}\,x_4 - 2\,\eta^{2}\,m_{B^+}^{2}\,x_3\,x_4\\
		&{}+2\,\eta\,m_p^{2}\,x_3\,x_4 + 2\,\eta^{2}\,m_{B^+}^{2}\,x_3^{2}\,x_4 + 20\,m_p^{2}\,x_2\,x_3\,x_4\\
		&{}-20\,\eta\,m_{B^+}^{2}\,x_2\,x_3\,x_4 - 20\,\eta\,m_p^{2}\,x_2\,x_3\,x_4^{2} + 8\,\eta\,m_{B^+}\,m_p\,x_2\,x_4\\
		&{}+20\,\eta^{2}\,m_{B^+}^{2}\,x_2\,x_3\,x_4^{2} + 40\,m_{B^+}\,m_p\,x_2\,x_3\,x_4\\
		&{}-40\,\eta\,m_{B^+}\,m_p\,x_2\,x_3\,x_4
	\end{aligned}\right)\nonumber\\[1pt]
	&+\,\psi_4^{0}\,\frac{2\,x_3}{\eta^{2}\,m_{B^+}}\,\left(\begin{aligned}
		&{}-2\,m_p^{3} + \eta^{2}\,m_{B^+}^{3} - 2\,m_p^{3}\,x_2 + 2\,m_p^{3}\,x_3 + 2\,m_p^{3}\,x_4\\
		&{}-\eta\,m_{B^+}\,m_p^{2} - \eta^{2}\,m_{B^+}^{3}\,x_3 - \eta^{3}\,m_{B^+}^{3}\,x_4 - 2\,\eta\,m_p^{3}\,x_4^{2}\\
		&{}-2\,\eta^{2}\,m_{B^+}^{2}\,m_p + 2\,\eta\,m_{B^+}^{2}\,m_p + 2\,\eta\,m_p^{3}\,x_4 + \eta\,m_{B^+}\,m_p^{2}\,x_3\\
		&{}+\eta^{2}\,m_{B^+}\,m_p^{2}\,x_4 + \eta^{3}\,m_{B^+}^{3}\,x_3\,x_4 - 2\,\eta\,m_{B^+}^{2}\,m_p\,x_3\\
		&{}-2\,\eta\,m_{B^+}^{2}\,m_p\,x_4 - 2\,\eta\,m_p^{3}\,x_3\,x_4 + 2\,\eta\,m_{B^+}^{2}\,m_p\,x_2\\
		&{}+2\,\eta\,m_p^{3}\,x_2\,x_4 + 2\,\eta^{2}\,m_{B^+}^{2}\,m_p\,x_3 + 2\,\eta^{2}\,m_{B^+}^{2}\,m_p\,x_4\\
		&{}-\eta^{2}\,m_{B^+}\,m_p^{2}\,x_3\,x_4 - 2\,\eta^{2}\,m_{B^+}^{2}\,m_p\,x_2\,x_4
	\end{aligned}\right)\nonumber\\[1pt]
	&+\,\psi_4^{-}\,\frac{2\,x_3}{\eta^{2}\,m_{B^+}}\,\left(\begin{aligned}
		&{}-2\,m_p^{3}\,x_2 - 2\,m_p^{3}\,x_2^{2} - 2\,m_p^{3}\,x_3^{2} - 2\,m_p^{3}\,x_4^{2} + 2\,m_p^{3}\,x_3\\
		&{}+2\,m_p^{3}\,x_4 + \eta^{2}\,m_{B^+}^{3}\,x_3 + \eta^{3}\,m_{B^+}^{3}\,x_4^{3} - \eta\,m_{B^+}^{2}\,m_p\\
		&{}-\eta^{2}\,m_{B^+}^{3}\,x_2^{2} - \eta^{2}\,m_{B^+}^{3}\,x_3^{2} - \eta^{2}\,m_{B^+}^{3}\,x_4^{2}\\
		&{}-4\,m_p^{3}\,x_3\,x_4 - 2\,\eta\,m_p^{3}\,x_4^{2} + 2\,\eta\,m_p^{3}\,x_4^{3} + 4\,m_p^{3}\,x_2\,x_3\\
		&{}+\eta\,m_{B^+}\,m_p^{2}\,x_2^{2} + \eta\,m_{B^+}\,m_p^{2}\,x_3^{2} + \eta\,m_{B^+}\,m_p^{2}\,x_4^{2}\\
		&{}+\eta^{2}\,m_{B^+}^{2}\,m_p\,x_3 + \eta^{3}\,m_{B^+}^{3}\,x_2^{2}\,x_4 + \eta^{3}\,m_{B^+}^{3}\,x_3^{2}\,x_4\\
		&{}-\eta\,m_{B^+}\,m_p^{2}\,x_3 - \eta\,m_{B^+}^{2}\,m_p\,x_3 - \eta\,m_{B^+}^{2}\,m_p\,x_4\\
		&{}-\eta^{2}\,m_{B^+}\,m_p^{2}\,x_4^{3} - \eta^{2}\,m_{B^+}^{2}\,m_p\,x_2 - \eta^{2}\,m_{B^+}^{2}\,m_p\,x_2^{2}\\
		&{}-\eta^{2}\,m_{B^+}^{2}\,m_p\,x_3^{2} - \eta^{2}\,m_{B^+}^{2}\,m_p\,x_4^{2} - \eta^{2}\,m_{B^+}^{2}\,m_p\,x_4^{3}\\
		&{}-\eta^{3}\,m_{B^+}^{3}\,x_3\,x_4 - 2\,\eta\,m_p^{3}\,x_3\,x_4 + 2\,\eta\,m_{B^+}^{2}\,m_p\,x_2^{2}\\
		&{}+2\,\eta\,m_{B^+}^{2}\,m_p\,x_3^{2} + 2\,\eta\,m_{B^+}^{2}\,m_p\,x_4^{2} + 2\,\eta\,m_p^{3}\,x_2\,x_4\\
		&{}+2\,\eta\,m_p^{3}\,x_2^{2}\,x_4 + 2\,\eta\,m_p^{3}\,x_3^{2}\,x_4 + 2\,\eta^{2}\,m_{B^+}^{2}\,m_p\,x_4\\
		&{}+3\,\eta\,m_{B^+}^{2}\,m_p\,x_2 + 4\,\eta\,m_p^{3}\,x_3\,x_4^{2} + \eta^{2}\,m_{B^+}\,m_p^{2}\,x_3\,x_4\\
		&{}-\eta^{2}\,m_{B^+}\,m_p^{2}\,x_2^{2}\,x_4 - \eta^{2}\,m_{B^+}\,m_p^{2}\,x_3^{2}\,x_4\\
		&{}-\eta^{2}\,m_{B^+}^{2}\,m_p\,x_2^{2}\,x_4 - \eta^{2}\,m_{B^+}^{2}\,m_p\,x_3^{2}\,x_4\\
		&{}-4\,\eta\,m_{B^+}^{2}\,m_p\,x_2\,x_3 - 4\,\eta\,m_p^{3}\,x_2\,x_3\,x_4 - 2\,\eta^{2}\,m_{B^+}^{2}\,m_p\,x_2\,x_4\\
		&{}-2\,\eta^{2}\,m_{B^+}^{2}\,m_p\,x_3\,x_4 - 2\,\eta^{2}\,m_{B^+}^{2}\,m_p\,x_3\,x_4^{2}\\
		&{}+2\,\eta^{2}\,m_{B^+}^{2}\,m_p\,x_2\,x_3 + 4\,\eta\,m_{B^+}^{2}\,m_p\,x_3\,x_4\\
		&{}+2\,\eta^{2}\,m_{B^+}^{2}\,m_p\,x_2\,x_3\,x_4
	\end{aligned}\right)\nonumber\\[1pt]
	&+\,\psi_4^{+}\,\frac{2\,x_3}{\eta^{2}\,m_{B^+}}\,\left(\begin{aligned}
		&{}-2\,m_p^{3} + \eta^{2}\,m_{B^+}^{3} - 2\,m_p^{3}\,x_2 + 2\,m_p^{3}\,x_3 + 2\,m_p^{3}\,x_4\\
		&{}-\eta\,m_{B^+}\,m_p^{2} - \eta^{2}\,m_{B^+}^{3}\,x_3 - \eta^{3}\,m_{B^+}^{3}\,x_4 - 2\,\eta\,m_p^{3}\,x_4^{2}\\
		&{}-2\,\eta^{2}\,m_{B^+}^{2}\,m_p + 2\,\eta\,m_p^{3}\,x_4 + 3\,\eta\,m_{B^+}^{2}\,m_p + 20\,m_p^{3}\,x_2\,x_4\\
		&{}+\eta\,m_{B^+}\,m_p^{2}\,x_3 + \eta\,m_{B^+}^{2}\,m_p\,x_2 + \eta^{2}\,m_{B^+}\,m_p^{2}\,x_4\\
		&{}+\eta^{2}\,m_{B^+}^{2}\,m_p\,x_4 + \eta^{2}\,m_{B^+}^{2}\,m_p\,x_4^{2} + \eta^{3}\,m_{B^+}^{3}\,x_3\,x_4\\
		&{}-20\,\eta\,m_p^{3}\,x_2\,x_4^{2} - 10\,\eta^{2}\,m_{B^+}^{3}\,x_2\,x_4 - 3\,\eta\,m_{B^+}^{2}\,m_p\,x_3\\
		&{}-3\,\eta\,m_{B^+}^{2}\,m_p\,x_4 - 2\,\eta\,m_p^{3}\,x_3\,x_4 + 2\,\eta\,m_p^{3}\,x_2\,x_4\\
		&{}+2\,\eta^{2}\,m_{B^+}^{2}\,m_p\,x_3 + 10\,\eta^{3}\,m_{B^+}^{3}\,x_2\,x_4^{2} + \eta^{2}\,m_{B^+}^{2}\,m_p\,x_3\,x_4\\
		&{}-\eta^{2}\,m_{B^+}\,m_p^{2}\,x_3\,x_4 - 20\,\eta\,m_{B^+}^{2}\,m_p\,x_2\,x_4\\
		&{}-10\,\eta^{2}\,m_{B^+}\,m_p^{2}\,x_2\,x_4^{2} + 9\,\eta^{2}\,m_{B^+}^{2}\,m_p\,x_2\,x_4\\
		&{}+10\,\eta\,m_{B^+}\,m_p^{2}\,x_2\,x_4 + 10\,\eta^{2}\,m_{B^+}^{2}\,m_p\,x_2\,x_4^{2}
	\end{aligned}\right)\nonumber\\[1pt]
	&-\,\xi_4^{0}\,\frac{1}{\eta^{2}}\,\left(\begin{aligned}
		&{}-4\,m_p^{2}\,x_3^{2} + 4\,m_p^{2}\,x_3 - 4\,m_p^{2}\,x_3\,x_4 - 2\,\eta\,m_p^{2}\,x_3\\
		&{}+2\,\eta\,m_p^{2}\,x_3^{2} + 4\,m_p^{2}\,x_2\,x_3 - 16\,\eta\,m_p^{2}\,x_2\,x_4 - 4\,\eta\,m_p^{2}\,x_2\,x_3\\
		&{}-2\,\eta^{2}\,m_{B^+}^{2}\,x_3\,x_4 - 2\,\eta^{2}\,m_p^{2}\,x_3\,x_4^{2} + 2\,\eta^{2}\,m_{B^+}^{2}\,x_2\,x_3\\
		&{}+2\,\eta^{3}\,m_{B^+}^{2}\,x_3\,x_4^{2} + 4\,\eta\,m_p^{2}\,x_3\,x_4 + 16\,\eta^{2}\,m_{B^+}^{2}\,x_2\,x_4\\
		&{}-2\,\eta^{3}\,m_{B^+}^{2}\,x_2\,x_3\,x_4 + 2\,\eta^{2}\,m_p^{2}\,x_2\,x_3\,x_4
	\end{aligned}\right)\nonumber\\[1pt]
	&+\,\xi_4^{-}\,\frac{2\,x_3}{\eta^{2}}\,\left(\begin{aligned}
		&{}-2\,m_p^{2}\,x_2 - 2\,m_p^{2}\,x_2^{2} - 2\,m_p^{2}\,x_3^{2} - 2\,m_p^{2}\,x_4^{2} + 2\,m_p^{2}\,x_3\\
		&{}+2\,m_p^{2}\,x_4 + \eta\,m_p^{2}\,x_2^{2} + \eta\,m_p^{2}\,x_3^{2} + \eta\,m_p^{2}\,x_4^{2}\\
		&{}+\eta^{2}\,m_{B^+}^{2}\,x_4 + \eta^{2}\,m_p^{2}\,x_4^{2} - \eta\,m_p^{2}\,x_3 - \eta^{2}\,m_{B^+}^{2}\,x_2\\
		&{}-\eta^{3}\,m_{B^+}^{2}\,x_4^{2} - 4\,m_p^{2}\,x_3\,x_4 - 2\,\eta\,m_p^{2}\,x_4 + 2\,\eta\,m_p^{2}\,x_2\\
		&{}+4\,m_p^{2}\,x_2\,x_3 + \eta^{3}\,m_{B^+}^{2}\,x_2\,x_4 - \eta^{2}\,m_p^{2}\,x_2\,x_4\\
		&{}-4\,\eta\,m_p^{2}\,x_2\,x_3 - 2\,\eta^{2}\,m_{B^+}^{2}\,x_3\,x_4 - 2\,\eta^{2}\,m_p^{2}\,x_3\,x_4^{2}\\
		&{}+2\,\eta^{2}\,m_{B^+}^{2}\,x_2\,x_3 + 2\,\eta^{3}\,m_{B^+}^{2}\,x_3\,x_4^{2} + 4\,\eta\,m_p^{2}\,x_3\,x_4\\
		&{}-2\,\eta^{3}\,m_{B^+}^{2}\,x_2\,x_3\,x_4 + 2\,\eta^{2}\,m_p^{2}\,x_2\,x_3\,x_4
	\end{aligned}\right)\nonumber\\[1pt]
	&-\,\xi_4^{+}\,\frac{1}{\eta^{2}}\,\left(\begin{aligned}
		&{}-4\,m_p^{2}\,x_3^{2} + 4\,m_p^{2}\,x_3 - 4\,m_p^{2}\,x_3\,x_4 - 2\,\eta\,m_p^{2}\,x_3\\
		&{}+2\,\eta\,m_p^{2}\,x_3^{2} + 4\,m_p^{2}\,x_2\,x_3 - 40\,m_p^{2}\,x_2\,x_3\,x_4 - 16\,\eta\,m_p^{2}\,x_2\,x_4\\
		&{}-4\,\eta\,m_p^{2}\,x_2\,x_3 - 2\,\eta^{2}\,m_{B^+}^{2}\,x_3\,x_4 - 2\,\eta^{2}\,m_p^{2}\,x_3\,x_4^{2}\\
		&{}+2\,\eta^{2}\,m_{B^+}^{2}\,x_2\,x_3 + 2\,\eta^{3}\,m_{B^+}^{2}\,x_3\,x_4^{2} + 4\,\eta\,m_p^{2}\,x_3\,x_4\\
		&{}+16\,\eta^{2}\,m_{B^+}^{2}\,x_2\,x_4 - 80\,\eta^{2}\,m_{B^+}^{2}\,x_2\,x_3\,x_4\\
		&{}-2\,\eta^{3}\,m_{B^+}^{2}\,x_2\,x_3\,x_4 + 2\,\eta^{2}\,m_p^{2}\,x_2\,x_3\,x_4 + 100\,\eta\,m_p^{2}\,x_2\,x_3\,x_4
	\end{aligned}\right)\nonumber\\[1pt]
	&-\,\phi_5^{0}\,\frac{m_p}{\eta^{2}\,m_{B^+}}\,\left(\begin{aligned}
		&{}+m_p^{2}\,x_3 - \eta\,m_{B^+}^{2}\,x_3 + 2\,m_{B^+}\,m_p\,x_4 + \eta^{2}\,m_{B^+}^{2}\,x_3\,x_4\\
		&{}-\eta\,m_p^{2}\,x_3\,x_4 - 2\,\eta\,m_{B^+}\,m_p\,x_4
	\end{aligned}\right)\nonumber\\[1pt]
	&+\,\phi_5^{-}\,\frac{m_p}{\eta^{2}\,m_{B^+}}\,\left(\begin{aligned}
		&{}+m_{B^+}\,m_p\,x_3 + m_p^{2}\,x_2\,x_3 - m_{B^+}\,m_p\,x_3^{2} - m_p^{2}\,x_3\,x_4\\
		&{}+\eta\,m_{B^+}\,m_p\,x_3^{2} + \eta\,m_{B^+}^{2}\,x_3\,x_4 + \eta\,m_p^{2}\,x_3\,x_4^{2} - \eta\,m_{B^+}\,m_p\,x_3\\
		&{}-\eta\,m_{B^+}^{2}\,x_2\,x_3 - \eta^{2}\,m_{B^+}^{2}\,x_3\,x_4^{2} - 2\,m_{B^+}\,m_p\,x_2\,x_4\\
		&{}+\eta^{2}\,m_{B^+}^{2}\,x_2\,x_3\,x_4 - \eta\,m_p^{2}\,x_2\,x_3\,x_4 + 2\,\eta\,m_{B^+}\,m_p\,x_2\,x_4
	\end{aligned}\right)\nonumber\\[1pt]
	&+\,\phi_5^{+}\,\frac{m_p}{\eta^{2}\,m_{B^+}}\,\left(\begin{aligned}
		&{}-m_{B^+}\,m_p - m_p^{2}\,x_3 + 2\,m_p^{2}\,x_3^{2} + \eta\,m_{B^+}\,m_p + \eta\,m_{B^+}^{2}\,x_3\\
		&{}+m_{B^+}\,m_p\,x_3 - 2\,\eta\,m_{B^+}^{2}\,x_3^{2} + 2\,m_{B^+}\,m_p\,x_2^{2} + 2\,m_{B^+}\,m_p\,x_4^{2}\\
		&{}+\eta\,m_p^{2}\,x_3\,x_4 - \eta\,m_{B^+}\,m_p\,x_3 - \eta^{2}\,m_{B^+}^{2}\,x_3\,x_4\\
		&{}-2\,\eta\,m_{B^+}\,m_p\,x_2^{2} - 2\,\eta\,m_{B^+}\,m_p\,x_4^{2} - 2\,\eta\,m_p^{2}\,x_3^{2}\,x_4\\
		&{}+2\,\eta^{2}\,m_{B^+}^{2}\,x_3^{2}\,x_4
	\end{aligned}\right)\nonumber\\[1pt]
	&-\,\psi_5^{0}\,\frac{m_p}{\eta^{2}\,m_{B^+}}\,\left(\begin{aligned}
		&{}+m_{B^+}\,m_p - 2\,m_p^{2}\,x_4 - \eta\,m_{B^+}\,m_p - m_{B^+}\,m_p\,x_3\\
		&{}-2\,\eta^{2}\,m_{B^+}^{2}\,x_4^{2} + 2\,\eta\,m_{B^+}^{2}\,x_4 + 2\,\eta\,m_p^{2}\,x_4^{2} + \eta\,m_{B^+}\,m_p\,x_3
	\end{aligned}\right)\nonumber\\[1pt]
	&-\,\psi_5^{-}\,\frac{m_p}{\eta^{2}\,m_{B^+}}\,\left(\begin{aligned}
		&{}-2\,m_p^{2}\,x_4 + 2\,m_p^{2}\,x_4^{2} + m_{B^+}\,m_p\,x_3 - m_{B^+}\,m_p\,x_3^{2}\\
		&{}-2\,\eta\,m_{B^+}^{2}\,x_4^{2} - 2\,\eta\,m_p^{2}\,x_4^{3} - 2\,m_{B^+}\,m_p\,x_4 - 2\,\eta^{2}\,m_{B^+}^{2}\,x_4^{2}\\
		&{}+2\,\eta\,m_{B^+}^{2}\,x_4 + 2\,\eta\,m_p^{2}\,x_4^{2} + 2\,m_{B^+}\,m_p\,x_4^{2} + 2\,\eta^{2}\,m_{B^+}^{2}\,x_4^{3}\\
		&{}+4\,m_p^{2}\,x_3\,x_4 + \eta\,m_{B^+}\,m_p\,x_3^{2} - \eta\,m_{B^+}\,m_p\,x_3 - 4\,\eta\,m_{B^+}^{2}\,x_3\,x_4\\
		&{}-4\,\eta\,m_p^{2}\,x_3\,x_4^{2} - 2\,\eta\,m_{B^+}\,m_p\,x_4^{2} + 2\,\eta\,m_{B^+}\,m_p\,x_4\\
		&{}+2\,m_{B^+}\,m_p\,x_3\,x_4 + 4\,\eta^{2}\,m_{B^+}^{2}\,x_3\,x_4^{2} - 2\,\eta\,m_{B^+}\,m_p\,x_3\,x_4
	\end{aligned}\right)\nonumber\\[1pt]
	&-\,\psi_5^{+}\,\frac{m_p}{\eta^{2}\,m_{B^+}}\,\left(\begin{aligned}
		&{}-m_{B^+}\,m_p - 2\,m_p^{2}\,x_4 + 4\,m_p^{2}\,x_4^{2} + \eta\,m_{B^+}\,m_p - 4\,\eta\,m_{B^+}^{2}\,x_4^{2}\\
		&{}-4\,\eta\,m_p^{2}\,x_4^{3} - 4\,m_{B^+}\,m_p\,x_4^{2} - 2\,m_{B^+}\,m_p\,x_3^{2}\\
		&{}-2\,\eta^{2}\,m_{B^+}^{2}\,x_4^{2} + 2\,\eta\,m_{B^+}^{2}\,x_4 + 2\,\eta\,m_p^{2}\,x_4^{2} + 3\,m_{B^+}\,m_p\,x_3\\
		&{}+4\,m_{B^+}\,m_p\,x_4 + 4\,\eta^{2}\,m_{B^+}^{2}\,x_4^{3} - 4\,\eta\,m_{B^+}\,m_p\,x_4\\
		&{}-4\,m_{B^+}\,m_p\,x_3\,x_4 - 3\,\eta\,m_{B^+}\,m_p\,x_3 + 2\,\eta\,m_{B^+}\,m_p\,x_3^{2}\\
		&{}+4\,\eta\,m_{B^+}\,m_p\,x_4^{2} + 4\,\eta\,m_{B^+}\,m_p\,x_3\,x_4
	\end{aligned}\right)\nonumber\\[1pt]
	&+\,\xi_5^{0}\,\frac{2\,m_p^{2}\,x_3\,\left(\eta - 1\right)\,\left(\eta\,x_4 - 1\right)}{\eta^{2}}\nonumber\\[1pt]
	&+\,\xi_5^{-}\,\frac{m_p^{2}\,x_3\,\left(\eta - 1\right)\,\left(- x_2 + x_4\right)}{\eta^{2}}\nonumber\\[1pt]
	&-\,\xi_5^{+}\,\frac{m_p^{2}\,\left(\eta - 1\right)\,\left(2\,x_3 - 1\right)\,\left(2\,\eta\,x_3\,x_4 - x_2 - 2\,x_3 + x_4\right)}{\eta^{2}}\bigr]\nonumber\\[5pt]
	&-b_3b_4\,E(t_c)h_3(x_1,x_3,x_4,b_1,b_3,b_4)\bigl[\nonumber\\
	&-\,\phi_3^{0}\,\frac{40\,x_2\,x_3\,x_4\,\left(\eta^{2}\,m_{B^+}^{2} - 2\,\eta\,m_p^{2} + 2\,m_p^{2}\right)\,\left(\eta^{2}\,m_{B^+}^{2} + \eta\,m_p^{2}\,x_4 - m_p^{2}\right)}{\eta^{3}\,m_{B^+}}\nonumber\\[1pt]
	&+\,\phi_3^{-}\,\frac{40\,x_3\,x_4\,\left(-x_2\right)\,\left(x_4-x_2\right)\,\left(\eta^{2}\,m_{B^+}^{2} - 2\,\eta\,m_p^{2} + 2\,m_p^{2}\right)\,\left(\eta^{2}\,m_{B^+}^{2} + \eta\,m_p^{2}\,x_4 - m_p^{2}\right)}{\eta^{3}\,m_{B^+}}\nonumber\\[1pt]
	&+\,\phi_3^{+}\,\frac{40\,x_2\,x_3\,x_4\,\left(3\,x_3 - 1\right)\,\left(\eta^{2}\,m_{B^+}^{2} - 2\,\eta\,m_p^{2} + 2\,m_p^{2}\right)\,\left(\eta^{2}\,m_{B^+}^{2} + \eta\,m_p^{2}\,x_4 - m_p^{2}\right)}{\eta^{3}\,m_{B^+}}\nonumber\\[1pt]
	&-\,\phi_4^{0}\,\frac{16\,m_p^{2}\,x_2\,x_4\,\left(\eta - 1\right)\,\left(\eta^{2}\,m_{B^+}^{2} + \eta\,m_p^{2}\,x_4 - m_p^{2}\right)}{\eta^{3}\,m_{B^+}}\nonumber\\[1pt]
	&-\,\phi_4^{-}\,\frac{16\,m_p^{2}\,x_2\,x_4\,\left(\eta - 1\right)\,\left(- x_2 + x_4\right)\,\left(\eta^{2}\,m_{B^+}^{2} + \eta\,m_p^{2}\,x_4 - m_p^{2}\right)}{\eta^{3}\,m_{B^+}}\nonumber\\[1pt]
	&+\,\phi_4^{+}\,\frac{16\,m_p^{2}\,x_2\,x_4\,\left(\eta - 1\right)\,\left(5\,x_3 - 1\right)\,\left(\eta^{2}\,m_{B^+}^{2} + \eta\,m_p^{2}\,x_4 - m_p^{2}\right)}{\eta^{3}\,m_{B^+}}\nonumber\\[1pt]
	&-\,\xi_4^{0}\,\frac{8\,m_p\,x_3\,\left(\eta\,x_4 - 2\right)\,\left(x_4-x_2\right)\,\left(\eta^{2}\,m_{B^+}^{2} - 2\,\eta\,m_p^{2} + 2\,m_p^{2}\right)}{\eta^{2}}\nonumber\\[1pt]
	&+\,\xi_4^{-}\,\frac{8\,m_p\,x_3\,\left(2\,x_3 - 1\right)\,\left(\eta\,x_4 - 2\right)\,\left(x_4-x_2\right)\,\left(\eta^{2}\,m_{B^+}^{2} - 2\,\eta\,m_p^{2} + 2\,m_p^{2}\right)}{\eta^{2}}\nonumber\\[1pt]
	&-\,\xi_4^{+}\,\frac{8\,m_p\,x_3\,\left(\eta\,x_4 - 2\right)\,\left(x_4-x_2\right)\,\left(\eta^{2}\,m_{B^+}^{2} - 2\,\eta\,m_p^{2} + 2\,m_p^{2}\right)}{\eta^{2}}\nonumber\\[1pt]
	&+\,\phi_5^{0}\,\frac{2\,m_p^{2}}{\eta^{3}\,m_{B^+}}\,\left(\begin{aligned}
		&{}-2\,m_p^{2}\,x_3 - \eta^{2}\,m_{B^+}^{2}\,x_3 - 4\,\eta^{2}\,m_{B^+}\,m_p + 2\,\eta\,m_p^{2}\,x_3\\
		&{}+4\,\eta\,m_{B^+}\,m_p + \eta^{3}\,m_{B^+}^{2}\,x_3\,x_4 - 8\,\eta\,m_{B^+}\,m_p\,x_4 - 4\,\eta\,m_{B^+}\,m_p\,x_3\\
		&{}-4\,\eta^{3}\,m_{B^+}\,m_p\,x_4^{2} - 2\,\eta^{2}\,m_p^{2}\,x_3\,x_4 + 2\,\eta\,m_p^{2}\,x_3\,x_4\\
		&{}+2\,\eta^{3}\,m_{B^+}\,m_p\,x_4 + 4\,\eta^{2}\,m_{B^+}\,m_p\,x_3 + 4\,\eta^{2}\,m_{B^+}\,m_p\,x_4^{2}\\
		&{}+6\,\eta^{2}\,m_{B^+}\,m_p\,x_4 - 2\,\eta^{3}\,m_{B^+}\,m_p\,x_3\,x_4 + 2\,\eta^{2}\,m_{B^+}\,m_p\,x_3\,x_4
	\end{aligned}\right)\nonumber\\[1pt]
	&+\,\phi_5^{-}\,\frac{2\,m_p^{2}\,x_3\,\left(- x_2 + x_4\right)\,\left(\eta\,x_4 - 1\right)\,\left(\eta^{2}\,m_{B^+}^{2} - 2\,\eta\,m_p^{2} + 2\,m_p^{2}\right)}{\eta^{3}\,m_{B^+}}\nonumber\\[1pt]
	&-\,\phi_5^{+}\,\frac{2\,m_p^{2}\,x_3\,\left(2\,x_3 - 1\right)\,\left(\eta\,x_4 - 1\right)\,\left(\eta^{2}\,m_{B^+}^{2} - 2\,\eta\,m_p^{2} + 2\,m_p^{2}\right)}{\eta^{3}\,m_{B^+}}\nonumber\\[1pt]
	&-\,\xi_5^{-}\,\frac{4\,m_p^{3}\,x_3\,\left(\eta - 1\right)\,\left(\eta\,x_4 - 2\right)\,\left(x_4-x_2\right)}{\eta^{2}}\nonumber\\[1pt]
	&+\,\xi_5^{+}\,\frac{4\,m_p^{3}\,\left(\eta - 1\right)\,\left(2\,x_3 - 1\right)\,\left(\eta\,x_4 - 2\right)\,\left(x_4-x_2\right)}{\eta^{2}}\nonumber\\[1pt]
	&+\,\phi_6^{0}\,\frac{4\,m_p^{4}\,\left(\eta - 1\right)\,\left(\eta\,x_4 - 1\right)}{3\,\eta^{3}\,m_{B^+}}\nonumber\\[1pt]
	&+\,\phi_6^{-}\,\frac{4\,m_p^{4}\,\left(\eta - 1\right)\,\left(- x_2 + x_4\right)\,\left(\eta\,x_4 - 1\right)}{3\,\eta^{3}\,m_{B^+}}\nonumber\\[1pt]
	&-\,\phi_6^{+}\,\frac{4\,m_p^{4}\,\left(\eta - 1\right)\,\left(3\,x_3 - 1\right)\,\left(\eta\,x_4 - 1\right)}{3\,\eta^{3}\,m_{B^+}}\bigr]\biggr).
\end{flalign}

\begin{flalign}
	\widetilde{F}_{B^+\to p_L}^{(d)}(q^2)={}&\frac{\pi\sqrt{3}\Lambda_{\mathrm{QCD}}C_F}{18}\,m_p^2\int_0^1 dx_1\,dx_3\,dx_4\int_0^{1/\Lambda_{\mathrm{QCD}}}db_1\,db_3\,db_4\nonumber\\
	&\times \phi_{B^+}\,\biggl(\frac{b_1b_4}{4\pi^2}\,E(t_a)h_1(x_1,x_4,b_1,b_4)\bigl[\nonumber\\
	&-\,\phi_3^{0}\,\frac{160\,x_2\,x_3\,x_4\,\left(\eta^{2}\,m_{B^+}^{2} + \eta\,m_p^{2}\,x_4 - m_p^{2}\right)}{\eta^{2}\,m_{B^+}}\nonumber\\[1pt]
	&-\,\phi_3^{-}\,\frac{160\,x_2\,x_3\,x_4\,\left(- x_2 + x_4\right)\,\left(\eta^{2}\,m_{B^+}^{2} + \eta\,m_p^{2}\,x_4 - m_p^{2}\right)}{\eta^{2}\,m_{B^+}}\nonumber\\[1pt]
	&+\,\phi_3^{+}\,\frac{160\,x_2\,x_3\,x_4\,\left(3\,x_3 - 1\right)\,\left(\eta^{2}\,m_{B^+}^{2} + \eta\,m_p^{2}\,x_4 - m_p^{2}\right)}{\eta^{2}\,m_{B^+}}\nonumber\\[1pt]
	&-\,\phi_4^{0}\,\frac{1}{\eta^{2}\,m_{B^+}}\,\left(\begin{aligned}
		&{}+32\,m_p^{2}\,x_2\,x_4 - 32\,\eta\,m_p^{2}\,x_2\,x_4^{2} - 32\,\eta^{2}\,m_{B^+}^{2}\,x_2\,x_4\\
		&{}-8\,\eta\,m_{B^+}\,m_p\,x_3 + 8\,\eta\,m_{B^+}\,m_p\,x_3^{2}
	\end{aligned}\right)\nonumber\\[1pt]
	&+\,\phi_4^{-}\,\frac{1}{\eta^{2}\,m_{B^+}}\,\left(\begin{aligned}
		&{}-32\,m_p^{2}\,x_2\,x_4^{2} + 32\,m_p^{2}\,x_2^{2}\,x_4 - 32\,\eta\,m_p^{2}\,x_2^{2}\,x_4^{2}\\
		&{}-32\,\eta^{2}\,m_{B^+}^{2}\,x_2^{2}\,x_4 - 8\,\eta\,m_{B^+}\,m_p\,x_3^{2} + 8\,\eta\,m_{B^+}\,m_p\,x_3^{3}\\
		&{}+32\,\eta\,m_p^{2}\,x_2\,x_4^{3} + 32\,\eta^{2}\,m_{B^+}^{2}\,x_2\,x_4^{2} + 8\,\eta\,m_{B^+}\,m_p\,x_2^{2}\,x_3\\
		&{}+8\,\eta\,m_{B^+}\,m_p\,x_3\,x_4^{2}
	\end{aligned}\right)\nonumber\\[1pt]
	&-\,\phi_4^{+}\,\frac{1}{\eta^{2}\,m_{B^+}}\,\left(\begin{aligned}
		&{}+32\,m_p^{2}\,x_2\,x_4 - 160\,m_p^{2}\,x_2\,x_3\,x_4 - 32\,\eta\,m_p^{2}\,x_2\,x_4^{2}\\
		&{}-32\,\eta^{2}\,m_{B^+}^{2}\,x_2\,x_4 - 8\,\eta\,m_{B^+}\,m_p\,x_3 + 8\,\eta\,m_{B^+}\,m_p\,x_3^{2}\\
		&{}+160\,\eta\,m_p^{2}\,x_2\,x_3\,x_4^{2} + 160\,\eta^{2}\,m_{B^+}^{2}\,x_2\,x_3\,x_4\\
		&{}+80\,\eta\,m_{B^+}\,m_p\,x_2\,x_3\,x_4
	\end{aligned}\right)\nonumber\\[1pt]
	&-\,\psi_4^{0}\,8\,m_p\,x_3\,\left(x_3 - 1\right) / \eta\nonumber\\[1pt]
	&-\,\psi_4^{-}\,8\,m_p\,x_3\,\left(x_2^{2} + x_3^{2} - x_3 + x_4^{2}\right) / \eta\nonumber\\[1pt]
	&-\,\psi_4^{+}\,8\,m_p\,x_3\,\left(10\,x_2\,x_4 + x_3 - 1\right) / \eta\nonumber\\[1pt]
	&-\,\xi_4^{0}\,8\,m_p\,x_3\,\left(3\,x_2-x_4-2\,\eta\,x_2\,x_4\right) / \eta\nonumber\\[1pt]
	&+\,\xi_4^{-}\,8\,m_p\,x_3 / \eta\,\left(\begin{aligned}
		&{}-3 - 6\,x_3^{2} - 2\,x_4^{2} + 6\,x_4 + 9\,x_3\\
		&{}-10\,x_3\,x_4 - 4\,\eta\,x_4^{2} + 2\,\eta\,x_4 + 2\,\eta\,x_4^{3}\\
		&{}-6\,\eta\,x_3\,x_4 + 4\,\eta\,x_3^{2}\,x_4 + 6\,\eta\,x_3\,x_4^{2}
	\end{aligned}\right)\nonumber\\[1pt]
	&-\,\xi_4^{+}\,8\,m_p\,x_3 / \eta\,\left(\begin{aligned}
		&{}+1 - x_3 - 2\,x_4 + 2\,x_2 + \eta\,x_4^{2} - \eta\,x_4\\
		&{}-10\,x_2\,x_4 + \eta\,x_3\,x_4 - \eta\,x_2\,x_4 + 10\,\eta\,x_2\,x_4^{2}
	\end{aligned}\right)\nonumber\\[1pt]
	&+\,\phi_5^{0}\,\frac{2\,m_p\,\left(2\,\eta^{2}\,m_{B^+}\,x_4^{2} - \eta\,m_{B^+}\,x_3 - 4\,\eta\,m_{B^+}\,x_4 + \eta\,m_{B^+} + 4\,\eta\,m_p\,x_3\,x_4 - 4\,m_p\,x_3\right)}{\eta^{2}\,m_{B^+}}\nonumber\\[1pt]
	&+\,\phi_5^{-}\,\frac{2\,m_p\,\left(\eta\,x_4 - 1\right)}{\eta^{2}\,m_{B^+}}\,\left(\begin{aligned}
		&{}+\eta\,m_{B^+}\,x_3^{2} - \eta\,m_{B^+}\,x_3 - 4\,m_p\,x_2\,x_3 + 4\,m_p\,x_3\,x_4\\
		&{}+2\,\eta\,m_{B^+}\,x_2\,x_4
	\end{aligned}\right)\nonumber\\[1pt]
	&-\,\phi_5^{+}\,\frac{2\,m_p\,\left(\eta\,x_4 - 1\right)}{\eta^{2}\,m_{B^+}}\,\left(\begin{aligned}
		&{}-\eta\,m_{B^+} - 4\,m_p\,x_3 + 8\,m_p\,x_3^{2} + \eta\,m_{B^+}\,x_3 + 2\,\eta\,m_{B^+}\,x_2^{2}\\
		&{}+2\,\eta\,m_{B^+}\,x_4^{2}
	\end{aligned}\right)\nonumber\\[1pt]
	&-\,\psi_5^{0}\,2\,m_p\,\left(x_3 - 1\right)\,\left(\eta\,x_4 - 1\right) / \eta\nonumber\\[1pt]
	&-\,\psi_5^{-}\,2\,m_p\,\left(\eta\,x_4 - 1\right)\,\left(2\,x_2\,x_4 + x_3^{2} - x_3\right) / \eta\nonumber\\[1pt]
	&-\,\psi_5^{+}\,2\,m_p\,\left(\eta\,x_4 - 1\right)\,\left(2\,x_2^{2} + x_3 + 2\,x_4^{2} - 1\right) / \eta\nonumber\\[1pt]
	&-\,\xi_5^{0}\,2\,m_p\,\left(x_3 - 1\right) / \eta\nonumber\\[1pt]
	&+\,\xi_5^{-}\,2\,m_p\,\left(- \eta\,x_2\,x_3\,x_4 + \eta\,x_3\,x_4^{2} + 2\,x_2\,x_3 + 2\,x_2\,x_4 + x_3^{2} - 2\,x_3\,x_4 - x_3\right) / \eta\nonumber\\[1pt]
	&-\,\xi_5^{+}\,2\,m_p / \eta\,\left(\begin{aligned}
		&{}-1 + x_3 - 2\,x_2 + 2\,x_4 + 2\,x_2^{2} + 2\,x_4^{2} - \eta\,x_4^{2} - 4\,x_3\,x_4\\
		&{}+4\,x_2\,x_3 + \eta\,x_2\,x_4 + 2\,\eta\,x_3\,x_4^{2} - 2\,\eta\,x_2\,x_3\,x_4
	\end{aligned}\right)\nonumber\\[1pt]
	&-\,\phi_6^{0}\,\frac{8\,m_p^{2}\,\left(\eta\,x_4 - 1\right)}{3\,\eta^{2}\,m_{B^+}}\nonumber\\[1pt]
	&-\,\phi_6^{-}\,\frac{8\,m_p^{2}\,\left(- x_2 + x_4\right)\,\left(\eta\,x_4 - 1\right)}{3\,\eta^{2}\,m_{B^+}}\nonumber\\[1pt]
	&+\,\phi_6^{+}\,\frac{8\,m_p^{2}\,\left(3\,x_3 - 1\right)\,\left(\eta\,x_4 - 1\right)}{3\,\eta^{2}\,m_{B^+}}\bigr]\nonumber\\[5pt]
	&-2\,b_3b_4\,E(t_b)h_2(x_1,x_3,x_4,b_1,b_3,b_4)\bigl[\nonumber\\
	&-\,\phi_3^{0}\,\frac{20\,x_2\,x_3\,x_4\,\left(\eta^{2}\,m_{B^+}^{2} + 4\,\eta\,m_{B^+}\,m_p + \eta\,m_p^{2}\,x_4 - m_p^{2}\right)}{\eta^{2}\,m_{B^+}}\nonumber\\[1pt]
	&+\,\phi_3^{-}\,\frac{20\,x_2\,x_3\,x_4}{\eta^{2}\,m_{B^+}}\,\left(\begin{aligned}
		&{}+m_p^{2}\,x_4 - m_p^{2}\,x_2 - \eta^{2}\,m_{B^+}^{2} + \eta^{2}\,m_{B^+}^{2}\,x_3 - \eta\,m_p^{2}\,x_4^{2}\\
		&{}-2\,\eta\,m_{B^+}\,m_p + 2\,\eta^{2}\,m_{B^+}^{2}\,x_2 + \eta\,m_p^{2}\,x_2\,x_4 + 6\,\eta\,m_{B^+}\,m_p\,x_3
	\end{aligned}\right)\nonumber\\[1pt]
	&+\,\phi_3^{+}\,\frac{20\,x_2\,x_3\,x_4\,\left(3\,x_3 - 1\right)\,\left(\eta^{2}\,m_{B^+}^{2} - 2\,\eta\,m_{B^+}\,m_p + \eta\,m_p^{2}\,x_4 - m_p^{2}\right)}{\eta^{2}\,m_{B^+}}\nonumber\\[1pt]
	&-\,\phi_4^{0}\,2\,\left(- 4\,\eta\,m_{B^+}\,x_2\,x_4 + \eta\,m_p\,x_3^{2}\,x_4 - \eta\,m_p\,x_3\,x_4 - m_p\,x_3^{2} + m_p\,x_3\right) / \eta\nonumber\\[1pt]
	&+\,\phi_4^{-}\,1 / \eta\,\left(\begin{aligned}
		&{}-2\,m_p\,x_3^{3} + 2\,m_p\,x_3^{2} - 2\,m_p\,x_2^{2}\,x_3 - 2\,m_p\,x_3\,x_4^{2}\\
		&{}-8\,\eta\,m_{B^+}\,x_2^{2}\,x_4 - 2\,\eta\,m_p\,x_3^{2}\,x_4 + 2\,\eta\,m_p\,x_3\,x_4^{3} + 2\,\eta\,m_p\,x_3^{3}\,x_4\\
		&{}+8\,\eta\,m_{B^+}\,x_2\,x_4^{2} + 2\,\eta\,m_p\,x_2^{2}\,x_3\,x_4
	\end{aligned}\right)\nonumber\\[1pt]
	&-\,\phi_4^{+}\,1 / \eta\,\left(\begin{aligned}
		&{}-2\,m_p\,x_3^{2} + 2\,m_p\,x_3 - 20\,m_p\,x_2\,x_3\,x_4 - 8\,\eta\,m_{B^+}\,x_2\,x_4\\
		&{}-2\,\eta\,m_p\,x_3\,x_4 + 2\,\eta\,m_p\,x_3^{2}\,x_4 + 20\,\eta\,m_p\,x_2\,x_3\,x_4^{2}\\
		&{}+40\,\eta\,m_{B^+}\,x_2\,x_3\,x_4
	\end{aligned}\right)\nonumber\\[1pt]
	&-\,\psi_4^{0}\,\frac{2\,x_3}{\eta^{2}\,m_{B^+}}\,\left(\begin{aligned}
		&{}+2\,m_p^{2} - 2\,m_p^{2}\,x_3 - 2\,m_p^{2}\,x_4 + 2\,m_p^{2}\,x_2 + 2\,\eta^{2}\,m_{B^+}^{2}\\
		&{}+\eta\,m_{B^+}\,m_p - 2\,\eta\,m_p^{2}\,x_4 - 2\,\eta^{2}\,m_{B^+}^{2}\,x_3 - 2\,\eta^{2}\,m_{B^+}^{2}\,x_4\\
		&{}+2\,\eta\,m_p^{2}\,x_4^{2} - \eta\,m_{B^+}\,m_p\,x_3 - \eta^{2}\,m_{B^+}\,m_p\,x_4 - 2\,\eta\,m_p^{2}\,x_2\,x_4\\
		&{}+2\,\eta\,m_p^{2}\,x_3\,x_4 + \eta^{2}\,m_{B^+}\,m_p\,x_3\,x_4
	\end{aligned}\right)\nonumber\\[1pt]
	&-\,\psi_4^{-}\,\frac{2\,x_3}{\eta^{2}\,m_{B^+}}\,\left(\begin{aligned}
		&{}-2\,m_p^{2}\,x_3 - 2\,m_p^{2}\,x_4 + 2\,m_p^{2}\,x_2 + 2\,m_p^{2}\,x_2^{2} + 2\,m_p^{2}\,x_3^{2}\\
		&{}+2\,m_p^{2}\,x_4^{2} + \eta^{2}\,m_{B^+}^{2}\,x_2 + \eta^{2}\,m_{B^+}^{2}\,x_2^{2} + \eta^{2}\,m_{B^+}^{2}\,x_3^{2}\\
		&{}+\eta^{2}\,m_{B^+}^{2}\,x_4^{2} - \eta^{2}\,m_{B^+}^{2}\,x_3 - \eta^{2}\,m_{B^+}^{2}\,x_4 - 4\,m_p^{2}\,x_2\,x_3\\
		&{}-2\,\eta\,m_p^{2}\,x_4^{3} + 2\,\eta\,m_p^{2}\,x_4^{2} + 4\,m_p^{2}\,x_3\,x_4 + \eta\,m_{B^+}\,m_p\,x_3\\
		&{}+\eta^{2}\,m_{B^+}\,m_p\,x_4^{3} - \eta\,m_{B^+}\,m_p\,x_2^{2} - \eta\,m_{B^+}\,m_p\,x_3^{2}\\
		&{}-\eta\,m_{B^+}\,m_p\,x_4^{2} - 4\,\eta\,m_p^{2}\,x_3\,x_4^{2} - 2\,\eta\,m_p^{2}\,x_2\,x_4\\
		&{}-2\,\eta\,m_p^{2}\,x_2^{2}\,x_4 - 2\,\eta\,m_p^{2}\,x_3^{2}\,x_4 - 2\,\eta^{2}\,m_{B^+}^{2}\,x_2\,x_3\\
		&{}+2\,\eta\,m_p^{2}\,x_3\,x_4 + 2\,\eta^{2}\,m_{B^+}^{2}\,x_3\,x_4 + \eta^{2}\,m_{B^+}\,m_p\,x_2^{2}\,x_4\\
		&{}+\eta^{2}\,m_{B^+}\,m_p\,x_3^{2}\,x_4 - \eta^{2}\,m_{B^+}\,m_p\,x_3\,x_4 + 4\,\eta\,m_p^{2}\,x_2\,x_3\,x_4
	\end{aligned}\right)\nonumber\\[1pt]
	&-\,\psi_4^{+}\,\frac{2\,x_3}{\eta^{2}\,m_{B^+}}\,\left(\begin{aligned}
		&{}+2\,m_p^{2} - 2\,m_p^{2}\,x_3 - 2\,m_p^{2}\,x_4 + 2\,m_p^{2}\,x_2 + 2\,\eta^{2}\,m_{B^+}^{2}\\
		&{}+\eta\,m_{B^+}\,m_p - 20\,m_p^{2}\,x_2\,x_4 - 2\,\eta\,m_p^{2}\,x_4 - 2\,\eta^{2}\,m_{B^+}^{2}\,x_3\\
		&{}-2\,\eta^{2}\,m_{B^+}^{2}\,x_4 + 2\,\eta\,m_p^{2}\,x_4^{2} - \eta\,m_{B^+}\,m_p\,x_3 - \eta^{2}\,m_{B^+}\,m_p\,x_4\\
		&{}-10\,\eta^{2}\,m_{B^+}^{2}\,x_2\,x_4 - 2\,\eta\,m_p^{2}\,x_2\,x_4 + 2\,\eta\,m_p^{2}\,x_3\,x_4\\
		&{}+20\,\eta\,m_p^{2}\,x_2\,x_4^{2} + \eta^{2}\,m_{B^+}\,m_p\,x_3\,x_4 - 10\,\eta\,m_{B^+}\,m_p\,x_2\,x_4\\
		&{}+10\,\eta^{2}\,m_{B^+}\,m_p\,x_2\,x_4^{2}
	\end{aligned}\right)\nonumber\\[1pt]
	&-\,\xi_4^{0}\,2\,m_p\,\left(\eta\,x_2\,x_3\,x_4 - \eta\,x_3\,x_4^{2} - 2\,x_2\,x_3 - 8\,x_2\,x_4 + x_3^{2} + 2\,x_3\,x_4 - x_3\right) / \eta\nonumber\\[1pt]
	&+\,\xi_4^{-}\,2\,m_p\,x_3 / \eta\,\left(\begin{aligned}
		&{}+x_2^{2} + x_3^{2} + x_4^{2} - x_3 - 2\,x_4 + 2\,x_2 + \eta\,x_4^{2} - 4\,x_2\,x_3\\
		&{}+4\,x_3\,x_4 - \eta\,x_2\,x_4 - 2\,\eta\,x_3\,x_4^{2} + 2\,\eta\,x_2\,x_3\,x_4
	\end{aligned}\right)\nonumber\\[1pt]
	&-\,\xi_4^{+}\,2\,m_p / \eta\,\left(\begin{aligned}
		&{}+x_3^{2} - x_3 - 8\,x_2\,x_4 - 2\,x_2\,x_3 + 2\,x_3\,x_4 - \eta\,x_3\,x_4^{2}\\
		&{}+50\,x_2\,x_3\,x_4 + \eta\,x_2\,x_3\,x_4
	\end{aligned}\right)\nonumber\\[1pt]
	&+\,\phi_5^{0}\,\frac{m_p\,\left(2\,\eta\,m_{B^+}\,x_4 + \eta\,m_p\,x_3\,x_4 - m_p\,x_3\right)}{\eta^{2}\,m_{B^+}}\nonumber\\[1pt]
	&+\,\phi_5^{-}\,\frac{m_p}{\eta^{2}\,m_{B^+}}\,\left(\begin{aligned}
		&{}+\eta\,m_{B^+}\,x_3^{2} + m_p\,x_2\,x_3 - \eta\,m_{B^+}\,x_3 - m_p\,x_3\,x_4 + \eta\,m_p\,x_3\,x_4^{2}\\
		&{}+2\,\eta\,m_{B^+}\,x_2\,x_4 - \eta\,m_p\,x_2\,x_3\,x_4
	\end{aligned}\right)\nonumber\\[1pt]
	&-\,\phi_5^{+}\,\frac{m_p}{\eta^{2}\,m_{B^+}}\,\left(\begin{aligned}
		&{}+m_p\,x_3 - \eta\,m_{B^+} - 2\,m_p\,x_3^{2} + \eta\,m_{B^+}\,x_3 + 2\,\eta\,m_{B^+}\,x_2^{2}\\
		&{}+2\,\eta\,m_{B^+}\,x_4^{2} - \eta\,m_p\,x_3\,x_4 + 2\,\eta\,m_p\,x_3^{2}\,x_4
	\end{aligned}\right)\nonumber\\[1pt]
	&-\,\psi_5^{0}\,\frac{m_p\,\left(\eta\,m_{B^+}\,x_3 - \eta\,m_{B^+} + 2\,\eta\,m_p\,x_4^{2} - 2\,m_p\,x_4\right)}{\eta^{2}\,m_{B^+}}\nonumber\\[1pt]
	&-\,\psi_5^{-}\,\frac{m_p}{\eta^{2}\,m_{B^+}}\,\left(\begin{aligned}
		&{}+m_p\,x_3 - m_p\,x_3^{2} + \eta\,m_{B^+}\,x_3^{2} + m_p\,x_3\,x_4 - \eta\,m_{B^+}\,x_3\\
		&{}-m_p\,x_2\,x_3 - 2\,m_p\,x_2\,x_4 + \eta\,m_p\,x_3^{2}\,x_4 - \eta\,m_p\,x_3\,x_4\\
		&{}-\eta\,m_p\,x_3\,x_4^{2} + 2\,\eta\,m_{B^+}\,x_2\,x_4 + 2\,\eta\,m_p\,x_2\,x_4^{2} + \eta\,m_p\,x_2\,x_3\,x_4
	\end{aligned}\right)\nonumber\\[1pt]
	&-\,\psi_5^{+}\,\frac{m_p}{\eta^{2}\,m_{B^+}}\,\left(\begin{aligned}
		&{}+\eta\,m_{B^+} - 2\,m_p\,x_4 + 4\,m_p\,x_4^{2} - 4\,\eta\,m_{B^+}\,x_4 - 4\,\eta\,m_p\,x_4^{3}\\
		&{}-3\,\eta\,m_{B^+}\,x_3 + 2\,\eta\,m_{B^+}\,x_3^{2} + 2\,\eta\,m_p\,x_4^{2} + 4\,\eta\,m_{B^+}\,x_4^{2}\\
		&{}+4\,\eta\,m_{B^+}\,x_3\,x_4
	\end{aligned}\right)\nonumber\\[1pt]
	&+\,\xi_5^{0}\,2\,m_p\,x_3\,\left(\eta\,x_4 - 1\right) / \eta\nonumber\\[1pt]
	&+\,\xi_5^{-}\,m_p\,x_3\,\left(- x_2 + x_4\right) / \eta\nonumber\\[1pt]
	&-\,\xi_5^{+}\,m_p\,\left(2\,x_3 - 1\right)\,\left(2\,\eta\,x_3\,x_4 - x_2 - 2\,x_3 + x_4\right) / \eta\bigr]\nonumber\\[5pt]
	&-b_3b_4\,E(t_c)h_3(x_1,x_3,x_4,b_1,b_3,b_4)\bigl[\nonumber\\
	&+\,\phi_3^{0}\,\frac{80\,x_2\,x_3\,x_4\,\left(\eta^{2}\,m_{B^+}^{2} + \eta\,m_p^{2}\,x_4 - m_p^{2}\right)}{\eta^{2}\,m_{B^+}}\nonumber\\[1pt]
	&+\,\phi_3^{-}\,\frac{80\,x_2\,x_3\,x_4\,\left(- x_2 + x_4\right)\,\left(\eta^{2}\,m_{B^+}^{2} + \eta\,m_p^{2}\,x_4 - m_p^{2}\right)}{\eta^{2}\,m_{B^+}}\nonumber\\[1pt]
	&-\,\phi_3^{+}\,\frac{80\,x_2\,x_3\,x_4\,\left(3\,x_3 - 1\right)\,\left(\eta^{2}\,m_{B^+}^{2} + \eta\,m_p^{2}\,x_4 - m_p^{2}\right)}{\eta^{2}\,m_{B^+}}\nonumber\\[1pt]
	&-\,\phi_4^{0}\,\frac{16\,x_2\,x_4\,\left(\eta^{2}\,m_{B^+}^{2} + \eta\,m_p^{2}\,x_4 - m_p^{2}\right)}{\eta^{2}\,m_{B^+}}\nonumber\\[1pt]
	&-\,\phi_4^{-}\,\frac{16\,x_2\,x_4\,\left(- x_2 + x_4\right)\,\left(\eta^{2}\,m_{B^+}^{2} + \eta\,m_p^{2}\,x_4 - m_p^{2}\right)}{\eta^{2}\,m_{B^+}}\nonumber\\[1pt]
	&+\,\phi_4^{+}\,\frac{16\,x_2\,x_4\,\left(5\,x_3 - 1\right)\,\left(\eta^{2}\,m_{B^+}^{2} + \eta\,m_p^{2}\,x_4 - m_p^{2}\right)}{\eta^{2}\,m_{B^+}}\nonumber\\[1pt]
	&+\,\xi_4^{0}\,16\,m_p\,x_3\,\left(\eta\,x_4 - 2\right)\,\left(x_4-x_2\right) / \eta\nonumber\\[1pt]
	&-\,\xi_4^{-}\,16\,m_p\,x_3\,\left(2\,x_3 - 1\right)\,\left(\eta\,x_4 - 2\right)\,\left(x_4-x_2\right) / \eta\nonumber\\[1pt]
	&+\,\xi_4^{+}\,16\,m_p\,x_3\,\left(\eta\,x_4 - 2\right)\,\left(x_4-x_2\right) / \eta\nonumber\\[1pt]
	&-\,\phi_5^{0}\,\frac{4\,m_p}{\eta^{2}\,m_{B^+}}\,\left(\begin{aligned}
		&{}-m_p\,x_3 + 2\,\eta\,m_{B^+} - \eta^{2}\,m_{B^+}\,x_4 - 4\,\eta\,m_{B^+}\,x_4 - 2\,\eta\,m_{B^+}\,x_3\\
		&{}+2\,\eta^{2}\,m_{B^+}\,x_4^{2} + \eta\,m_p\,x_3\,x_4 + \eta^{2}\,m_{B^+}\,x_3\,x_4
	\end{aligned}\right)\nonumber\\[1pt]
	&-\,\phi_5^{-}\,\frac{4\,m_p^{2}\,x_3\,\left(- x_2 + x_4\right)\,\left(\eta\,x_4 - 1\right)}{\eta^{2}\,m_{B^+}}\nonumber\\[1pt]
	&+\,\phi_5^{+}\,\frac{4\,m_p^{2}\,x_3\,\left(2\,x_3 - 1\right)\,\left(\eta\,x_4 - 1\right)}{\eta^{2}\,m_{B^+}}\nonumber\\[1pt]
	&-\,\xi_5^{-}\,4\,m_p\,x_3\,\left(\eta\,x_4 - 2\right)\,\left(x_4-x_2\right) / \eta\nonumber\\[1pt]
	&+\,\xi_5^{+}\,4\,m_p\,\left(2\,x_3 - 1\right)\,\left(\eta\,x_4 - 2\right)\,\left(x_4-x_2\right) / \eta\nonumber\\[1pt]
	&+\,\phi_6^{0}\,\frac{4\,m_p^{2}\,\left(\eta\,x_4 - 1\right)}{3\,\eta^{2}\,m_{B^+}}\nonumber\\[1pt]
	&+\,\phi_6^{-}\,\frac{4\,m_p^{2}\,\left(- x_2 + x_4\right)\,\left(\eta\,x_4 - 1\right)}{3\,\eta^{2}\,m_{B^+}}\nonumber\\[1pt]
	&-\,\phi_6^{+}\,\frac{4\,m_p^{2}\,\left(3\,x_3 - 1\right)\,\left(\eta\,x_4 - 1\right)}{3\,\eta^{2}\,m_{B^+}}\bigr]\biggr).
\end{flalign}


\begin{references}
\bibitem{Elor:2018twp}
G.~Elor, M.~Escudero and A.~Nelson,
Phys. Rev. D \textbf{99} (2019) no.3, 035031
doi:10.1103/PhysRevD.99.035031
[arXiv:1810.00880 [hep-ph]].

\bibitem{Alonso-Alvarez:2021qfd}
G.~Alonso-{\'A}lvarez, G.~Elor and M.~Escudero,
Phys. Rev. D \textbf{104} (2021) no.3, 035028
doi:10.1103/PhysRevD.104.035028
[arXiv:2101.02706 [hep-ph]].

\bibitem{Elahi:2021jia}
F.~Elahi, G.~Elor and R.~McGehee,
Phys. Rev. D \textbf{105} (2022) no.5, 055024
doi:10.1103/PhysRevD.105.055024
[arXiv:2109.09751 [hep-ph]].

\bibitem{Nelson:2019fln}
A.~E.~Nelson and H.~Xiao,
Phys. Rev. D \textbf{100} (2019) no.7, 075002
doi:10.1103/PhysRevD.100.075002
[arXiv:1901.08141 [hep-ph]].

\bibitem{Miro:2024fid}
C.~Mir{\'o}, M.~Escudero and M.~Nebot,
Phys. Rev. D \textbf{110} (2024) no.11, 115033
doi:10.1103/PhysRevD.110.115033
[arXiv:2410.13936 [hep-ph]].

\bibitem{Zheng:2024tkj}
Y.~Zheng, J.~N.~Ding, D.~H.~Li, L.~Y.~Li, C.~D.~L{\"u} and F.~S.~Yu,
Chin. Phys. C \textbf{48} (2024) no.8, 083109
doi:10.1088/1674-1137/ad4afa
[arXiv:2404.04337 [hep-ph]].

\bibitem{Khodjamirian:2023wol}
A.~Khodjamirian, B.~Meli{\'c} and Y.~M.~Wang,
Eur. Phys. J. ST \textbf{233} (2024) no.2, 271-298
doi:10.1140/epjs/s11734-023-01046-6
[arXiv:2311.08700 [hep-ph]].

\bibitem{Lenz:2022rbq}
A.~Lenz, M.~L.~Piscopo and A.~V.~Rusov,
JHEP \textbf{01} (2023), 004
doi:10.1007/JHEP01(2023)004
[arXiv:2208.02643 [hep-ph]].

\bibitem{Azatov:2021irb}
A.~Azatov, M.~Vanvlasselaer and W.~Yin,
JHEP \textbf{10} (2021), 043
doi:10.1007/JHEP10(2021)043
[arXiv:2106.14913 [hep-ph]].

\bibitem{Bodeker:2020ghk}
D.~Bodeker and W.~Buchmuller,
Rev. Mod. Phys. \textbf{93} (2021) no.3, 3
doi:10.1103/RevModPhys.93.035004
[arXiv:2009.07294 [hep-ph]].

\bibitem{Alonso-Alvarez:2019fym}
G.~Alonso-{\'A}lvarez, G.~Elor, A.~E.~Nelson and H.~Xiao,
JHEP \textbf{03} (2020), 046
doi:10.1007/JHEP03(2020)046
[arXiv:1907.10612 [hep-ph]].

\bibitem{Bringmann:2018sbs}
T.~Bringmann, J.~M.~Cline and J.~M.~Cornell,
Phys. Rev. D \textbf{99} (2019) no.3, 035024
doi:10.1103/PhysRevD.99.035024
[arXiv:1810.08215 [hep-ph]].

\bibitem{Gu:2010ft}
P.~H.~Gu, M.~Lindner, U.~Sarkar and X.~Zhang,
Phys. Rev. D \textbf{83} (2011), 055008
doi:10.1103/PhysRevD.83.055008
[arXiv:1009.2690 [hep-ph]].

\bibitem{Ellis:1978xg}
J.~R.~Ellis, M.~K.~Gaillard and D.~V.~Nanopoulos,
Phys. Lett. B \textbf{80} (1979), 360
[erratum: Phys. Lett. B \textbf{82} (1979), 464]
doi:10.1016/0370-2693(79)91190-0

\bibitem{Riotto:1999yt}
A.~Riotto and M.~Trodden,
Ann. Rev. Nucl. Part. Sci. \textbf{49} (1999), 35-75
doi:10.1146/annurev.nucl.49.1.35
[arXiv:hep-ph/9901362 [hep-ph]].

\bibitem{BESIII:2025sfl}
M.~Ablikim \textit{et al.} [BESIII],
[arXiv:2505.22140 [hep-ex]].

\bibitem{Ai:2024nmn}
X.~Ai, W.~Altmannshofer, P.~Athron, X.~Bai, L.~Calibbi, L.~Cao, Y.~Che, C.~Chen, J.~Y.~Chen and L.~Chen, \textit{et al.}
Chin. Phys. C \textbf{49} (2025) no.10, 103003
doi:10.1088/1674-1137/adf1f0
[arXiv:2412.19743 [hep-ex]].

\bibitem{BaBar:2024qqx}
J.~P.~Lees \textit{et al.} [BaBar],
Phys. Rev. D \textbf{111} (2025) no.3, L031101
doi:10.1103/PhysRevD.111.L031101
[arXiv:2412.06950 [hep-ex]].

\bibitem{BaBar:2023dtq}
J.~P.~Lees \textit{et al.} [BaBar],
Phys. Rev. Lett. \textbf{131} (2023) no.20, 201801
doi:10.1103/PhysRevLett.131.201801
[arXiv:2306.08490 [hep-ex]].

\bibitem{BaBar:2023rer}
J.~P.~Lees \textit{et al.} [BaBar],
Phys. Rev. D \textbf{107} (2023) no.9, 092001
doi:10.1103/PhysRevD.107.092001
[arXiv:2302.00208 [hep-ex]].

\bibitem{Shi:2022rjn}
X.~Shi [BESIII],
PoS \textbf{EPS-HEP2021} (2022), 663
doi:10.22323/1.398.0663

\bibitem{Belle:2021gmc}
C.~Hadjivasiliou \textit{et al.} [Belle],
Phys. Rev. D \textbf{105} (2022) no.5, L051101
doi:10.1103/PhysRevD.105.L051101
[arXiv:2110.14086 [hep-ex]].

\bibitem{Khodjamirian:2022vta}
A.~Khodjamirian and M.~Wald,
Phys. Lett. B \textbf{834} (2022), 137434
doi:10.1016/j.physletb.2022.137434
[arXiv:2206.11601 [hep-ph]].

\bibitem{Boushmelev:2023huu}
A.~Boushmelev and M.~Wald,
Phys. Rev. D \textbf{109} (2024) no.5, 055049
doi:10.1103/PhysRevD.109.055049
[arXiv:2311.13482 [hep-ph]].

\bibitem{Elor:2022jxy}
G.~Elor and A.~W.~M.~Guerrera,
JHEP \textbf{02} (2023), 100
doi:10.1007/JHEP02(2023)100
[arXiv:2211.10553 [hep-ph]].

\bibitem{Biswas:2026oxq}
A.~Biswas, A.~Khodjamirian and A.~Mohamed,
[arXiv:2603.01723 [hep-ph]].

\bibitem{Shi:2023riy}
Y.~J.~Shi, Y.~Xing and Z.~P.~Xing,
Eur. Phys. J. C \textbf{83} (2023) no.8, 744
doi:10.1140/epjc/s10052-023-11930-z
[arXiv:2305.17622 [hep-ph]].

\bibitem{Shi:2024uqs}
Y.~J.~Shi, Y.~Xing and Z.~P.~Xing,
Eur. Phys. J. C \textbf{84} (2024) no.3, 306
doi:10.1140/epjc/s10052-024-12663-3
[arXiv:2401.14120 [hep-ph]].

\bibitem{Li:2024htn}
L.~Y.~Li, C.~D.~L{\"u}, J.~Wang and Y.~B.~Wei,
Phys. Rev. D \textbf{109} (2024) no.11, 116012
doi:10.1103/PhysRevD.109.116012
[arXiv:2401.11978 [hep-ph]].

\bibitem{Xing:2025pfw}
Y.~Xing, Y.~J.~Shi and X.~H.~Hu,
Phys. Rev. D \textbf{112} (2025) no.11, 116018
doi:10.1103/xr47-p5n2
[arXiv:2508.05181 [hep-ph]].


\bibitem{Xing:2019xti}
Y.~Xing and Z.~P.~Xing,
Chin. Phys. C \textbf{43} (2019) no.7, 073103
doi:10.1088/1674-1137/43/7/073103
[arXiv:1903.04255 [hep-ph]].

\bibitem{Han:2022srw}
J.~J.~Han, Y.~Li, H.~n.~Li, Y.~L.~Shen, Z.~J.~Xiao and F.~S.~Yu,
Eur. Phys. J. C \textbf{82} (2022) no.8, 686
doi:10.1140/epjc/s10052-022-10642-0
[arXiv:2202.04804 [hep-ph]].

\bibitem{He:2006ud}
X.~G.~He, T.~Li, X.~Q.~Li and Y.~M.~Wang,
Phys. Rev. D \textbf{74} (2006), 034026
doi:10.1103/PhysRevD.74.034026
[arXiv:hep-ph/0606025 [hep-ph]].

\bibitem{Li:2008tk}
R.~H.~Li, C.~D.~Lu, W.~Wang and X.~X.~Wang,
Phys. Rev. D \textbf{79} (2009), 014013
doi:10.1103/PhysRevD.79.014013
[arXiv:0811.2648 [hep-ph]].

\bibitem{Rui:2022sdc}
Z.~Rui, C.~Q.~Zhang, J.~M.~Li and M.~K.~Jia,
Phys. Rev. D \textbf{106} (2022) no.5, 053005
doi:10.1103/PhysRevD.106.053005
[arXiv:2206.04501 [hep-ph]].

\bibitem{Ou-Yang:2025ije}
J.~Ou-Yang, R.~H.~Li and S.~H.~Zhou,
Phys. Rev. D \textbf{112} (2025) no.5, 056005
doi:10.1103/vl97-y4ql
[arXiv:2506.14675 [hep-ph]].

\bibitem{Chang:2025kfa}
Q.~Chang, D.~H.~Yao and X.~Liu,
Eur. Phys. J. C \textbf{85} (2025) no.3, 292
doi:10.1140/epjc/s10052-025-13939-y
[arXiv:2501.01075 [hep-ph]].

\bibitem{Zhang:2024rmb}
C.~Q.~Zhang, J.~Sun, Z.~P.~Xing and R.~L.~Zhu,
Phys. Rev. D \textbf{111} (2025) no.11, 113003
doi:10.1103/6g56-v496
[arXiv:2501.00512 [hep-ph]].

\bibitem{Ren:2023ebq}
J.~L.~Ren, M.~Q.~Li, X.~Liu, Z.~T.~Zou, Y.~Li and Z.~J.~Xiao,
Eur. Phys. J. C \textbf{84} (2024) no.4, 358
doi:10.1140/epjc/s10052-024-12702-z
[arXiv:2311.16824 [hep-ph]].

\bibitem{Chai:2025xuz}
J.~Chai and S.~Cheng,
JHEP \textbf{06} (2025), 229
doi:10.1007/JHEP06(2025)229
[arXiv:2501.08783 [hep-ph]].

\bibitem{Li:1992nu}
H.~n.~Li and G.~F.~Sterman,
Nucl. Phys. B \textbf{381} (1992), 129-140
doi:10.1016/0550-3213(92)90643-P

\bibitem{Li:1994cka}
H.~n.~Li and H.~L.~Yu,
Phys. Rev. Lett. \textbf{74} (1995), 4388-4391
doi:10.1103/PhysRevLett.74.4388
[arXiv:hep-ph/9409313 [hep-ph]].

\bibitem{Li:2012cfa}
H.~n.~Li, C.~D.~Lu and F.~S.~Yu,
Phys. Rev. D \textbf{86} (2012), 036012
doi:10.1103/PhysRevD.86.036012
[arXiv:1203.3120 [hep-ph]].

\bibitem{Savage:1989ub}
M.~J.~Savage and M.~B.~Wise,
Phys. Rev. D \textbf{39} (1989), 3346
[erratum: Phys. Rev. D \textbf{40} (1989), 3127]
doi:10.1103/PhysRevD.39.3346

\bibitem{Chiang:2004nm}
C.~W.~Chiang, M.~Gronau, J.~L.~Rosner and D.~A.~Suprun,
Phys. Rev. D \textbf{70} (2004), 034020
doi:10.1103/PhysRevD.70.034020
[arXiv:hep-ph/0404073 [hep-ph]].

\bibitem{Li:2023kcl}
N.~Li, Y.~Xing and X.~H.~Hu,
Eur. Phys. J. C \textbf{83} (2023) no.11, 1013
doi:10.1140/epjc/s10052-023-12188-1
[arXiv:2303.08008 [hep-ph]].

\bibitem{Xing:2019wil}
Y.~Xing,
Eur. Phys. J. C \textbf{80} (2020) no.1, 57
doi:10.1140/epjc/s10052-020-7625-3
[arXiv:1910.11593 [hep-ph]].

\bibitem{Shi:2017dto}
Y.~J.~Shi, W.~Wang, Y.~Xing and J.~Xu,
Eur. Phys. J. C \textbf{78} (2018) no.1, 56
doi:10.1140/epjc/s10052-018-5532-7
[arXiv:1712.03830 [hep-ph]].


\bibitem{Abri:2026rqj}
M.~A.~Abri, N.~Hajirasouliha and K.~Azizi,
[arXiv:2605.13701 [hep-ph]].

\bibitem{King:1986wi}
I.~D.~King and C.~T.~Sachrajda,
Nucl. Phys. B \textbf{279} (1987), 785-803
doi:10.1016/0550-3213(87)90019-8

\bibitem{Braun:2000kw}
V.~Braun, R.~J.~Fries, N.~Mahnke and E.~Stein,
Nucl. Phys. B \textbf{589} (2000), 381-409
[erratum: Nucl. Phys. B \textbf{607} (2001), 433-433]
doi:10.1016/S0550-3213(00)00516-2
[arXiv:hep-ph/0007279 [hep-ph]].

\bibitem{RQCD:2019hps}
G.~S.~Bali \textit{et al.} [RQCD],
Eur. Phys. J. A \textbf{55} (2019) no.7, 116
doi:10.1140/epja/i2019-12803-6
[arXiv:1903.12590 [hep-lat]].

\bibitem{Liu:2013bxa}
Y.~L.~Liu, C.~Y.~Cui and M.~Q.~Huang,
Phys. Rev. D \textbf{89} (2014) no.3, 035005
doi:10.1103/PhysRevD.89.035005
[arXiv:1311.5960 [hep-ph]].

\bibitem{Ali:2007ff}
A.~Ali, G.~Kramer, Y.~Li, C.~D.~Lu, Y.~L.~Shen, W.~Wang and Y.~M.~Wang,
Phys. Rev. D \textbf{76} (2007), 074018
doi:10.1103/PhysRevD.76.074018
[arXiv:hep-ph/0703162 [hep-ph]].

\bibitem{Belle:2026kum}
M.~Abumusabh \textit{et al.} [Belle and Belle-II],
Phys. Rev. Lett. \textbf{136} (2026) no.23, 231802
doi:10.1103/1rl4-j3np
[arXiv:2601.07104 [hep-ex]].


\end{references}
\end{document}